\documentclass{article}
\pdfoutput=1

\usepackage{arxiv}

\usepackage[utf8]{inputenc} 
\usepackage[T1]{fontenc}    
\usepackage{hyperref}       
\usepackage{url}            
\usepackage{booktabs}       
\usepackage{amsfonts}       
\usepackage{nicefrac}       
\usepackage{microtype}      
\usepackage{lipsum}
\usepackage{graphicx}
\usepackage{amssymb}
\usepackage{lineno}
\usepackage{amsmath,amssymb}
\usepackage{xcolor}
\usepackage{mathrsfs}
\usepackage{float}
\usepackage[flushleft]{threeparttable}
\usepackage{caption}
\graphicspath{{Figures/}}
\usepackage{subcaption}

\title{Finite element solution of nonlocal Cahn-Hilliard equations with feedback control time step size adaptivity}

\author{
  Gabriel F. Barros\\
  COPPE\\
  Federal University of Rio de Janeiro\\
   Rio de Janeiro, RJ 21941-598 Brazil \\
  \texttt{gabriel.barros@nacad.ufrj.br} \\
   \And
 Adriano M. A. Cortes \\
 NUMPEX-COMP\\
  Federal University of Rio de Janeiro, Brazil\\
  Duque de Caxias, RJ 25240-005 Brazil \\
  \texttt{adriano@nacad.ufrj.br} \\
\And
 Alvaro L. G. A. Coutinho \\
 COPPE\\
 Federal University of Rio de Janeiro\\
 Rio de Janeiro, RJ 21941-598 Brazil \\
  \texttt{alvaro@nacad.ufrj.br} \\
}

\begin{document}
\maketitle
\begin{abstract}
In this study, we evaluate the performance of feedback control-based time step adaptivity schemes for the nonlocal Cahn-Hilliard equation derived from the Ohta-Kawasaki free energy functional. The temporal adaptivity scheme is recast under the linear feedback control theory equipped with an error estimation that extrapolates the solution obtained from an energy-stable, fully implicit time marching scheme. We test three time step controllers with different properties: a simple Integral controller, a complete Proportional-Integral-Derivative controller, and the PC11 predictive controller. We assess the performance of the adaptive schemes for the nonlocal Cahn-Hilliard equation in terms of the number of time steps required for the complete simulation and the computational effort measured by the required number of nonlinear and linear solver iterations. We also present numerical evidence of mass conservation and free energy decay for simulations with the three different time step controllers. The PC11 predictive controller is the best in all three-dimensional test cases.
\end{abstract}


\section{Introduction}
\label{S:1}

\par 
The Cahn-Hilliard equation (or simply the CH equation) was derived in 1958 to model the phase separation in binary alloys \cite{CH1,CH2}.
Since then the CH equation appears in several different physical contexts \cite{Kim} such as diblock copolymers \cite{ohtakawasaki,choksi}, image inpainting \cite{bertozzi}, binary fluid flow \cite{hohenberg,gurtin,lowengrubtruskinovsky,abels2012,adriano}, fracture propagation \cite{hughesfracture,duda}, tumour growth \cite{wise,wu} and topology optimization \cite{zhou}, to mention a few. The CH equation is
\begin{equation}
\label{eq:ch}
\frac{\partial \phi}{\partial t} = \nabla \cdot \bigg[ M(\phi)\nabla\bigg(\frac{\partial \Psi}{\partial \phi} - \epsilon^2\Delta\phi\bigg)\bigg],
\end{equation}
where $\phi(\mathbf{x},t)$ is a field evolving in time and space, $M(\phi) > 0$ is the mobility, related to the diffusion process, $\Psi(\phi)$ is the bulk free energy, and $\epsilon^2$ is a parameter related to the interfacial energy. From a strict mathematical point of view, considering no-flux or periodic boundary conditions, the equation can also be seen as the gradient flow \cite{shin} of the Ginzburg-Landau functional described by
\begin{equation}
\label{eq:ginzburg}
F(\phi) = \int_\Omega \bigg(\Psi(\phi) + \dfrac{\epsilon^2}{2}|\nabla\phi|^2\bigg) d\Omega.
\end{equation}

\par The derivation of (\ref{eq:ch}) considers only short-ranged microforces. Therefore, the use of the CH equation becomes restricted to model physical phenomena where only local interactions of particles are taken into account. The development of a nonlocal Cahn-Hilliard equation (NCH for short)  \cite{giacomin1,giacomin2, Gajewski2003} fills this gap through the derivation of a phase-separation model that also considers long-range interactions. Among the various NCH applications, we highlight the modeling of the diblock copolymer self-assembly. Diblock copolymers are a specific class of copolymers, in which two chemically distinct monomer units are grouped in discrete blocks along the polymer chain \cite{copbook}. The large variety of morphologies obtained in the resulting self-assembled copolymer by manipulating different molecular parameters, added to the crescent research interest in nanotechnology, lead to the development of block copolymers related discoveries in advanced materials, drug delivery, patterning, porous materials, and many others over the last decades \cite{kimcop}. Since we are interested in the patterns formed by the solution of the NCH equation, one of the key questions in the diblock copolymer context that remains open is if it is possible to find which pattern minimizes the energy over all possible patterns for a given set of parameters.  The Ohta-Kawasaki  (O--K) free energy functional models pattern morphologies via energy minimization involving the competition of both short and long-range microforces. In \cite{Alberti2008, Ren2003, ohnsh}, we find analytical solutions in one dimension for a global minimizer for the O--K free energy functional. However, they are of limited use. Numerical simulations of the NCH equation for different parameter sets became fundamental in exploring the associated phase diagrams for diblock copolymer melts \cite{choksi, Cristoferi2018, vanderberg, Choksi2011, jeong2}. However, the use of these numerical simulations is nontrivial and can present difficulties related to the stiffness of the equations and demand large computational power \cite{Farrell2017}. In this sense, the development of computational techniques that improve the performance and accuracy of the computational modeling of the NCH equation is an active topic of research.
\par 
This paper evaluates time step adaptive schemes for the NCH equation in the linear feedback control theory context with proper error estimation and time step controller. Several works presented adaptive time-stepping using error estimation and time step controllers in the CH equation \cite{cueto,gomezhughes,vignals,wodo,stogner}, the use of more sophisticated controllers and error estimation techniques are not widely addressed in the nonlocal case. In this study, we use three different time-step controllers with different properties and behavior, together with a proper error estimation method that prevents the calculation of the same time step multiple times. We evaluate the controllers in terms of performance in different examples in terms of required time steps for the completion of the simulations and the total number of nonlinear and linear iterations required by the solvers. We also assess the employment of an implicit time-marching scheme, mathematically and numerically proven to be energy-stable in standard phase-field functionals. Our results reveal numerical evidence of mass conservation and free energy decay for the nonlocal case as anticipated theoretically in \cite{choksi, Parsons, Gal2018}. 
The paper is structured as follows. Section \ref{S:2} describes the NCH formulation. Numerical and computational implementation details are shown in Section \ref{S:3}. Section \ref{S:4} introduces the temporal adaptivity schemes, and in Section \ref{S:5}, we show several numerical examples for solving the CH/NCH equations with adaptive time step control. The paper ends with the conclusions we drew from the comparison of the three controllers in different contexts.

\section{Governing equations}
\label{S:2}
\par 
We consider the following nonlocal extension of the standard CH equation, that is, the NCH equation,
\begin{equation}
\label{eq:nch}
\dfrac{\partial \phi}{\partial t} = \nabla \cdot \bigg[ M(\phi)\nabla\bigg(\frac{\partial \Psi}{\partial \phi} - \epsilon^2\Delta\phi\bigg)\bigg] - \sigma(\phi - \bar{\phi}),
\end{equation}
where $\sigma$ represents the nonlocal parameter, responsible for modeling the magnitude of the long-range microforces between the phases. The parameter $\bar{\phi}$ is the mean value of $\phi$ in the domain  $\Omega\in\mathrm{R}^{n_{sd}}$ with boundary $\partial \Omega$ and $n_{sd} = 2,3$, such that,
\begin{equation}
	\bar{\phi} = \dfrac{1}{|\Omega|} \int_{\Omega}\phi d\Omega.
\end{equation}
\par 
One way of deriving Eq. (\ref{eq:nch}) is under the $\mathcal{H}^{-1}(\Omega)$ gradient flow from the  O--K free energy functional \cite{Li2018}. The  O--K functional can be written \cite{vanderberg,choksi} as,
\begin{equation}
\label{eq:okfunc}
    F(\phi) = \int_\Omega \bigg(\Psi(\phi) + \dfrac{\epsilon^2}{2}|\nabla\phi|^2 \bigg)d\Omega  + \int_\Omega \dfrac{\sigma}{2} |\nabla v|^2 d\Omega,
\end{equation}
where $v$ is related to $\phi$ via the boundary value problem $-\Delta v = (\phi - \bar{\phi})$.

\par The O--K free energy functional is derived from the mean field theory in the context of diblock copolymers \cite{ohtakawasaki}. In the case where $\sigma = 0$, the O--K functional (\ref{eq:okfunc}) becomes the Ginzburg-Landau free energy functional (\ref{eq:ginzburg}). Consequently, in this case, the NCH equation (\ref{eq:nch}) becomes the CH equation (\ref{eq:ch}). Both CH \cite{wodo} and NCH \cite{Choksi2008} equations minimize the interfaces between phases, solving the isoperimetric problem. The difference, however, lies in the fact that the Ginzburg-Landau free energy functional is minimized through the separation of phases due to short-range microforces while the O--K free energy functional models pattern morphologies via energy minimization involving the competition of both short and long-range microforces, where the latter is modeled by the magnitude of the nonlocal parameter $\sigma$. The competition between local and nonlocal microforces in the O--K free energy functional leads to many different pattern formations in the equilibrium configuration of a copolymer melt, such as lamellae, spheres, gyroids, and cylinders \cite{copbook}. It is possible to map a given copolymer structure to a set of NCH parameters, such as $\epsilon$, $\bar{\phi}$, and $\sigma$ by a phase diagram \cite{choksi, vanderberg, Choksi2011}.  
\par 
Equation (\ref{eq:nch}) is solved on a bounded domain $\Omega$ with Lipschitz-continuous boundaries $\partial\Omega$, and on the time interval $[0,T]$ with prescribed initial conditions $\phi(\mathbf{x},0) = \phi_0$. Regarding the boundary conditions the usual ones are the no-flux boundary conditions, that is, $\nabla\phi \cdot \mathbf{n} = 0$ and $ M(\phi)\nabla (\frac{\partial \Psi}{\partial \phi} - \epsilon^2\nabla^2\phi) \cdot \boldsymbol{n} = 0$, and the periodic boundary conditions. The use of these boundary conditions implies on mass conservation and free energy decay for both CH \cite{elliott3} and NCH \cite{Parsons, Gal2018} equations. 
In the past decades, several studies investigated numerical strategies to solve the CH equation. Considering spatial discretization, there are models based on finite differences \cite{eyre}, finite volumes \cite{cueto}, finite elements \cite{elliot,elliot2,elliott3} and spectral methods \cite{he}, all with their advantages and drawbacks. In the finite element method, the presence of fourth-order spatial derivatives in the CH equation (in the strong form) requires the use of $C^1$-continuous elements in the primal variational formulation of the equation. Stogner et al. \cite{stogner} presented a finite element formulation using $C^1$-continuous elements in two dimensions for rectangular grids, while other studies circumvented this situation with different techniques such as NURBS-based isogeometric analysis, \cite{gomezhughes}, variable splitting technique (mixed formulation) \cite{elliot2} and discontinuous Galerkin methods \cite{wells}. 
The CH equation temporal discretization is also nontrivial. Besides being a stiff and nonlinear equation, which makes it practically unsolvable by explicit methods \cite{eyre}, the time integration method must generally obey an energy decay property since, in most cases, e.g., considering no-flux or periodic boundary conditions, the free energy functional is a Lyapunov functional. Many studies developed different time integration methods for the CH equation preserving this property \cite{vignals,gomezhughes}. In terms of numerical strategies explicitly designed for the NCH equation, the literature is not as rich as the case of the  CH equation. We highlight the development of preconditioners \cite{Li2018, Li2020, Farrell2017}, computational implementation details, and theoretical proofs regarding stability, boundedness, and mass conservation using the finite element method \cite{Parsons}.
\par 
Several strategies have been proposed to increase the accuracy and performance of the simulations. For example, the spatial discretization of the CH equation must be fine enough to consider the smooth transition of the interface that arises between different phases \cite{wodo} while the bulk domain does not require fine meshes. Therefore, it is common to track the interface areas to refine the mesh while coarsening the bulk domain \cite{cueto}. When considering temporal discretization, some physical phenomena described by the CH equation require small time step sizes to capture fast dynamics. However, there are stages where the dynamics are slow, and consequently, the use of larger time steps is allowed. The use of smaller time step sizes in these stages is translated into unnecessary computational costs. Nevertheless, an adaptive time-stepping strategy can help determine whether or not the time step size can be enlarged or reduced. The same problems also appear in the nonlocal case and, although several studies present strategies to circumvent these difficulties in the CH context, this issue is not widely addressed in the NCH literature. The present paper contributes to fill this gap.

\section{Numerical Methodology}
\label{S:3}

In this study, the finite element method is employed to discretize in space the NCH equation. We use for temporal integration, an implicit, second-order, unconditionally energy-stable method originally proposed for the CH equation and other traditional phase-field equations \cite{vignals}. This method enables the use of larger time steps obtained by time adaptivity without affecting the numerical stability of the simulations. All the numerical solutions are computed using the \verb|FEniCS| framework version 2019.1.0 \cite{fenicsbook,fenics2}, a high-performance finite element library written in Python/C++.

\subsection{Spatial discretization}
\par 
The formulation of the NCH equation contains a biharmonic operator. Thus standard $C^0$-continuous finite elements are not suitable for its primal variational formulation. Nevertheless, a splitting strategy, also called mixed formulation, is employed to avoid the continuity constraint and enable the use of $C^0$-continuous elements to approximate the solution of the CH \cite{elliot2} and the NCH \cite{Parsons, Farrell2017} equations, converting the nonlinear equation into a coupled nonlinear system, with two degrees of freedom per node. 
\par
The split form can be achieved by introducing the chemical potential $\mu$ as an unknown field. Given a spatial domain $\Omega\in\mathrm{R}^{n_{sd}}$ with boundaries $\partial \Omega$ and $n_{sd} = 2,3$ and the primal NCH equation on (\ref{eq:ch}), its split version is given by
\begin{flalign}
    \label{eq:split_strong}
    \frac{\partial \phi}{\partial t} &= \nabla \cdot (M(\phi)\nabla \mu) - \sigma(\phi - \bar{\phi}), \\
    \mu &= \frac{\partial \Psi}{\partial \phi} - \epsilon^2 \nabla^2\phi.  \nonumber
\end{flalign}

\par 
The weak form of the system can be obtained by integrating both equations (\ref{eq:split_strong}) in their strong form against weighting functions $q,w \in H^1(\Omega)$, where $H^1(\Omega)$ is the Sobolev space of the square integrable functions with an integrable first weak derivative, and applying the divergence theorem. 
The Galerkin method approximates the unknown fields through functions in a finite dimension space. Considering a partition of the form $\Omega = \bigcup_e \Omega^e$, and being $P^k(\Omega^e)$ the space of polynomials of degree equal or less than $k$ over $\Omega^e$, the function spaces are defined as: 
\begin{flalign}
    \label{eq:spaces}
    S_t^h &= \{\phi^h(\cdot,t),\mu^h(\cdot,t) \in H^1(\Omega) ~|~ \phi^h(\cdot,t)|_{\Omega_e},\mu^h(\cdot,t)|_{\Omega_e}\in P^k(\Omega^e), \forall e\}, \\
    W^h &= \{w^h,q^h\in H^1(\Omega) ~|~ w^h|_{\Omega_e},q^h|_{\Omega_e}\in P^k(\Omega^e), \forall e\}.
\end{flalign}

For a standard finite element discretization, the semi-discrete finite element formulation for the NCH nonlinear system is: Given $\phi(\mathbf{x},0) = \phi^h_0(\mathbf{x})$, find $\phi^h(t), \mu^h(t) \in S_t^h, \forall w^h, q^h \in W^h$, so that:
\begin{flalign}
    \label{eq:split_weak}
    \bigg( w^h, \frac{\partial \phi^h}{\partial t}\bigg) + (\nabla w^h, M(\phi^h)\nabla \mu^h) + (w^h,\sigma \phi^h) - (w^h,\sigma \bar{\phi^h}) &= 0,\\
     (q^h,\mu^h) - \bigg( q^h, \frac{\partial \Psi}{\partial \phi}\bigg) - (\nabla q^h,\epsilon^2\nabla \phi^h) &= 0, \nonumber
\end{flalign}
where $(\cdot,\cdot)$ is the $L^2$ inner product. After splitting the NCH equation, the chemical potential $\mu$ becomes another solvable field. Periodic boundary conditions as well as no-flux boundary conditions are considered.

\subsection{Temporal integration}
\par
The NCH equation is a time-dependent equation, so a proper time integration method must be chosen. It is essential to use an energy stable integration method since both the Ginzburg Landau and the Ohta-Kawasaki free energy functionals are Lyapunov functionals when no-flux or periodic boundary conditions are applied to the domain \cite{gomezzee, choksi, Gal2018}. 

The choice of a time integration method for the CH/NCH equations is not a trivial task. Explicit methods are often prohibitive due to severe restrictions on the time step size, which is around $\mathcal{O}(\Delta x^4)$, arising from the stiffness of the equations. Fully implicit methods, because of the nonlinear nature of the CH/NCH equations, require nonlinear solvers, increasing memory and computational requirements. Implicit methods allow larger time steps, but if the time step is too large, the nonlinear discrete systems emanating from the CH equation can present multiple solutions \cite{eyre,gomezzee}. An intermediate approach is provided by semi-implicit time-stepping algorithms, where some terms are implicitly treated while others remain explicit. An important example of this family of methods is the convex-concave additive decomposition of the free energy introduced by Eyre~\cite{eyre97} for general gradient flows, particularly for the CH equation. In this case, since the gradient term is quadratic, it contributes to the convex part of the decomposition, the challenge being the decomposition of $\Psi$, the bulk free energy, generally a non-convex function like a double-well. The additive decomposition of $\Psi$ is done by splitting concave and convex terms of the functional such that the convex terms are treated implicitly, and the concave terms can remain explicit. However, the additive decomposition of $\Psi$ is not unique for all the possible functions of $\Psi$ \cite{gomezzee}. The splitting method proposed in \cite{eyre97} yields an unconditionally energy-stable method and is unconditionally uniquely solvable, although the proved local truncation error is just second-order accurate in the timestep size, rendering a first-order accurate in time method. Some studies developed second-order convex-splitting time integration schemes for the CH equation \cite{calo2020, gomezhughes}. However, unconditionally-energy stability and unconditionally uniquely solvability properties are not yet achieved for a general form of $\Psi$, within the free energy decay original approach or without the use of numerical stabilization \cite{gomezzee}. The development of second-order, unconditionally-energy stable, and unconditionally uniquely solvable integration methods is still an active research topic in the present context.
\par 
Vignal et al.~\cite{vignals} introduced another approach to derive second-order unconditionally energy-stable time integrators for phase-field models. The main idea is to use Taylor's series expansions of the bulk free energy function $\Psi$ to derive an approximation for $\Psi' = \frac{\partial\Psi}{\partial\phi}$. This method is mathematically proven unconditionally energy-stable regardless of mesh and time step size and second-order accurate in time for quartic potentials. Although there is no proof that this method is unconditionally uniquely solvable, it works well with adaptive time-stepping \cite{vignals}. Due to its simplicity and desirable properties, this is the method chosen in this study. Applying Vignal et al.~\cite{vignals} time integration method to the semi-discrete variational formulation of the problem given in equation (\ref{eq:split_weak}), where the subindex $n$ is the time step number and considering the initial conditions $\phi(\mathbf{x},0) = \phi_0$, the fully discrete system is as follows:

\begin{flalign}
    \left( w^h,\dfrac{ [\![\phi]\!]}{\Delta t_{n+1}} \right)_{\Omega} + (\nabla w^h, M(\phi^h)\nabla \mu^h_{n+1})_{\Omega}  + (w^h,\sigma \{\phi \}  )_{\Omega} - (w^h,\sigma \bar{\phi^h})_{\Omega}  &= 0, \\
    \label{eq:chtime}
    (q^h,\mu^h_{n+1})_{\Omega} - (q^h,\tilde{\Psi}^{'})_{\Omega} - (\nabla q^h,  \epsilon^2\nabla\{\phi\})_{\Omega} &= 0, 
\end{flalign}
 
\noindent where $[\![\phi]\!] = \phi^h_{n+1} - \phi^h_{n}$, $\Delta t_{n+1} = t_{n+1} - t_{n}$, $\{\phi\} = \dfrac{\phi^h_{n+1} + \phi^h_{n}}{2}$,
and the approximation $\tilde{\Psi}^{'}$ is defined as

\begin{equation}
    \label{eq:tildepsi}
    \tilde{\Psi}^{'} = \dfrac{\partial \Psi_{n+1}}{\partial \phi} - \dfrac{\partial^2 \Psi_{n+1}}{\partial \phi^2} \dfrac{[\![\phi]\!]}{2} + \dfrac{\partial^3 \Psi_{n+1}}{\partial \phi^3} \dfrac{[\![\phi]\!]^2}{6}.
\end{equation}

By being fully implicit, this scheme suits the proposed adaptive time-stepping strategy, allowing larger time steps without compromising the stability of the method. It is essential to mention that the Taylor expansion of $\Psi$ presented in Eq. (\ref{eq:tildepsi}) has guaranteed convergence for quartic potentials representing the homogeneous free energy function. In this study, we apply the time integration scheme for the NCH equation derived from the O--K functional. Despite being mathematically designed for the CH equation, the employment of the described temporal integration method on the NCH equation holds the properties of free energy decay and mass conservation, as seen in Section \ref{S:5}.

\section{Temporal adaptivity}
\label{S:4}
The choice of a proper time integration method for the NCH equations is a difficult task, since, in several physical situations, the equation has different time scales, creating a tradeoff between accuracy and performance. For instance, the initial stage of the phase segregation of a mixture is dictated by fast dynamics, requiring small step sizes while the latter stages reveal slow dynamics, allowing large time steps. Thus, to improve the efficiency of the computations, a time adaptivity scheme is often used to automatically change the time step size to capture both fast and slow dynamics of the equations, as well as the nonlocal dynamics inherent to the NCH equation, improving the performance of the simulations without any accuracy loss. \par 
Studies in the literature discuss time adaptivity schemes for the CH equation. Some schemes rely on the evaluation of the  Ginzburg-Landau free energy \cite{guillen,zhang}, requiring, however, the tuning of very sensitive empirical parameters \cite{zhang}. Different approaches are seen in \cite{gomezhughes, wodo, cueto, vignals, calo2020, stogner} are based on simple time-step controllers and present more robustness and better results. In this study, we assess different controllers applied to the NCH equation. 

\par 
\subsection{The control theory on adaptive time-stepping}
The NCH equation can be expressed as a dynamical system of the form:
\begin{flalign}
    \label{eq:flow}
    \begin{split}
   	\dot{\phi} = F(\phi), \\  \phi(0) = \phi_0,
    \end{split}
\end{flalign}
where $\phi \in \mathrm{R}^{n_{sd}}$ and $F:\mathrm{R}^{n_{sd}} \rightarrow \mathrm{R}^{n_{sd}}$ is a Lipschitz map. Since the time integration method used in this study is a one-step method, considering a step size $\Delta t$, there is a map $\Phi:\mathrm{R}^{n_{sd}} \rightarrow \mathrm{R}^{n_{sd}}$ such that:
\begin{flalign}
    \label{eq:map1}
    \begin{split}
    \phi_{n+1} &= \Phi(\phi_n), \\  \phi(0) &= \phi_0.
    \end{split}
\end{flalign}
\par 
Equation (\ref{eq:map1}) is a discrete-time dynamical system that approximates Equation (\ref{eq:flow}). It is possible to use the same approach for an additional map $\Xi:\mathrm{R} \rightarrow \mathrm{R}$ to vary the step size:
\begin{equation}
    \label{eq:map2}
    \Delta t_{n+1} = \Xi(\Delta t_n).
\end{equation}
where $\Delta t_{n+1} = t_{n+1} - t_n$ and $\Delta t_{n} = t_{n} - t_{n-1}$. 
\par 
The map $\Xi$ uses information about the numerical solution $\phi_n$ when defining the new step size ($\Delta t_{n+1}$) while the map $\Phi$ is based on the time step $\Delta t_{n+1}$. An adaptive time–stepping method can be expressed as the following recursions:
\begin{flalign}
    \label{eq:ic1}
    \begin{split}
    \phi_{n+1} &= \Phi(\phi_n),\\
    \Delta t_{n+1} &= \Xi(\Delta t_n).
    \end{split}
\end{flalign}
\par 
We assume that the relation between the error and the step size is asymptotic, that is:
\begin{equation}
    \label{eq:asymp}
    r_n = |\zeta_n| \Delta t^\kappa_n
\end{equation}
where $r_n$ is the norm of the local error estimate, $|\zeta_n|$ is the norm of the principal error function, and $\kappa$ is related to the order of the method. In our case, the principal error function can be viewed as a disturbance in the system, such as a Newton solver residual, and the integration method is second-order accurate, so $\kappa = 2$. It can be seen that $r_n \rightarrow 0$ if $\Delta t \rightarrow 0$. \par
\begin{figure}
	\centering
	\includegraphics[scale=0.5]{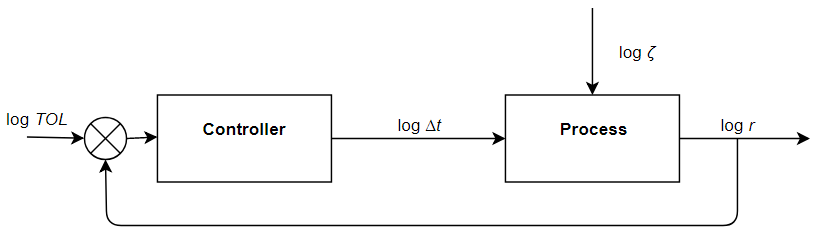}
	\caption{Adaptive time-stepping viewed as a feedback control system. Adapted from \cite{soderlind2}.}
	\label{fig:closedloop}
\end{figure}

 
The idea behind the use of control theory on adaptive time-stepping is that the map $\Xi$ controls an estimated numerical error within a prescribed tolerance, $TOL$. The mathematical background containing the detailed description of the use of control theory in temporal adaptivity in ODEs is given in  \cite{soderlind1, soderlind2, soderlind3}.  The recursion can be translated into a closed-loop, a  common dynamic structure in Control Theory, as seen in Figure \ref{fig:closedloop}. In this sense, the use of a linear feedback controller on step size adaptivity is translated in: given a present time step $t_n$, the controller defines a future time step $\Delta t_{n+1}$ such that the local error of the present time step $r_n$ is controlled within a given tolerance $TOL$ by a controller whose properties and tuning are defined in the mapping $\Xi$. The time step is evaluated within the process (which is solving the NCH equation in that given time) and feeding the error $r_{n+1}$ to the controller, restarting the loop. If the estimated error is not within the prescribed tolerance, the controller reevaluates a new, smaller time step size until this condition is satisfied. Besides the expectancy of reducing the number of linear/nonlinear systems to be solved, Control Theory provides smoother step size sequences (which improves the solution regarding smoothness \cite{soderlind1}), improved computational stability, and a regular, tight tolerance proportionality.

\subsection{Error estimation}
\par 
The literature presents various methods to estimate the temporal error on ordinary differential equations (ODEs). The interested reader can see in \cite{hairer, hairer2}. The usual strategy to estimate the temporal error is to evaluate an ODE solution at a given time step with integration methods of a different order of accuracy and compute the norm of the difference of the solutions relative to the norm of the solution obtained by the higher-order method. This strategy is seen in the CH context in \cite{gomezhughes, wodo, cueto} and it can be mathematically written as,
\begin{equation}
r = \dfrac{||\phi_{n+1} - \hat{\phi}_{n+1}||}{||\phi_{n+1}||},
\end{equation} 
where $\phi_{n+1}$ is obtained through an integration method of a higher order than the solution obtained in $\hat{\phi}_{n+1}$. Although this is considered a common error estimator, the calculation of the time step $n + 1$ twice is required to obtain $\phi_{n+1}$ and $\hat{\phi}_{n+1}$. In \cite{stogner}, $\phi_{n+1}$ is obtained by taking two time steps of size $\Delta t_{n+1} / 2$ while $\hat{\phi}_{n+1}$ is obtained by considering the time step $\Delta t$. In this case, the authors use the same integration method for both solutions, but each time step needs to be computed three times.

\par 
A different approach is seen in  \cite{vignals}. Consider an error estimation where the solutions at $t_n$ and $t_{n-1}$ are stored and the error is estimated \textit{a posteriori} by extrapolation by a lower-order time integration method, since the solution $t_{n+1}$ is obtained with a second-order scheme. This estimation is done through variable step-size backward differentiation, where the error obtained in the lower-order method is controlled. In the present work, we consider the lower-order method to be the backward-Euler method. Therefore, the local truncation error of the backward-Euler method is:
\begin{equation}
    \label{eq:taube}
    \tau^{BE}(t_{n+1}) = -\dfrac{\Delta t^2}{2}\phi^{''}(t_{n+1}) + {O}(\Delta t^3).
\end{equation}
Given the stored solutions $\phi_{n+1}$, $\phi_n$ and $\phi_{n-1}$ at times $t_{n+1}$, $t_n$ and $t_{n-1}$ respectively and neglecting the effects of the ${O}(\Delta t^3)$ terms, equation (\ref{eq:taube}) can be approximated by the variable step-size backward difference formula. So the error estimation is now:
\begin{equation}
    \label{eq:error}
    E_{n+1} = -\dfrac{1}{\eta}\mathbf{\phi}_{n+1} + \dfrac{1}{\eta - 1}\mathbf{\phi}_{n} - \dfrac{1}{\eta(\eta - 1)}\mathbf{\phi}_{n-1}
\end{equation}
where $\eta = (\Delta t_{n+1} + \Delta t_{n})/\Delta t_{n+1}$.
\par 
With the error function, the weighted local truncation error (WLTE) can be written as:
\begin{equation}
    r = \sqrt{\dfrac{1}{n_{nodes}} \sum_{i=1}^{n_{nodes}} \bigg( \dfrac{E_{n+1}^{i}}{\tau^{abs} + \tau^{rel} \text{max}(|\phi_{n+1}^{i}|,|\phi_{n+1}^{i} + E_{(n+1)}^{i}|)}\bigg)^2}
\end{equation}
where $\tau^{abs}$ and $\tau^{rel}$ define tunable absolute and relative tolerances, respectively, and the index $i = 1, 2, ... n_{nodes}$ refers to the nodal index.
\par 
The weighted local truncation error $r$ is used to control the error at each time step. By definition, values of $r\leq1$ mean that the local truncation error is within the user-prescribed tolerances. In this case, the step just taken can be accepted, and the time integration can move forward with either the same or a larger time step size. On the contrary, values of WLTE larger than one imply unacceptable errors. That said, the step taken is then rejected and retaken with a smaller time step size. We also constrain the time step such that $\Delta  t_{n+1} \in [\Delta  t_{min}, \Delta  t_{max}]$, that is, $\Delta  t_{n+1}=max(\Delta  t_{n+1}, \Delta  t_{min})$ and $\Delta  t_{n+1}=min(\Delta  t_{n+1}, \Delta  t_{max})$, where $\Delta  t_{min}$ and $\Delta  t_{max}$  are user supplied parameters.\par 

\subsection{Timestep controllers}
There are several timestep controllers in the literature and many ways to classify them \cite{soderlind2}. In this study, we consider three controllers: an integral controller, a PID controller, and the PC11 predictive controller. The three controllers have unique properties and have been used in the context of time adaptivity of PDEs such as the Navier-Stokes \cite{valli}  and the convection-diffusion equation \cite{ahmed}. The integral controller is the simplest and controls the relationship between the error in the present and past time. This simplicity is known to grant the integral controller a large number of rejected steps \cite{cueto}.  The PID controller has three controlling terms - proportional, integral, and derivative - that adjust the time step to changes in the estimated errors of the last three time steps. The predictive controller PC11 is suggested for time adaptivity in stiff equations \cite{soderlind1}, which is the case for the CH/NCH equations and has a different structure compared to the other two controllers. The three controllers can be written as:
\begin{equation}
    \Delta t_{n+1} = \rho \bigg(\frac{r_n}{r_{n+1}}\bigg)^{\kappa_P}  \bigg(\frac{1}{r_{n+1}}\bigg)^{\kappa_I} \bigg(\frac{r^2_n}{r_{n+1}r_{n-1}}\bigg)^{\kappa_D} \bigg(\frac{\Delta t_n}{\Delta t_{n-1}}\bigg)^{\kappa_T} \Delta t_n
\end{equation}
where the parameter $\rho$ is a safety factor used to smooth the time step growth. In the literature, it is common to see $\rho = 0.9$, although in \cite{calo2020} other values for $\rho$ and tolerances for other phase-field computations were proven better \cite{calo2020}. Here, we adopt $\rho=0.9$ unless stated otherwise. We evaluate these parameters for a nonlocal case in the numerical validation section and investigate the accuracy and performance results. The parameters $\kappa_P$, $\kappa_I$, $\kappa_D$ and $\kappa_T$ for each controller are shown in Table \ref{table:controle}. To avoid tuning the controller parameters, which can be very time consuming, the parameters for the I controller \cite{soderlind1}, the PID controller \cite{valli} and the PC11 controller \cite{ahmed} are taken from the literature.

\begin{table}
    \centering
    \caption{Controllers parameters.}
    \begin{tabular}{|c | c| c |c| c|}
        \hline
        Controller      & $\kappa_P$   & $\kappa_I$           & $\kappa_D$         &      $\kappa_T$\\
        \hline
        Integral (I)    & $0.0$    &     $0.5$      &     $0.0$     &      $0.0$\\
        PID             & $0.075$  &  $0.175$       &  $0.01$       &      $0.0$\\
        PC11            & $0.333$  &     $0.333$    &      $0.0$    &      $1.0$\\
        \hline
    \end{tabular}
    \label{table:controle}
\end{table}
\vspace{1cm} 

\par
\textbf{\textit{Remark:}} Although the use of time adaptivity schemes based on the linear feedback control theory has not yet been explicitly mentioned in the CH equation literature, the integral controller has been used by others \cite{cueto,gomezhughes,vignals,wodo, calo2020}. Even when the PID controller is used for CH equation \cite{cueto}, the integration method for the PID error estimation is not guaranteed to be energy stable. The use of a predictive controller for the NCH equation is unprecedented. Moreover, the error estimation is based on solution norms of time integration methods with different accuracy, requiring the calculation of the same step twice and, therefore, can be time-consuming. In the present work, we employ an error estimation method based on extrapolation that avoids computing the same time step twice.

\section{Numerical Experiments}
\label{S:5}
\subsection{Local Cahn-Hilliard simulation of spinodal decomposition}

Numerical simulations are made to validate the time adaptivity strategy. Initially, we consider a case with no-flux boundary conditions, $\bar{\phi} = 0.3$ and $\sigma = 0$, that is, a standard CH simulation of spinodal decomposition. We consider four simulations: one for each controller, and a fixed time step simulation. We compare the frames in all simulations to check if they all represent the same physical stages. We consider a square domain with a $129^2$ nodes. The square domain is divided into $128^2$ cells, each cell discretized by two linear triangles. For our simulations, we consider $\tau^{abs} = \tau^{rel} = 10^{-4}$, $\Psi(\phi) = \frac{1}{4}(\phi^2 - 1)^2$, $M(\phi) = 1$  and the initial time step $\Delta t_0 = 1 \times 10^{-9}$. We also constrain the time step size to the limits $\Delta t_{min} =  1 \times 10^{-12}$ and $\Delta t_{max} =  5 \times 10^{-3}$. Preliminary simulations showed that when the simulation approaches the steady-state, the controller allows time step sizes of $\mathcal{O}(10^{-2})$. Although accepted by the controllers, the use of time step sizes of this magnitude increases the number of linear iterations significantly in the later stages, increasing the computational cost. We also note that the number of rejected steps for the three controllers are significantly reduced, suggesting that limiting the time step improves the controllers' behavior. The \verb|FEniCS| framework v2019.1.0 invokes several linear algebra backend packages to solve the linear and nonlinear systems arising from the finite element method. In our case, we choose the PETSc package, inheriting Newton's method to solve the nonlinear systems with a relative tolerance $\eta_{NL}=10^{-5}$ and the GMRES solver with Block-Jacobi ILU(0) preconditioner with relative tolerance $\eta_r=10^{-5}$ and absolute tolerance $\eta_a=10^{-8}$ for the linear systems. 
\begin{figure}
	\centering
	\includegraphics[width=0.90\linewidth]{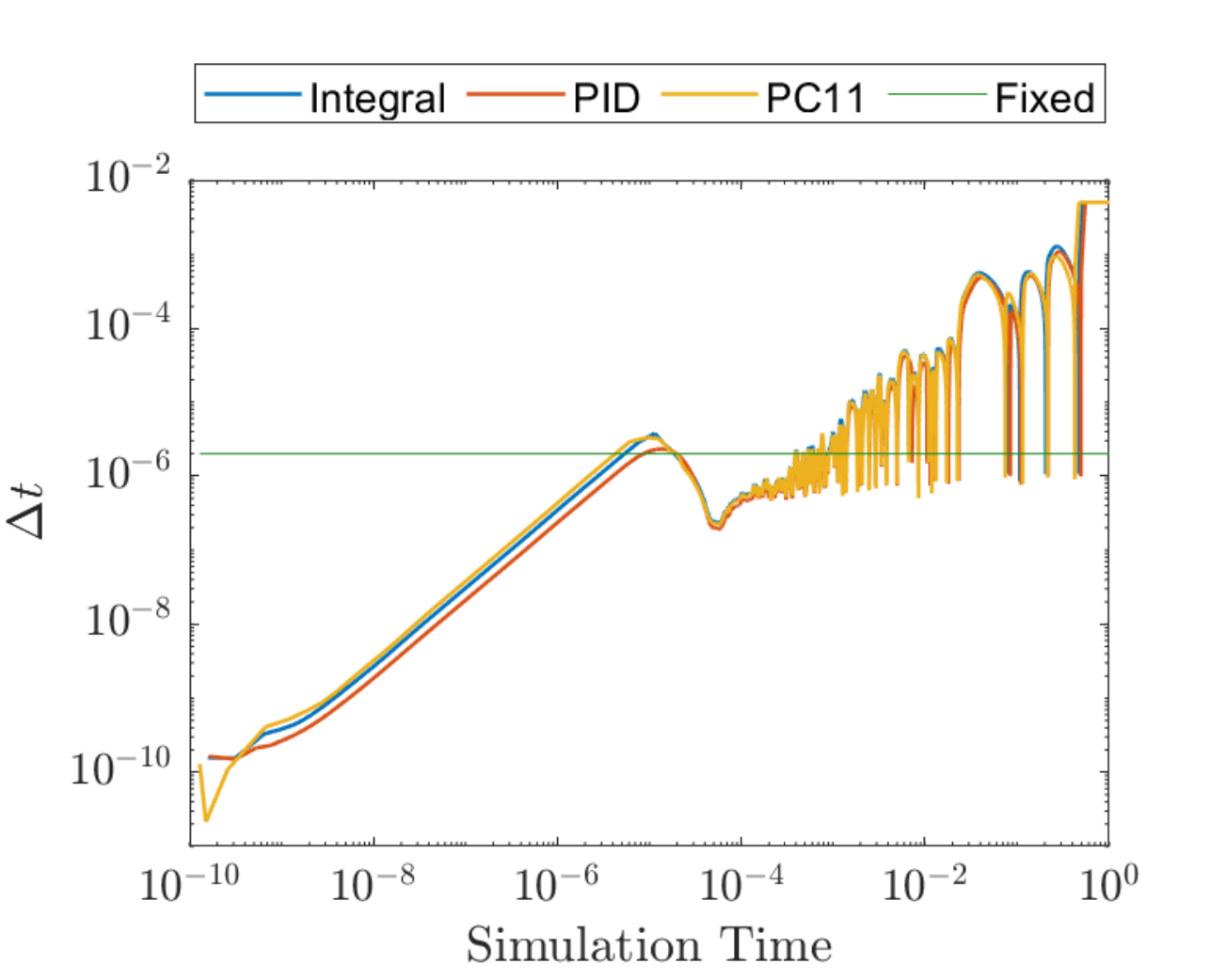}
    \caption{Time history of the time step $\Delta t$ during the 2D spinodal decomposition simulation.}
    \label{fig:sd_2d_dt}
\end{figure}

\par
\begin{figure}
	\centering
	\includegraphics[width=0.90\linewidth]{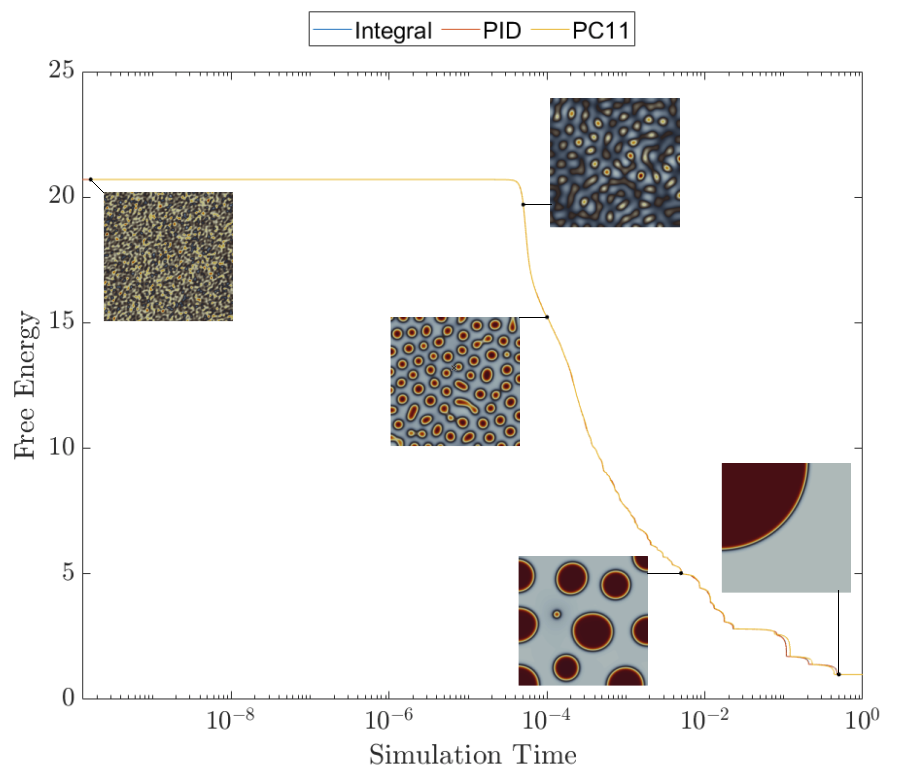}
	\caption{Free energy for the three simulations and some snapshots describing phase separation.}
	\label{fig:freeenergy2d}
\end{figure}

 
Figure \ref{fig:sd_2d_dt} shows the time step history for the four simulations. We can see in this figure the multiscale nature of the CH equation in the spinodal decomposition. In the initial stage where the phases are being defined, The time step increases by orders of magnitude and then decays (from $t = 10^{-10}$ to $t = 10^{-4}$, approximately). This strong variation occurs due to the rapid dynamics in the early stages of the spinodal decomposition. Consequently, the controller produces smaller time steps to keep the estimated error within the prescribed tolerance. The intermediate stage begins when all bubbles are approximate of the same size, leading to simultaneous  Ostwald's ripening events in the domain, preventing the time step from growing. The controller keeps the oscillations in this interval (from $t = 10^{-5}$ to $t = 10^{-3}$) to capture the bubble shrinkage. The final stages are where the time step has larger values, indicating slow dynamics involving surface motion. The final stage, however, still requires small time steps when an Ostwald ripening is occurring, but the controller allows the time step size to grow by several orders of magnitude since no rapid dynamics are seen in this stage when there is no shrinkage. A few simulation snapshots at the different stages and the free energy decay for the three controllers are seen in Fig. \ref{fig:freeenergy2d}.

\begin{figure}
	\centering
	\begin{minipage}{.5\textwidth}
		\centering
		\includegraphics[trim={25.00cm 0.00cm 25.00cm 0.00cm},clip,width=.85\linewidth]{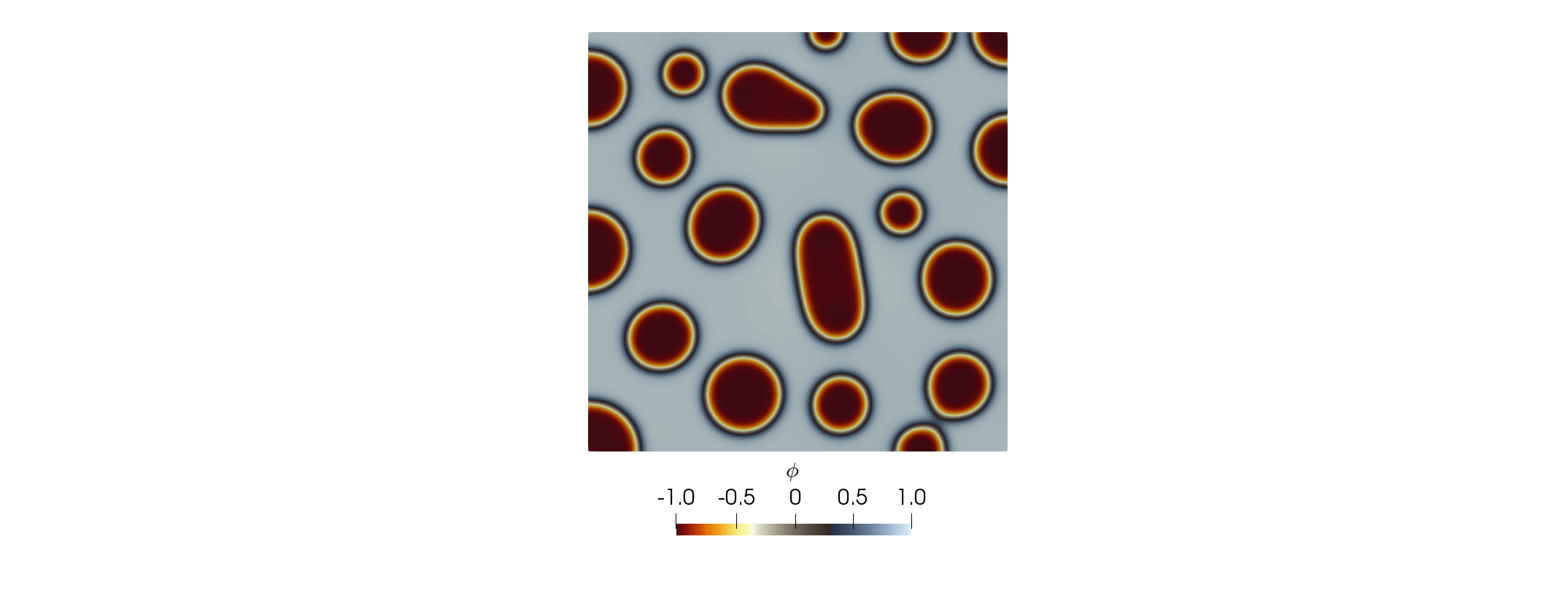}
		\subcaption{Fixed time step size.}
		\label{fig:fixed}
	\end{minipage}%
	\begin{minipage}{.5\textwidth}
		\centering
		\includegraphics[trim={25.00cm 0.00cm 25.00cm 0.00cm},clip,width=.85\linewidth]{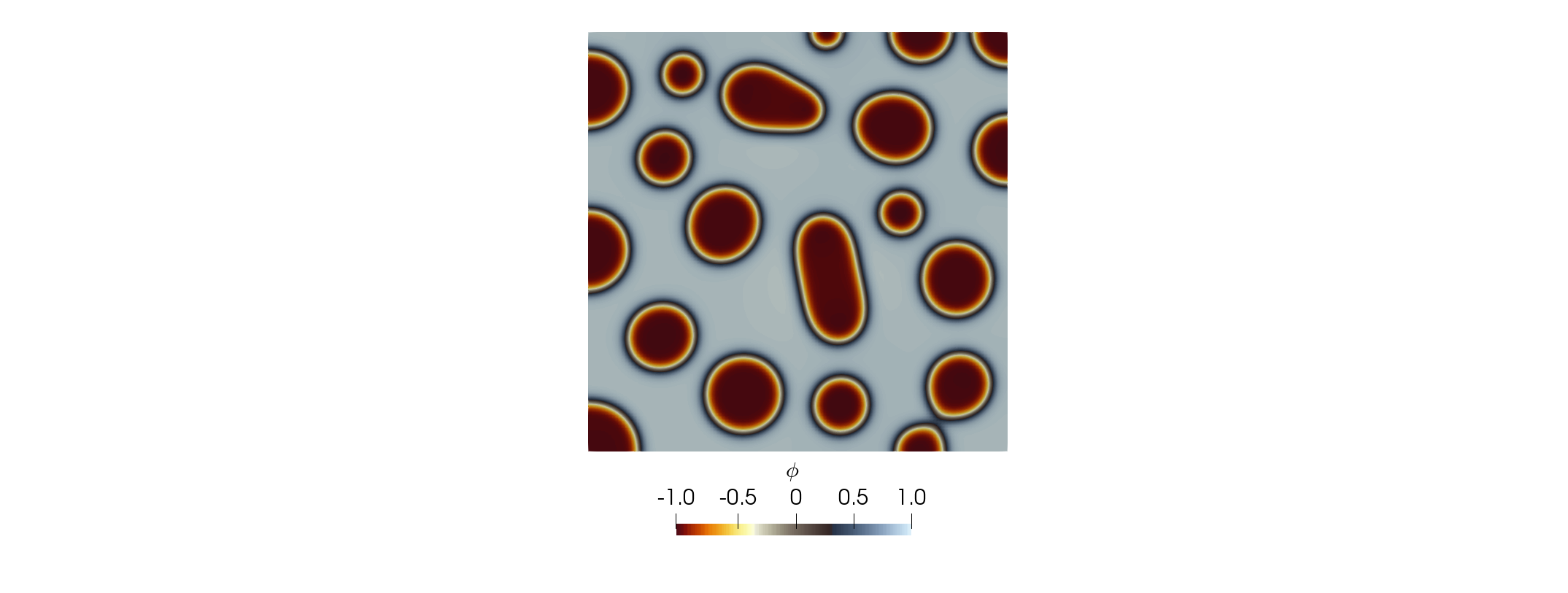}
		\subcaption{Integral controller.}
		\label{fig:integral}
	\end{minipage}\\
	\begin{minipage}{.5\textwidth}
		\centering
		\includegraphics[trim={25.00cm 0.00cm 25.00cm 0.00cm},clip,width=.85\linewidth]{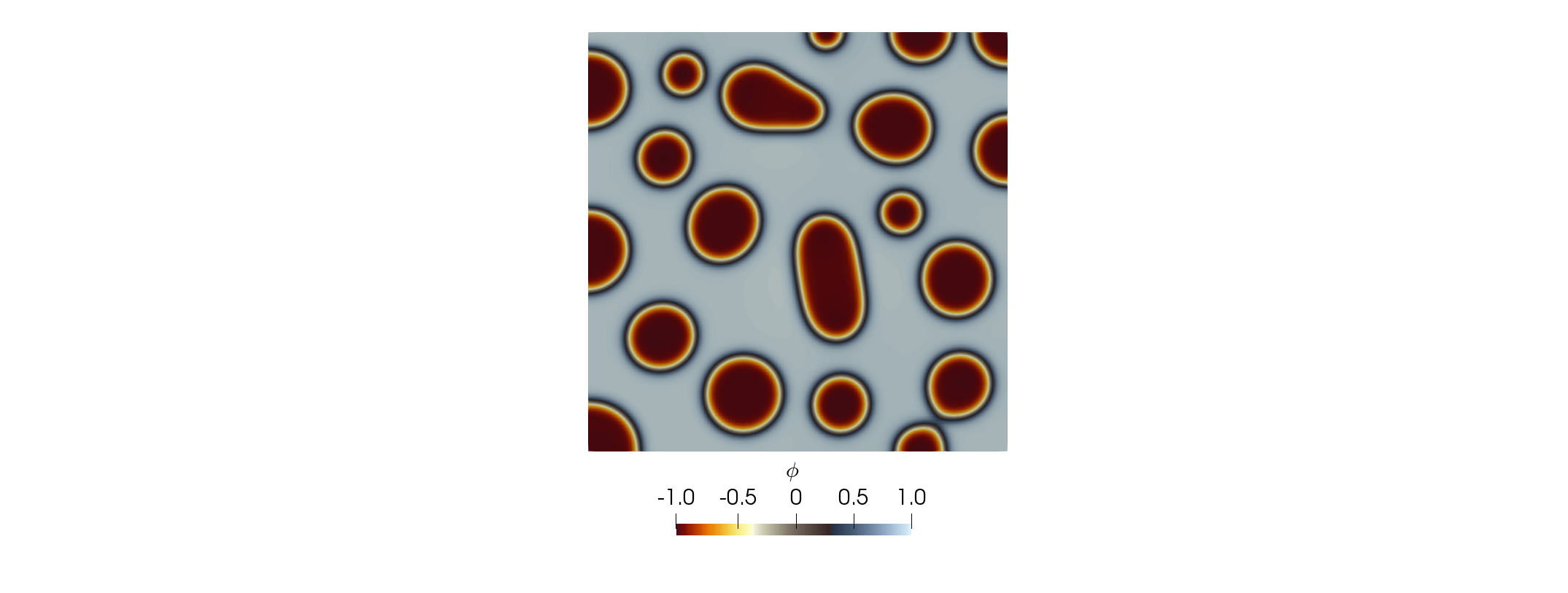}
		\subcaption{PID controller.}
		\label{fig:pid}
	\end{minipage}%
	\begin{minipage}{.5\textwidth}
		\centering
		\includegraphics[trim={25.00cm 0.00cm 25.00cm 0.00cm},clip,width=.85\linewidth]{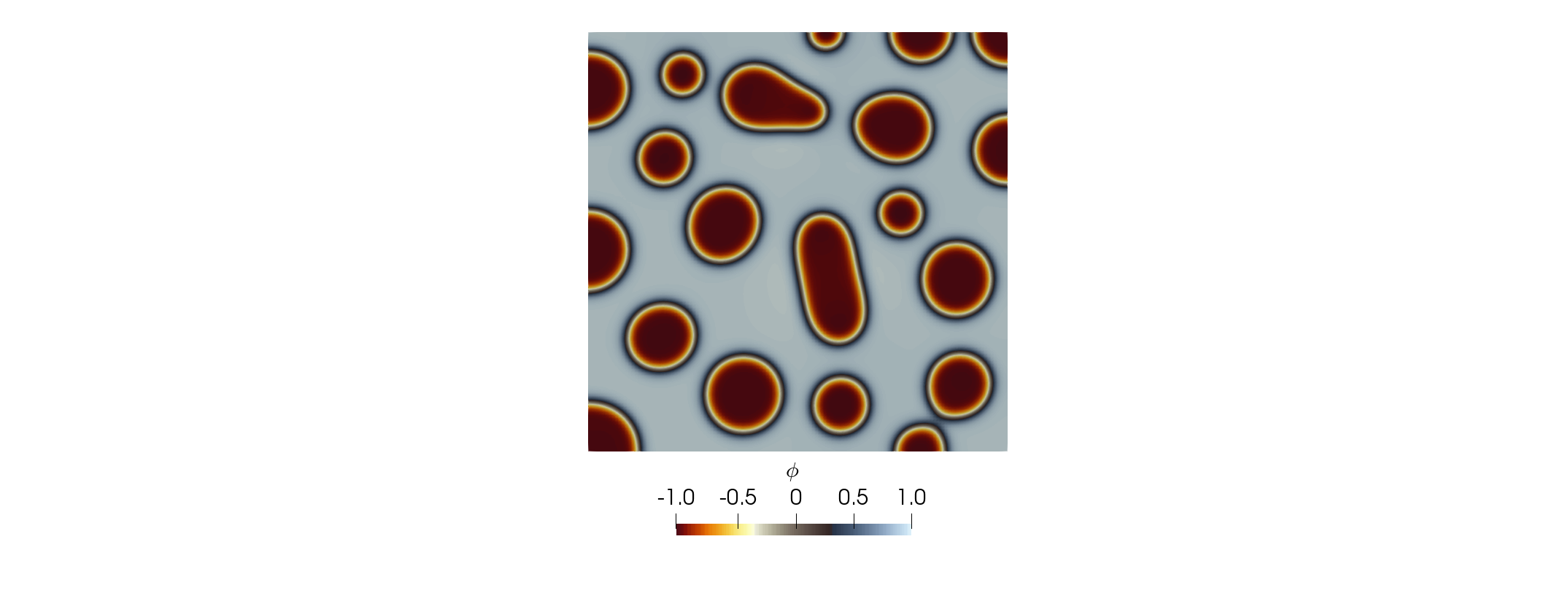}
		\subcaption{PC11 controller.}
		\label{fig:pc11}
	\end{minipage}\\
	\caption{Comparison of the four simulations at $t = 0.001 \pm 0.0000003$.}
	\label{fig:simultaneous}
\end{figure}

We now evaluate the use of time step controllers in terms of physical accuracy and performance. In terms of physical accuracy, we observe from both Figs \ref{fig:sd_2d_dt} and \ref{fig:freeenergy2d} that the three curves regarding the adaptive time stepping simulations are practically overlapping in terms of free energy and time step size. The overlapping is a strong indicator that the simulations are practically identical, meaning that the results are physically consistent for all three controllers. Figure \ref{fig:simultaneous} presents the results for all the simulations at  $t = 0.001 \pm 0.0000003$, showing that all the controllers yield the same solution.

\begin{figure}
	\centering
		\centering
		\includegraphics[width=.95\linewidth]{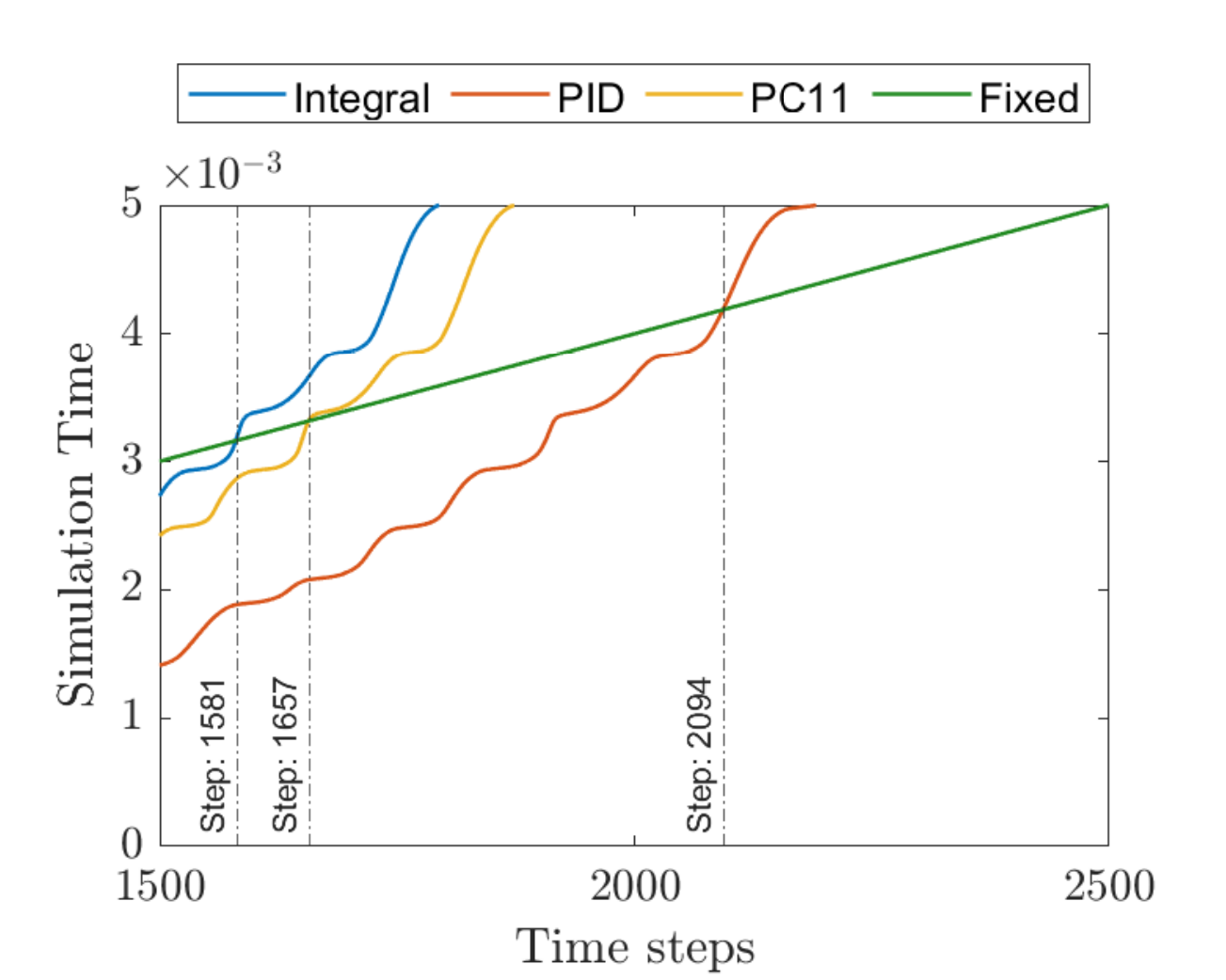}
	\caption{Time steps required for each controller to reach the same physical stage as the fixed time step simulation.}
	\label{fig:dt_step}
\end{figure}

\begin{figure}
	\centering
	\begin{minipage}{.49\textwidth}
		\centering
		\includegraphics[width=.95\linewidth]{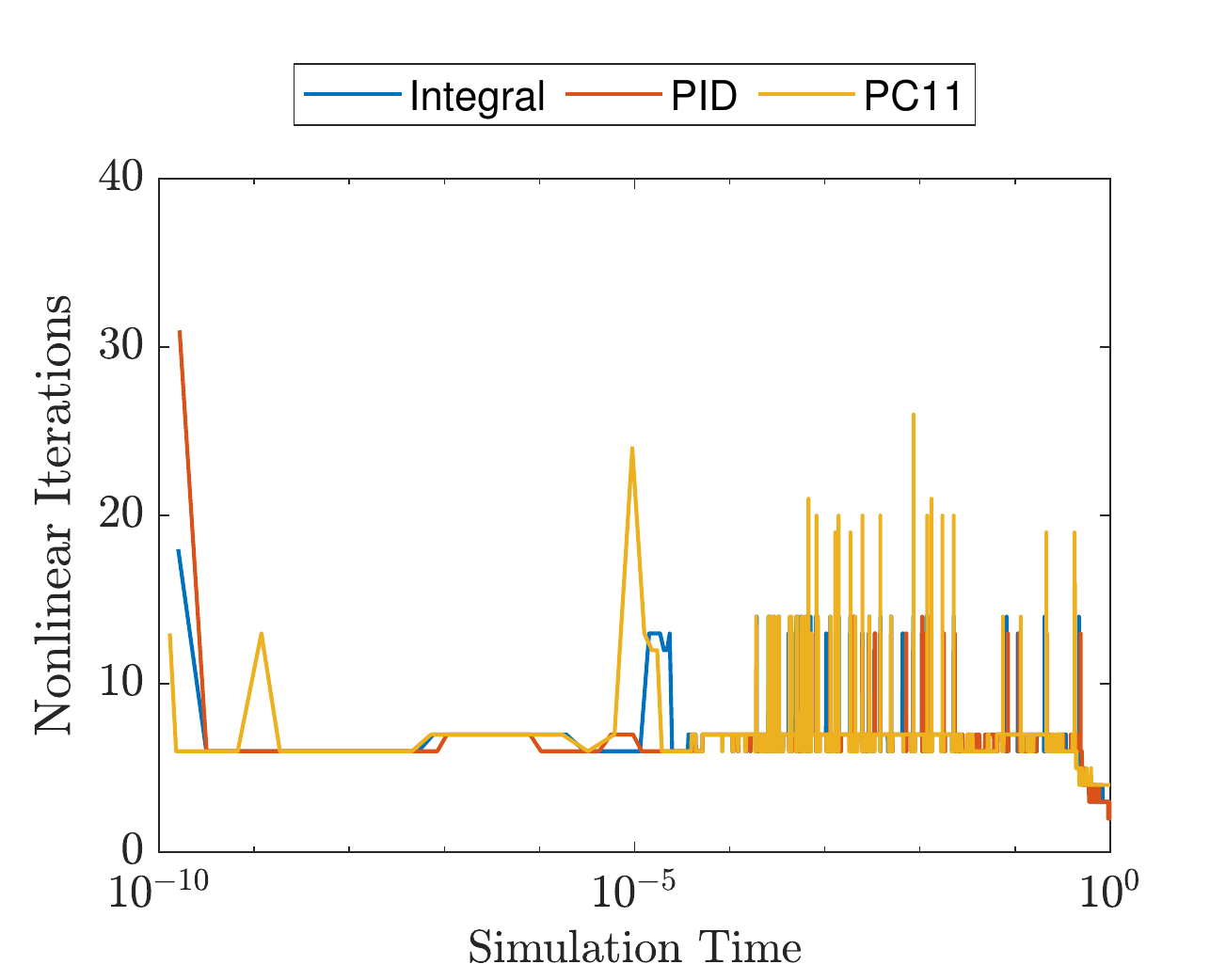}
		\label{fig:nitlocal}
	\end{minipage}
	\begin{minipage}{.49\textwidth}
		\centering
		\includegraphics[width=.95\linewidth]{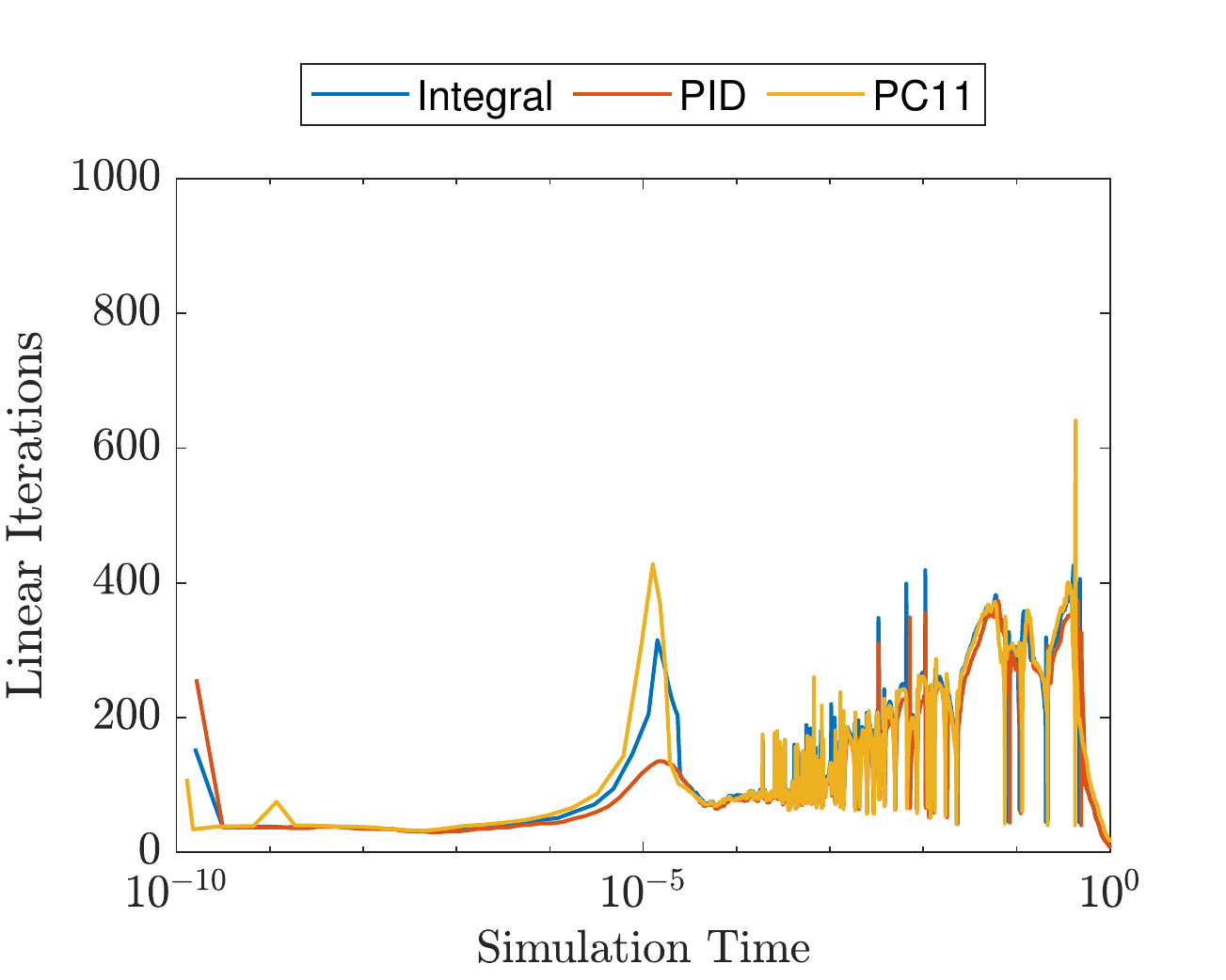}
		\label{fig:litlocal}
	\end{minipage}%
	\caption{Number of nonlinear (left) and linear iterations (right) for each time step during the adaptive simulations. Rejected steps included. Solver tolerances: $\eta_{NL}=10^{-5}$, $\eta_r=10^{-5}$, $\eta_a=10^{-8}$}.
	\label{fig:nitlit}
\end{figure}

We note that the early stages of the spinodal decomposition demand a smaller time step size than the maximum allowed for the fixed time step simulation. This fact means that the early stages for the fixed time step simulation are obtained with a smaller amount of time steps than the adaptive simulations. However, this is valid only to the initial simulation stage. Afterward, adaptive simulations become more efficient. Figure \ref{fig:dt_step} shows how many time steps are needed for each method to reach the point where the fixed time step is no longer more efficient. We observe that the PID demands more time steps to reach the same simulation stage as the fixed time step simulation compared to the integral and PC11 controllers. It is expected that the PID controller behaves more conservatively, since its formulation carries the estimated error in three different time steps, making it a more rigid controller than the others. Table \ref{tab:ta} show a comparison of the performance results for the three controllers, and Figure \ref{fig:nitlit} presents the evolution of linear and nonlinear iterations during the simulation.

All methods significantly improved the spinodal decomposition simulation since it is possible to reach larger simulation times with a smaller number of steps. We observe from Fig. \ref{fig:freeenergy2d} that the steady-state is reached at around $t=0.5$ in our simulations. To reach the steady-state using the fixed time step scheme, considering that $\Delta t = 2 \times 10^{-6}$ is the largest possible fixed time step that would not introduce unacceptable errors in the simulation, it would be necessary $2.5\times10^5$ time steps, while using the time step adaptivity, it is reached with circa $5,000$ time steps, as seen in Table \ref{tab:ta}. In relative terms, adaptive simulations reach the steady-state in approximately $1.4\%$ of the simulation time needed for a fixed time step simulation, reinforcing the importance of temporal adaptivity. Observe that, in terms of required time steps, the PC11 required fewer time steps than the other two controllers while the PID controller solved the larger amount of time steps. The Integral controller, the simplest controller, has less control over the growth of the time steps, presenting more rejected steps. We also evaluate the performance in terms of linear and nonlinear iterations. We consider the absolute CPU effort calculated as the total number of linear iterations during the simulation, considering accepted and rejected steps. The controller with a larger absolute CPU effort becomes the reference for calculating the relative CPU effort. Comparing in Table \ref{tab:ta} and Figure \ref{fig:nitlit} the three adaptive simulations, the PC11 controller has the smaller number of total linear iterations with an improvement of $11\%$ over the amount of the same quantity for the PID controller and $7\%$ in comparison with the I controller. Even though the PID solution presents the lower average of nonlinear and linear iterations, it requires more time steps. By computing the total CPU effort, that is, the total number of iterations evaluated, we see that the other two controllers have a better performance.

\begin{table}
    \centering
    \caption{Performance results for the time adaptivity schemes for each time step controller in the 2D spinodal decomposition.}
    \begin{tabular}{|c|c|c|c|c|c|}
\hline
         Step size          & Accepted      &  Rejected  & Avg. Nonlinear & Avg. linear & Relative CPU \\ 
           Controller       &     Steps     &   Steps    &  Iterations    & Iterations  & Effort \\
        \hline
         I                  &  $4675$   &  $276$     & $7.0033$       & $175.4065$& $0.96$ \\
         PID                &  $5433$   &  $105$     & $6.7533$       & $163.3379$& $1.00$ \\
         PC11               &  $4676$   &   $86$     & $6.7718$       & $169.7915$& $0.89$\\
         \hline
    \end{tabular}
    \label{tab:ta}
\end{table}

\subsection{Nonlocal Cahn-Hilliard simulation of diblock copolymer melts}
\label{S:6}

\begin{figure}
	\centering
	\begin{minipage}{.95\textwidth}
		\centering
		\includegraphics[width=.85\linewidth]{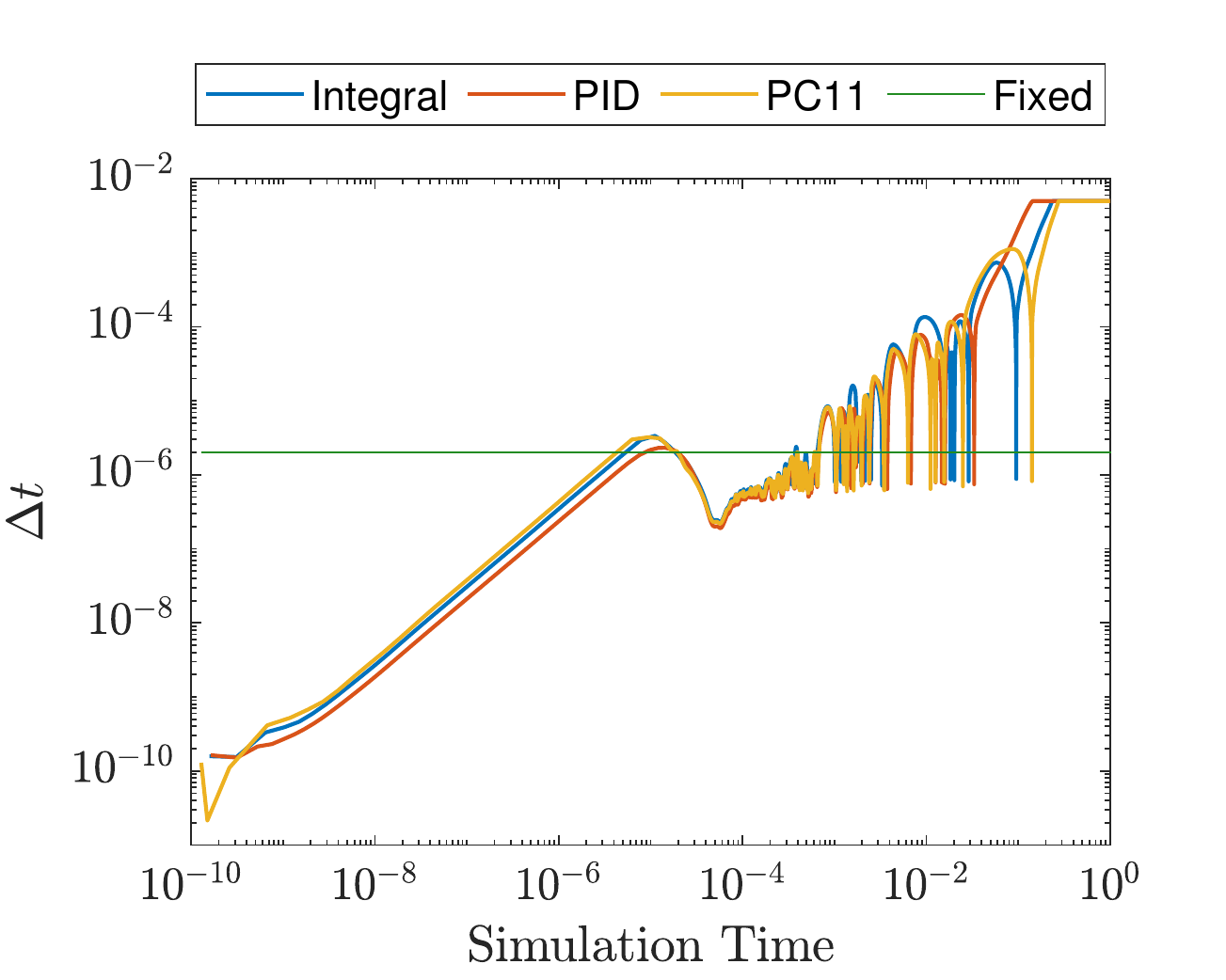}
		\label{fig:dtloose}
	\end{minipage}\\
	\begin{minipage}{.95\textwidth}
		\centering
		\includegraphics[width=.85\linewidth]{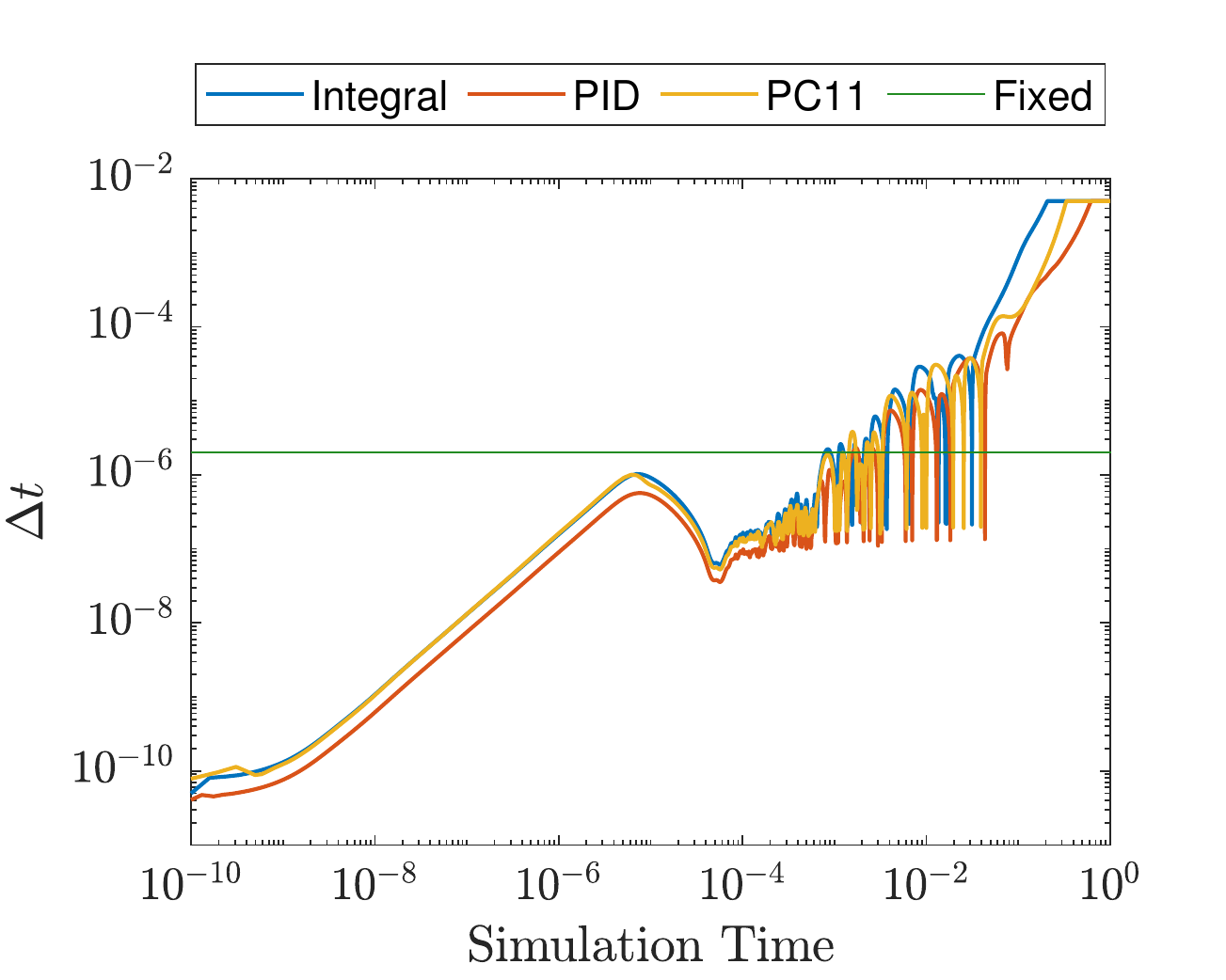}
		\label{fig:dttight}
	\end{minipage}%
	\caption{Time step size history for Case 1 (top) and Case 2 (bottom).}
	\label{fig:dtmissmatch}
\end{figure}

\begin{figure}
	\centering
	\begin{minipage}{.45\textwidth}
		\centering
		\includegraphics[trim={25.00cm 0.00cm 25.00cm 0.00cm},clip,width=.75\linewidth]{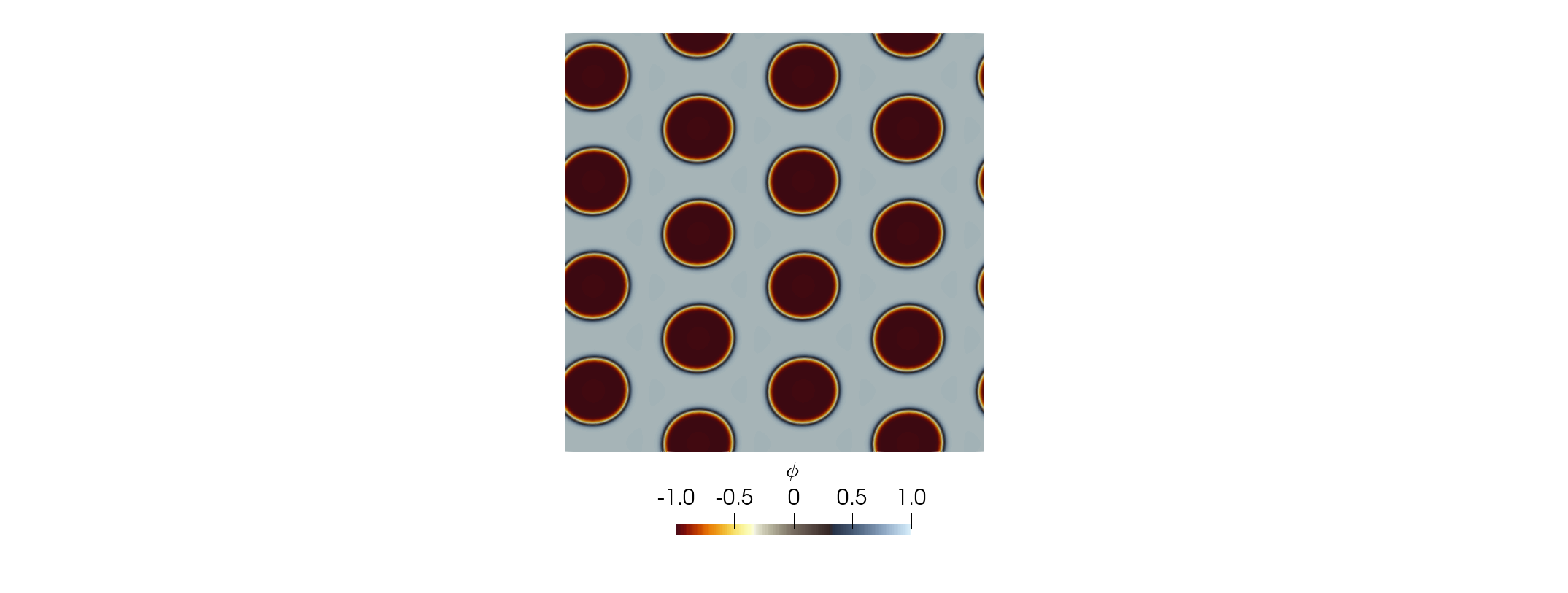}
		\subcaption{I controller.}
		\label{fig:mm1}
	\end{minipage}
	\begin{minipage}{.45\textwidth}
		\centering
		\includegraphics[trim={25.00cm 0.00cm 25.00cm 0.00cm},clip,width=.75\linewidth]{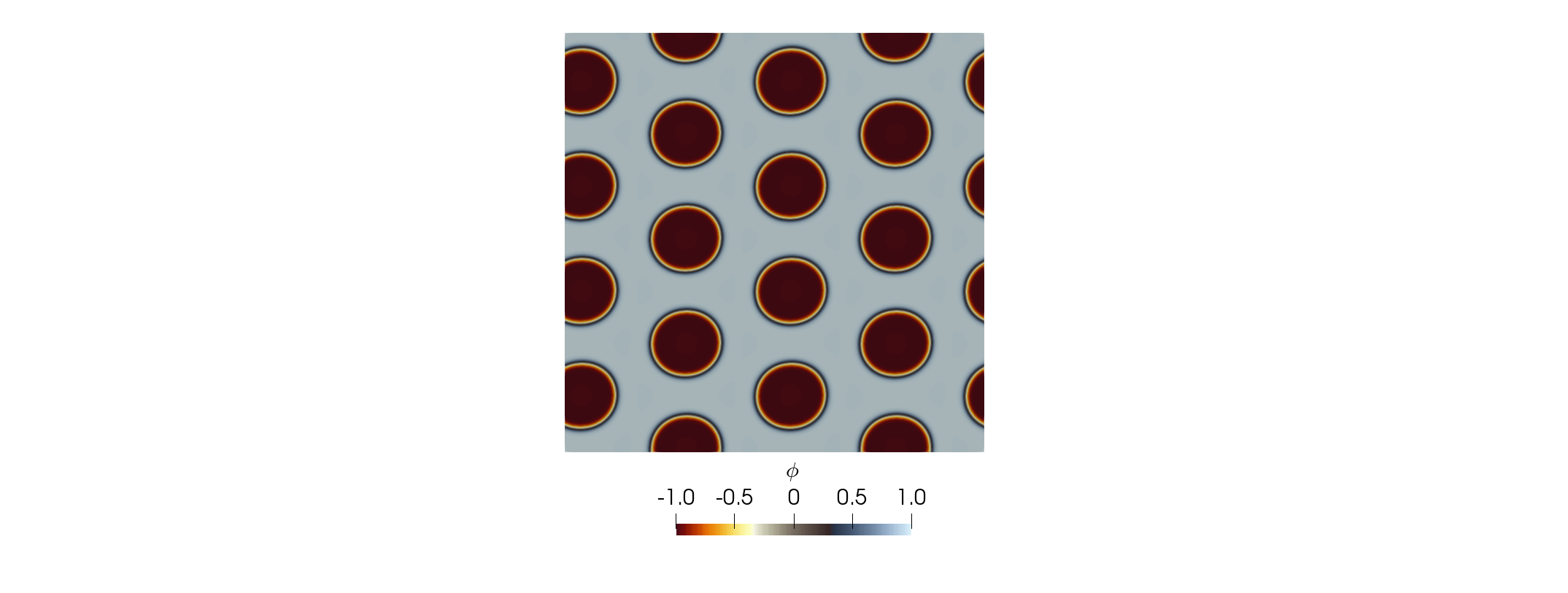}
		\subcaption{I controller.}
		\label{fig:mm2}
	\end{minipage}\\
	\begin{minipage}{.45\textwidth}
		\centering
		\includegraphics[trim={25.00cm 0.00cm 25.00cm 0.00cm},clip,width=.75\linewidth]{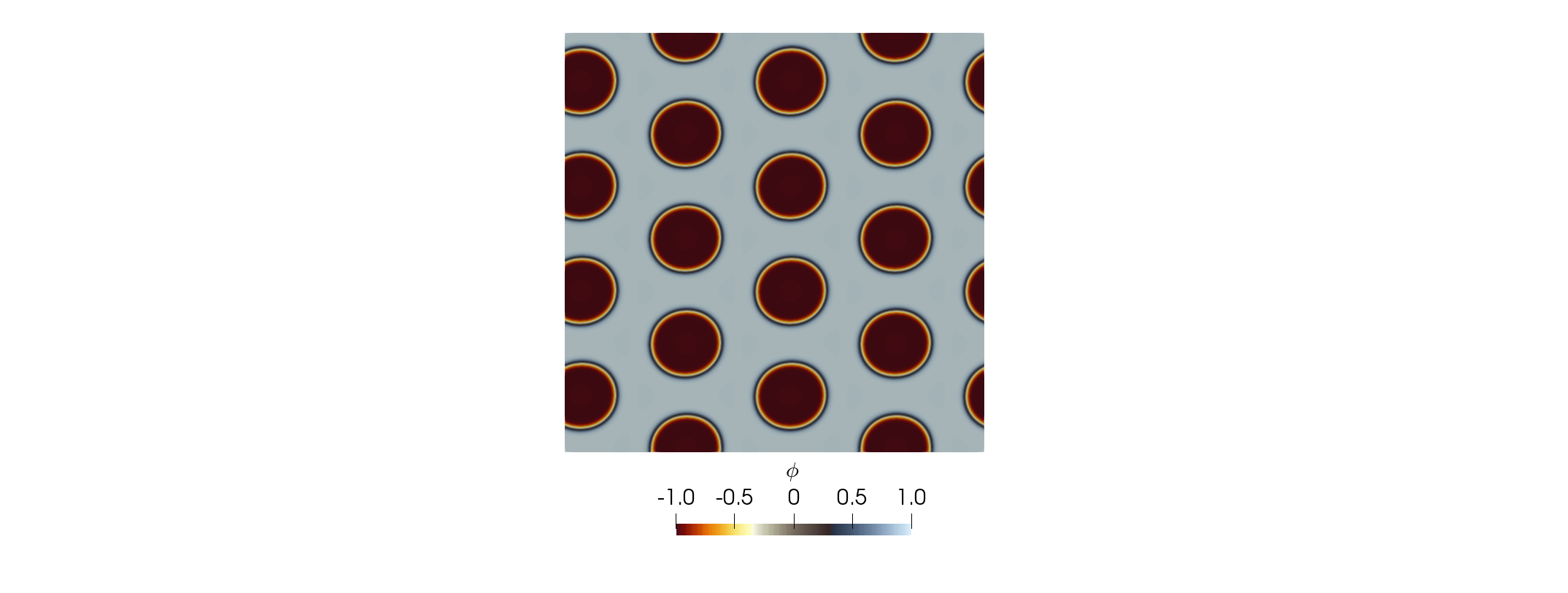}
		\subcaption{PID controller.}
		\label{fig:mm3}
	\end{minipage}
	\begin{minipage}{.45\textwidth}
		\centering
		\includegraphics[trim={25.00cm 0.00cm 25.00cm 0.00cm},clip,width=.75\linewidth]{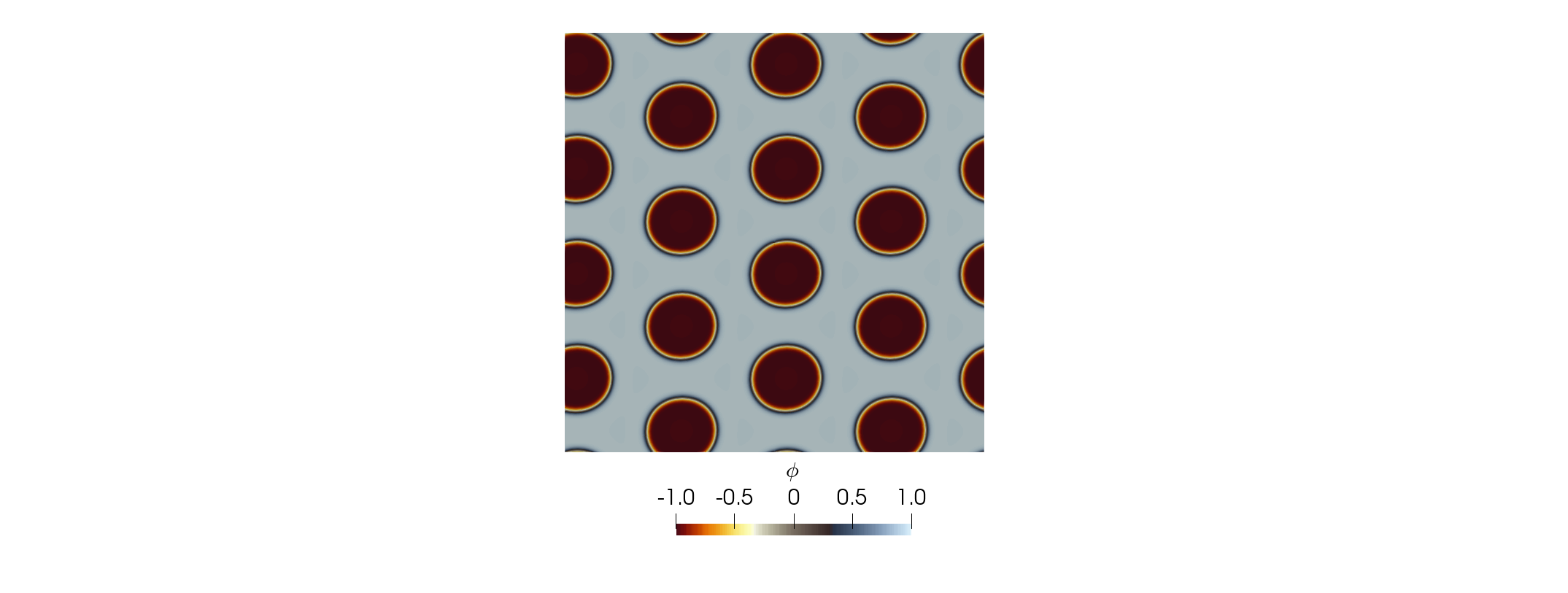}
		\subcaption{PID controller.}
		\label{fig:mm4}
	\end{minipage}\\
	\begin{minipage}{.45\textwidth}
		\centering
		\includegraphics[trim={25.00cm 0.00cm 25.00cm 0.00cm},clip,width=.75\linewidth]{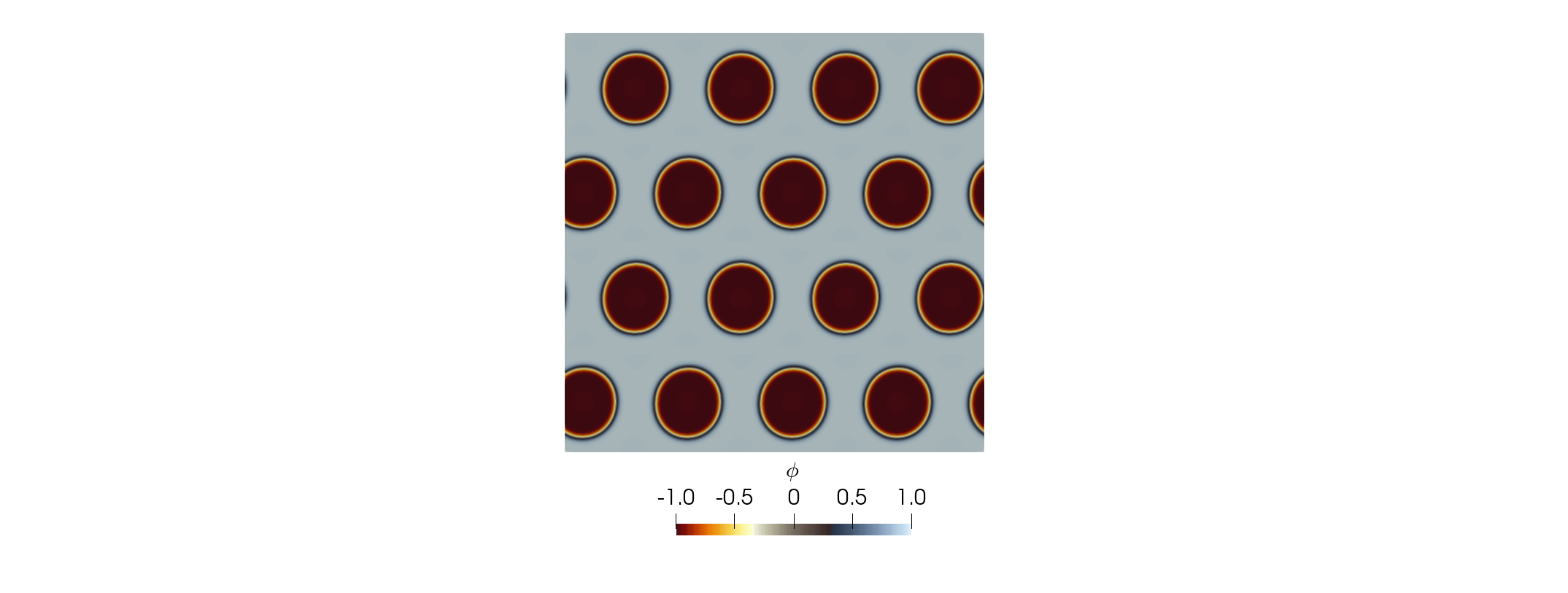}
		\subcaption{PC11 controller.}
		\label{fig:mm5}
	\end{minipage}
	\begin{minipage}{.45\textwidth}
		\centering
		\includegraphics[trim={25.00cm 0.00cm 25.00cm 0.00cm},clip,width=.75\linewidth]{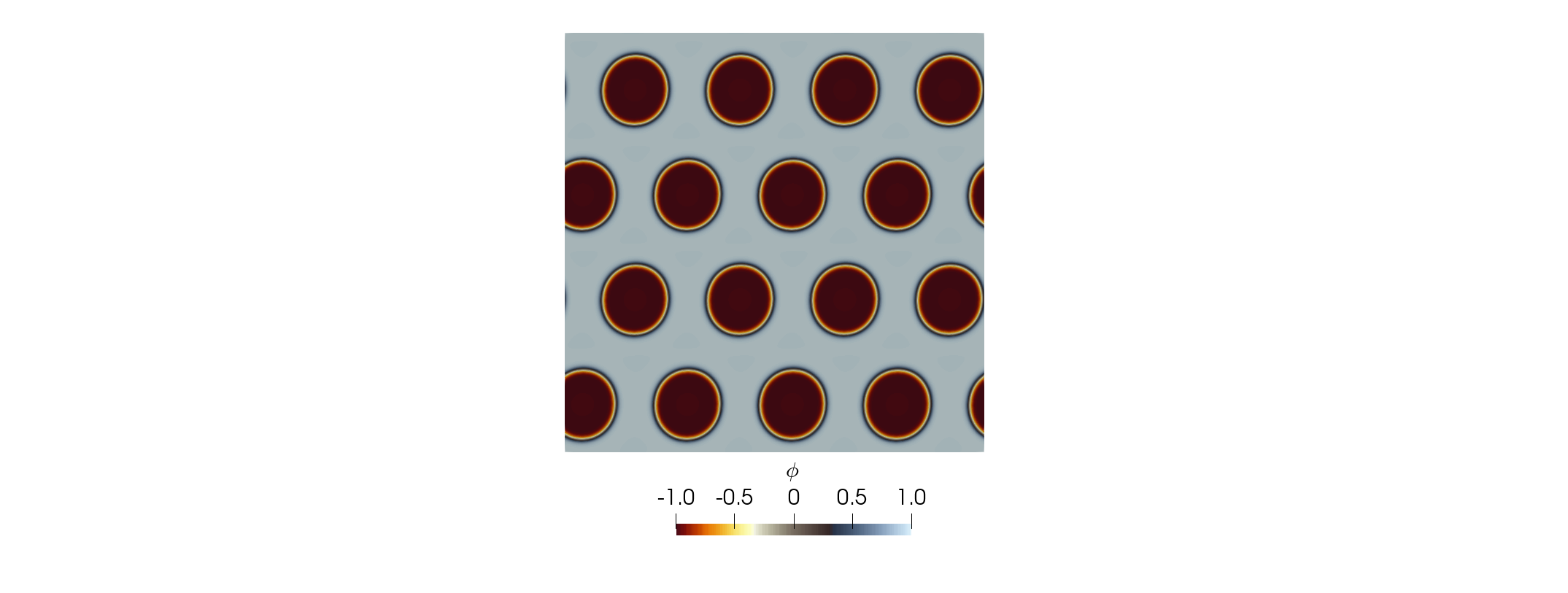}
		\subcaption{PC11 controller.}
		\label{fig:mm6}
	\end{minipage}
	\caption{Steady-state for the nonlocal Cahn-Hilliard equation in two dimensions for the three controllers for Case 1 (left) and Case 2 (right).}
	\label{fig:missmatch}
\end{figure}

In this section, we solve the NCH equation in two and three dimensions to evaluate the performance and accuracy of the time step controllers. The same parameters regarding the domain, mesh size, interface thickness, free energy homogeneous function, and other numerical parameters are extended from the previous examples to the nonlocal cases. The boundary conditions, however, are considered periodical to preserve the NCH equation pattern formation. It is known that the variation of the parameters $\epsilon$, $\sigma$, $\bar{\phi}$, and the domain size interfere directly with the steady-state structure of the NCH equation. We define these parameters such that the minimizers are situated on a locally stable region of the phase diagram and a domain size large enough compared to the intrinsic length scale of the minimizers of the O-K functional \cite{choksi}. Initially, we consider a 2D case, where the copolymers steady-state has a hexagonally packed spots structure, as seen in the phase diagrams in \cite{vanderberg,Choksi2011}. We consider $\bar{\phi} = 0.3$, $\epsilon = 0.1$ and $\sigma = 500$. For this first nonlocal example, we consider an assessment of the controllers' parameters. In \cite{calo2020}, there is a remark that the use of controllers for time step size adaptivity for the Swift-Hohenberg equation \cite{Cross1993, Swift1977} with a smaller safety coefficient and tighter tolerances yields a smaller percentage of rejected time steps and, according to \cite{Choksi2011}, the NCH can be viewed as a hybrid of the Swift–Hohenberg equation and the CH equation. Therefore, we compare the results of the standard controller parameters with the results obtained by considering $\rho = 0.75$ and $\tau_{abs} = \tau_{rel} = 1 \times 10^{-5}$. We label the simulations for the controller parameters used in the CH example as Case 1 and Case 2 for the new proposed values. Figure \ref{fig:dtmissmatch} shows the time step size history for the six simulations, while the corresponding steady-state configurations are seen in Figure \ref{fig:missmatch}. Table \ref{tab:missmatch} shows the results for the controllers' performance in all cases.

\begin{table}
	\centering
	\caption{Results for the time adaptivity schemes for each time step controller in the NCH simulations.}
	\begin{tabular}{|c|c|c|c|c|c|c|}
		\hline
		& Time Step           & Accepted      &  Rejected  &  Avg. Nonlinear & Avg. Linear  & Relative CPU\\ 
		&Controller     	  &     Steps     &   Steps    &  Iterations     & Iterations   & Effort\\
		\hline
		&I                  &  $2660$ &  $157$ &  $6.9357$ & $136.8090$ & $0.29$\\
		Case 1&PID          &  $3045$ &  $39$  &  $6.6860$ & $119.2141$ & $0.29$\\
		&PC11               &  $2523$ &  $65$  &  $6.7741$ & $128.2715$ & $0.26$\\
		\hline
		&I                  &  $10036$ &  $2$ &  $7.1842$ & $88.6024$ & $0.71$\\
		Case 2&PID          &  $16981$ &  $3$ &  $7.0434$ & $73.6844$  & $1.00$\\
		&PC11               &  $11324$ &   $4$ &  $7.1161$ & $83.1459$ & $0.75$\\		
		\hline
	\end{tabular}
	\label{tab:missmatch}
\end{table}

\par
\begin{figure}
	\centering
	\begin{minipage}{.49\textwidth}
		\centering
		\includegraphics[width=.95\linewidth]{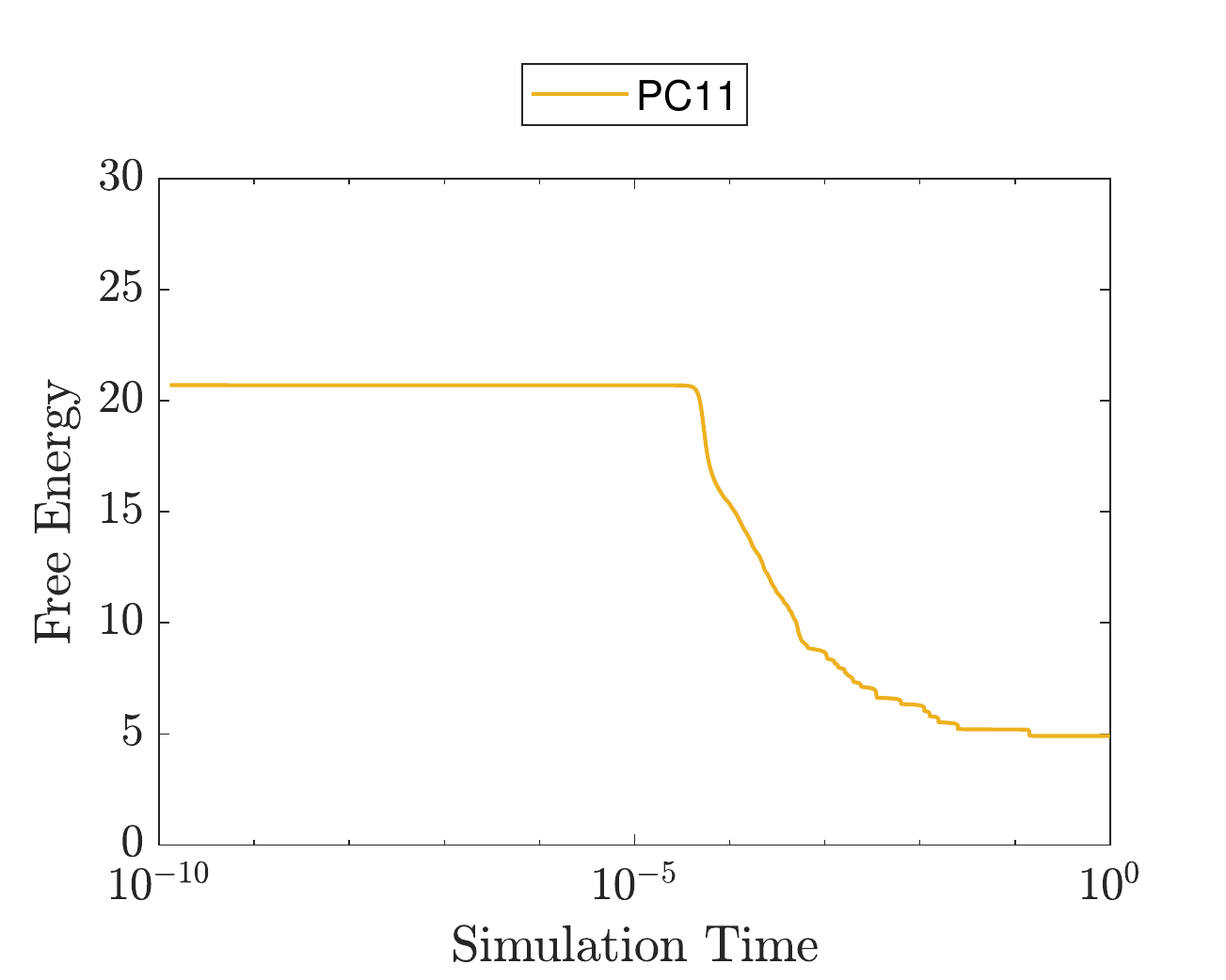}
		\label{fig:fe_loose}
	\end{minipage}
	\begin{minipage}{.49\textwidth}
		\centering
		\includegraphics[width=.95\linewidth]{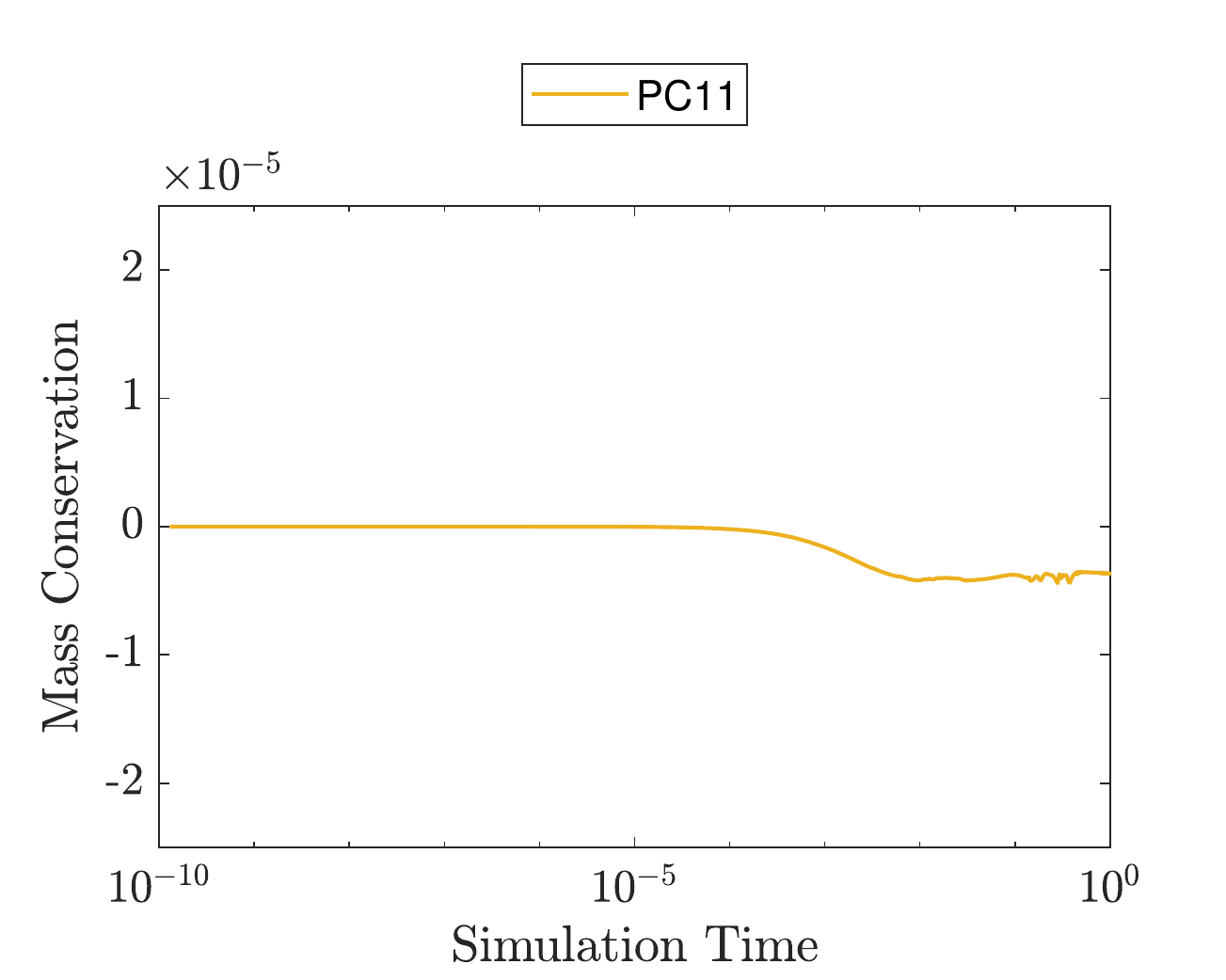}
		\label{fig:mass_loose}
	\end{minipage}\\
	\begin{minipage}{.49\textwidth}
		\centering
		\includegraphics[width=.95\linewidth]{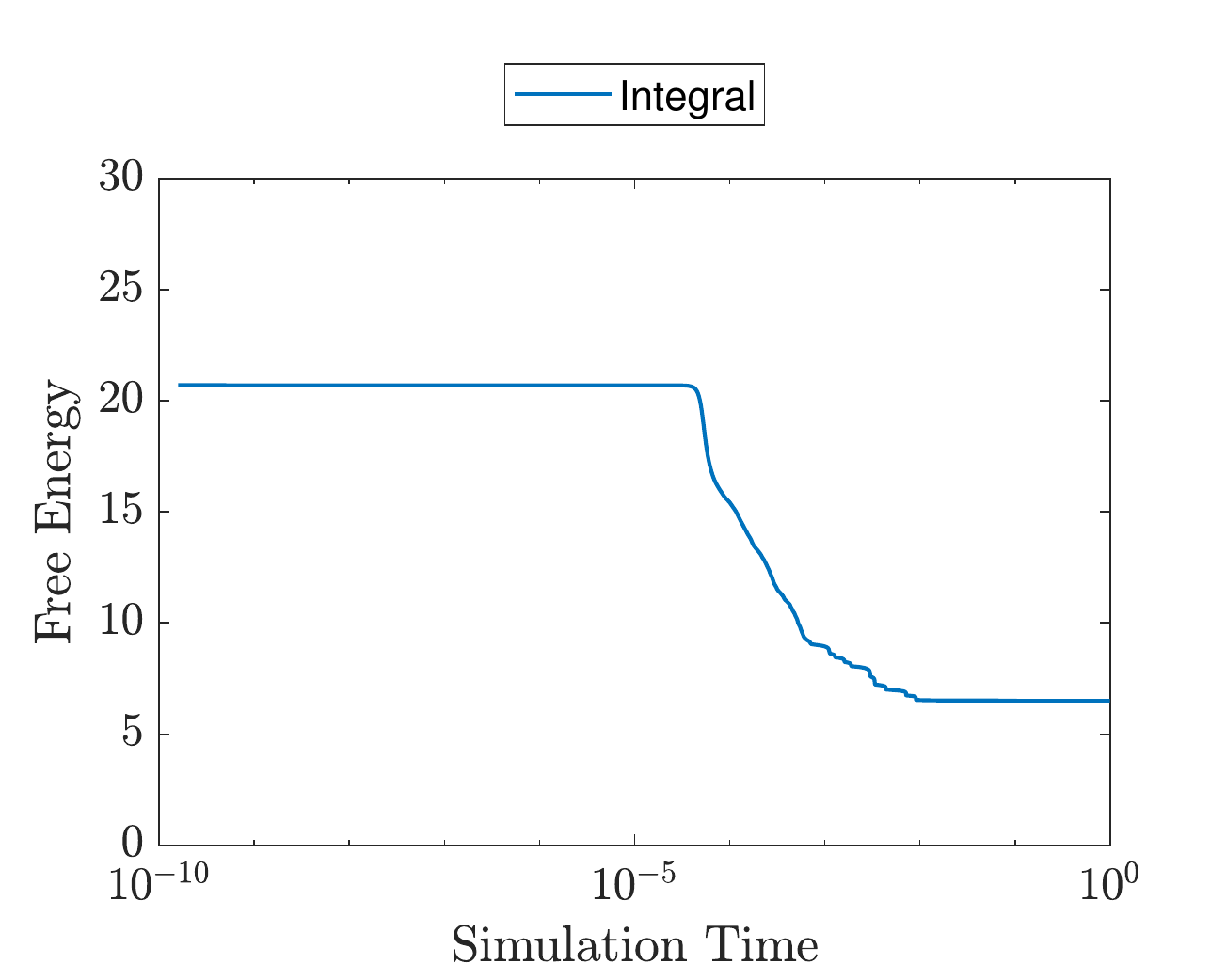}
		\label{fig:fe_tight}
	\end{minipage}
	\begin{minipage}{.49\textwidth}
		\centering
		\includegraphics[width=.95\linewidth]{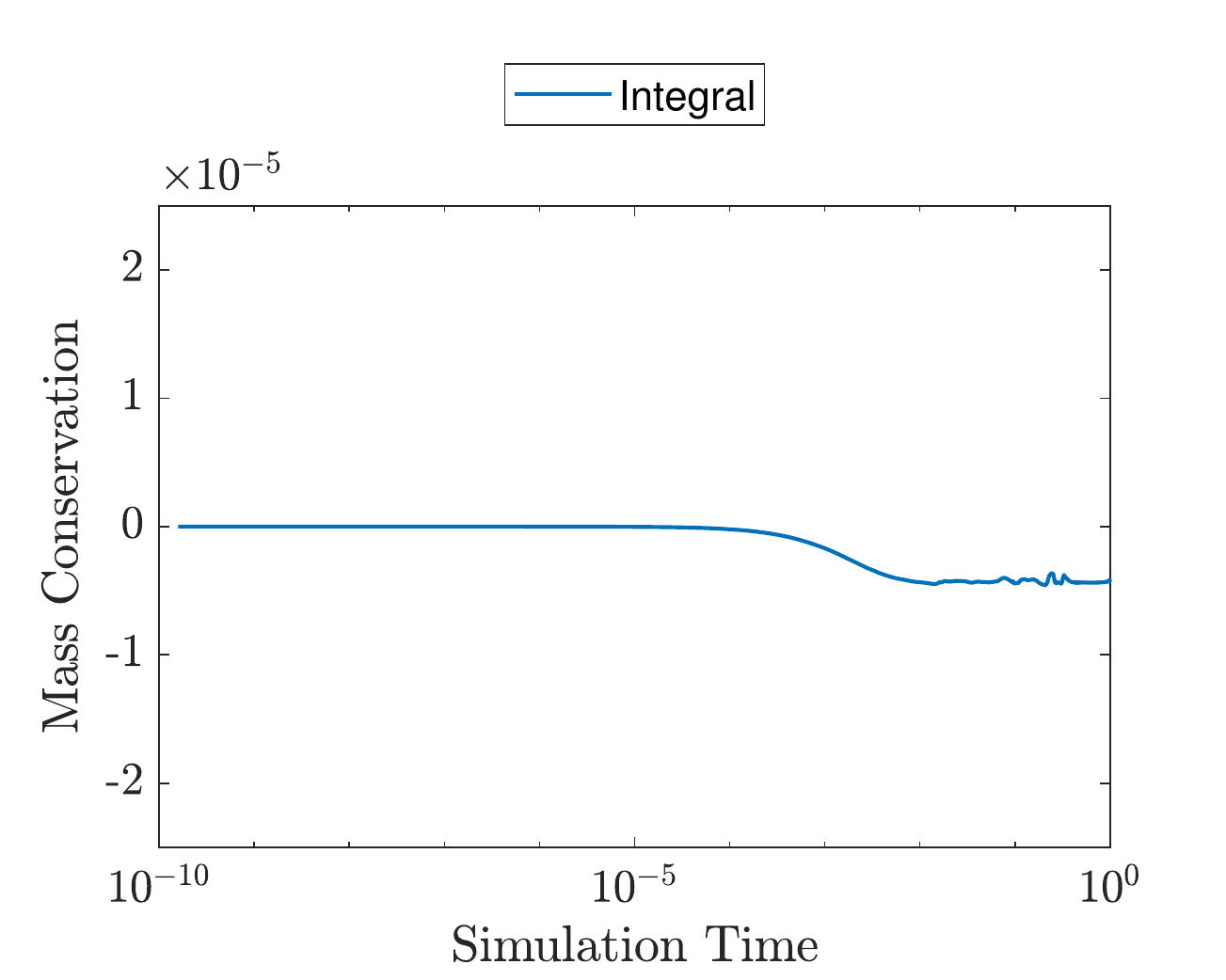}
		\label{fig:mass_tight}
	\end{minipage}%
	\caption{Free energy decay and mass conservation for the best performing controllers for Case 1 (top) and Case 2 (bottom) simulations.}
	\label{fig:massfecop}
\end{figure}

 
We note in both cases in Fig. \ref{fig:dtmissmatch} that the time step size curves are not overlapping, as in the previous examples. Therefore, for a given instant where the curves do not match, the phenomenon in Fig. \ref{fig:simultaneous} is not observed in the diblock copolymer context, meaning that different controllers lead to different observations in the time evolution of the diblock copolymer melt. However, this does not affect the formation of the steady-state structures related to the selected set of parameters, which is the information of interest in most cases. We observe the steady-state for all six simulations of Fig. \ref{fig:missmatch} and note that the six simulations converged to hexagonally packed spots, as initially predicted by the phase diagrams \cite{vanderberg,Choksi2011}, despite minor differences that arise due to the periodic boundary conditions. We also observe in Figure \ref{fig:massfecop} that our simulations do not present any unphysical properties in the free energy decay and the mass conservation for the best performing controllers.
\par Regarding the performance, we can see in Table \ref{tab:missmatch} that reducing the prescribed tolerances leads to a significant increase in the required number of time steps to reach the steady-state while the rejected steps decreased. We observe that the best performing controller in our simulations is PC11 for Case 1, while in Case 2, the I controller shows better performance. One possible explanation is that the I controller's aggressive behavior combined with the tighter tolerances in Case 2 leads to a more controlled environment where the number of rejected steps is not significant compared to the increased number of time steps required for the simulations to reach the steady-state. Nevertheless,  for both cases, the I controller presents the largest average number of linear iterations while the PID controller has the smaller.

\begin{figure}
	\centering
	\begin{minipage}{.45\textwidth}
		\centering
		\includegraphics[trim={25.00cm 0.00cm 25.00cm 0.00cm},clip,width=.85\linewidth]{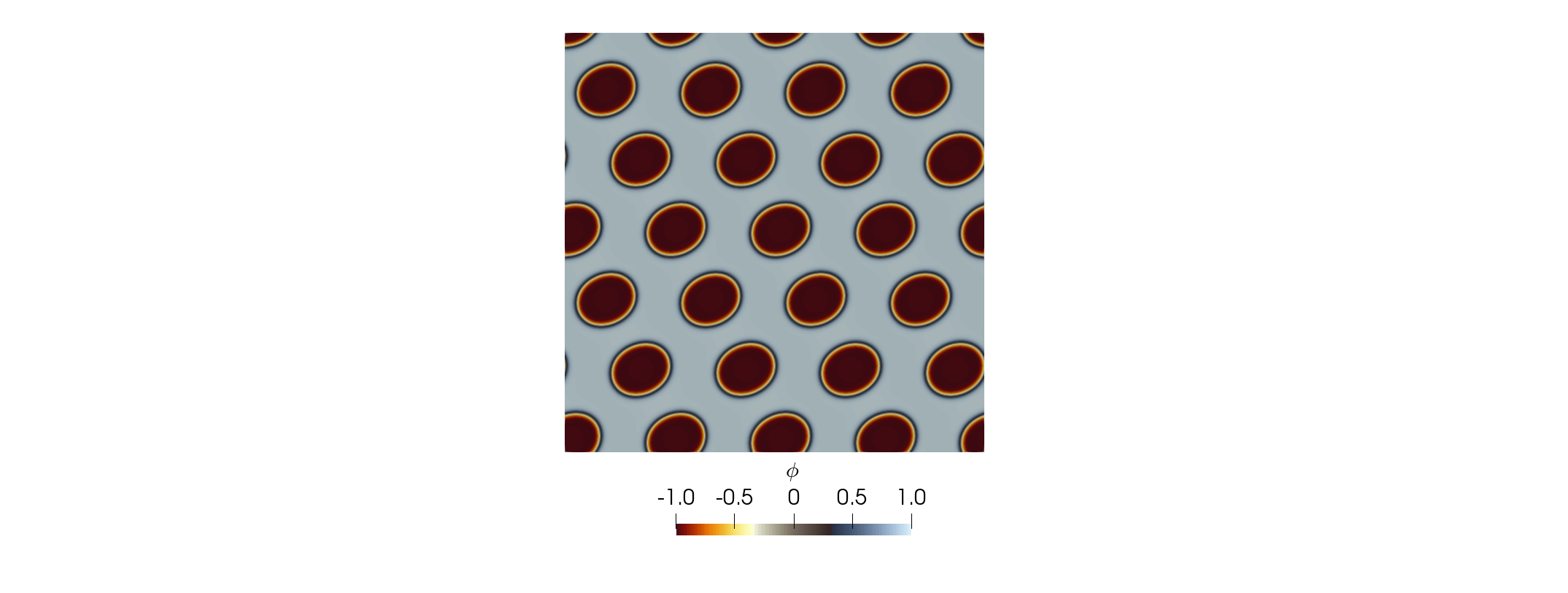}
		\subcaption{Test case A}
		\label{fig:dc2d2}
	\end{minipage}
	\begin{minipage}{.45\textwidth}
		\centering
		\includegraphics[trim={25.00cm 0.00cm 25.00cm 0.00cm},clip,width=.85\linewidth]{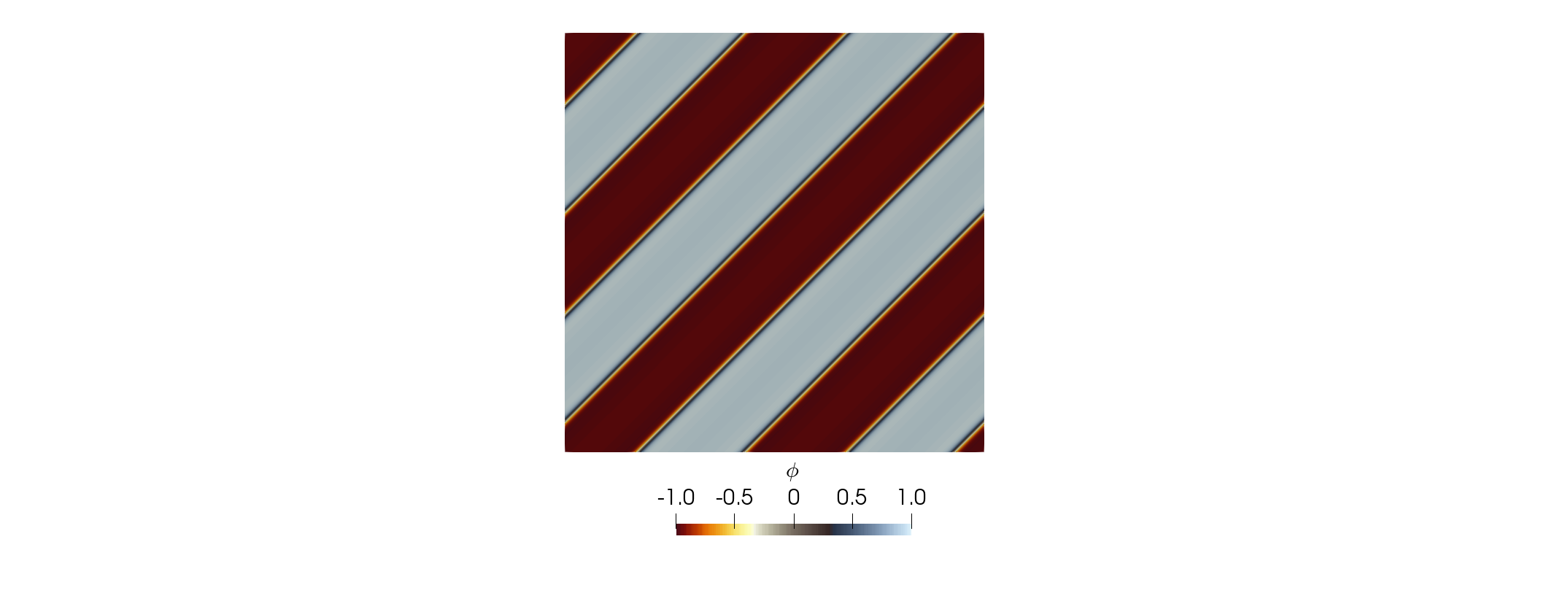}
		\subcaption{Test case B}
		\label{fig:dc2d3}
	\end{minipage}\\
	\begin{minipage}{.45\textwidth}
		\centering
		\includegraphics[trim={25.00cm 0.00cm 25.00cm 0.00cm},clip,width=.85\linewidth]{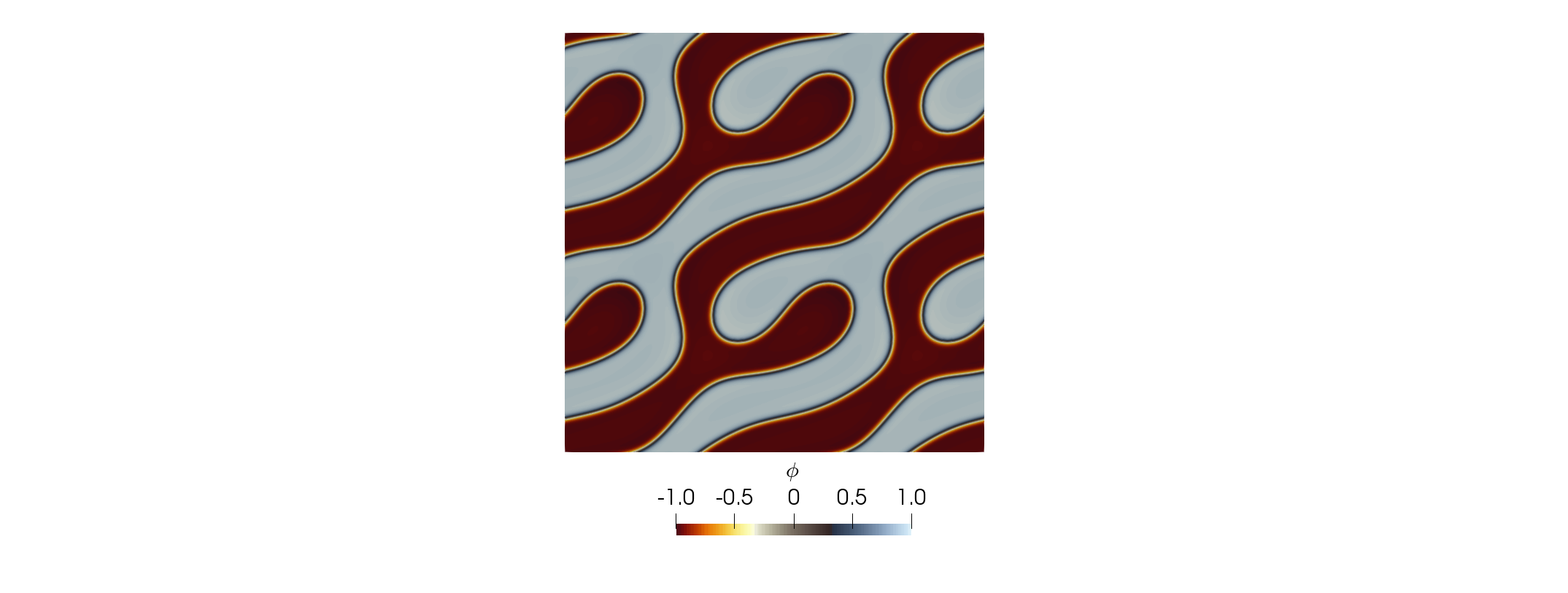}
		\subcaption{Test case C}
		\label{fig:dc2d4}
	\end{minipage}
	\caption{Steady-state for the NCH equation in two dimensions.}
	\label{fig:dc2d}
\end{figure}

After observing the effects of the controllers' parameters on the NCH equation, we note that the use of the standard values used in Case 1 requires less computational effort and does not influence the steady-state evaluation. Therefore, we extend our analysis using Case 1 for different parameter sets. We consider three examples: test case A, where $\bar{\phi} = 0.3$ and $\sigma = 1000$, test case B where $\bar{\phi} = 0.0$ and $\sigma = 500$ and test case C where $\bar{\phi} = 0.0$ and $\sigma = 1000$. Figure \ref{fig:dc2d} shows the steady-state these test cases. The minimizing structure of the melt in two dimensions are hexagonally packed spots, stripes and mixed states \cite{Choksi2011, vanderberg}. In the figure, we can see spots (Fig. \ref{fig:dc2d2}), stripes (Figs. \ref{fig:dc2d3}) and mixed structure (Figs. \ref{fig:dc2d4}) melts. All the simulations reached the steady-state at around $t = 0.15$.

\begin{figure}
	\centering
	\begin{minipage}{.6\textwidth}
		\centering
		\includegraphics[width=.85\linewidth]{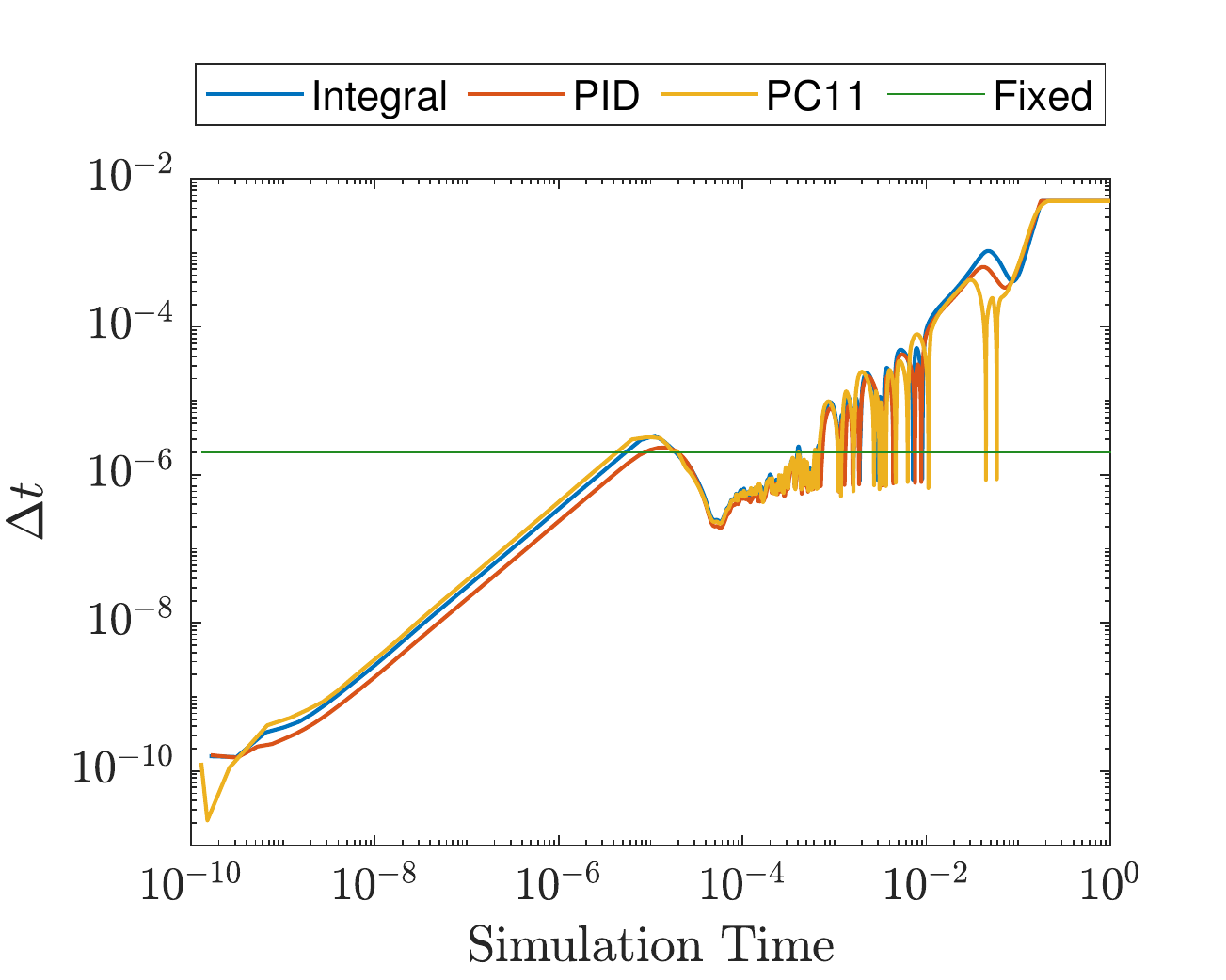}
		\subcaption{Test case A}
	\end{minipage}\\
	\begin{minipage}{.6\textwidth}
		\centering
		\includegraphics[width=.85\linewidth]{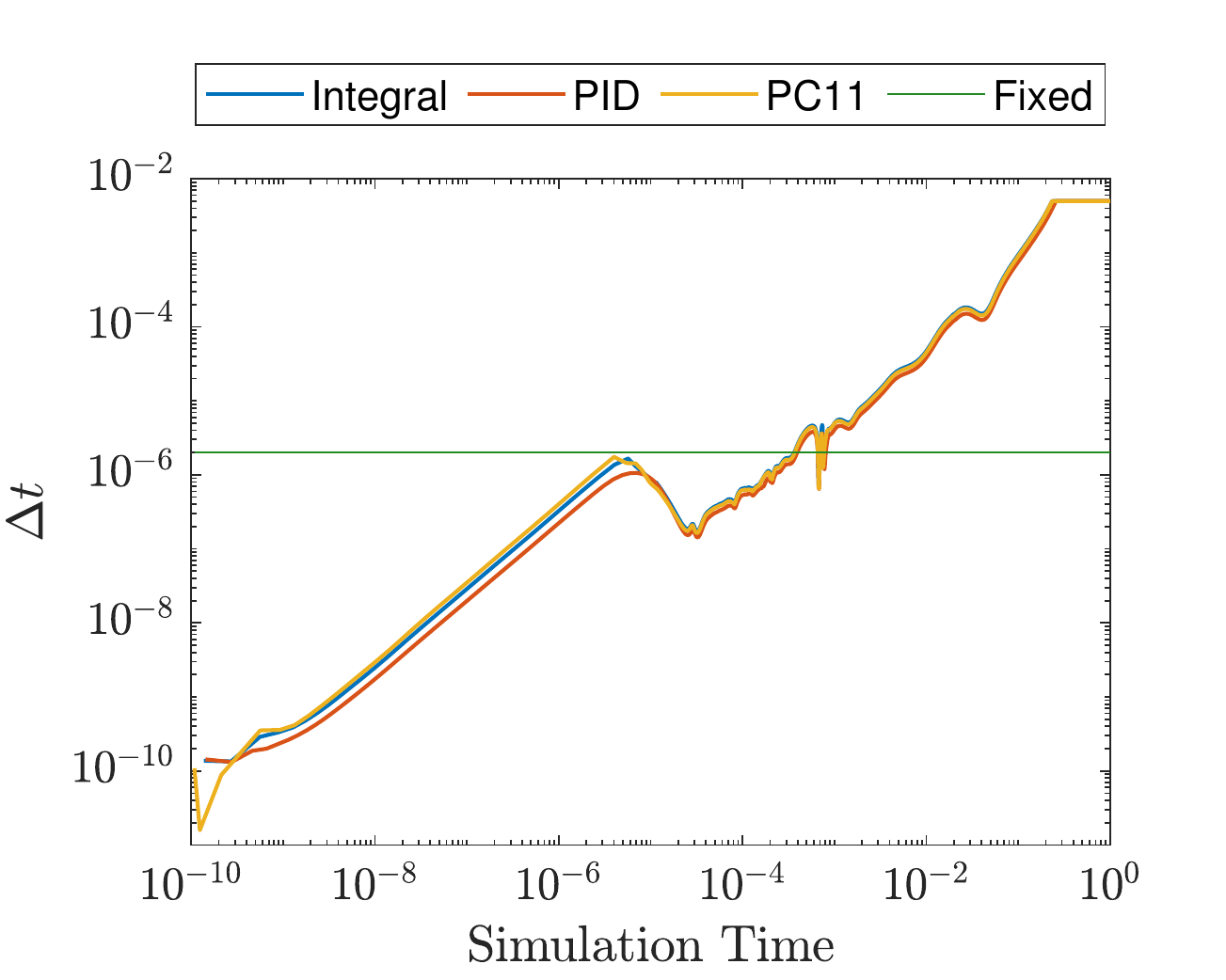}
		\subcaption{Test case B}
	\end{minipage}\\
	\begin{minipage}{.6\textwidth}
		\centering
		\includegraphics[width=.85\linewidth]{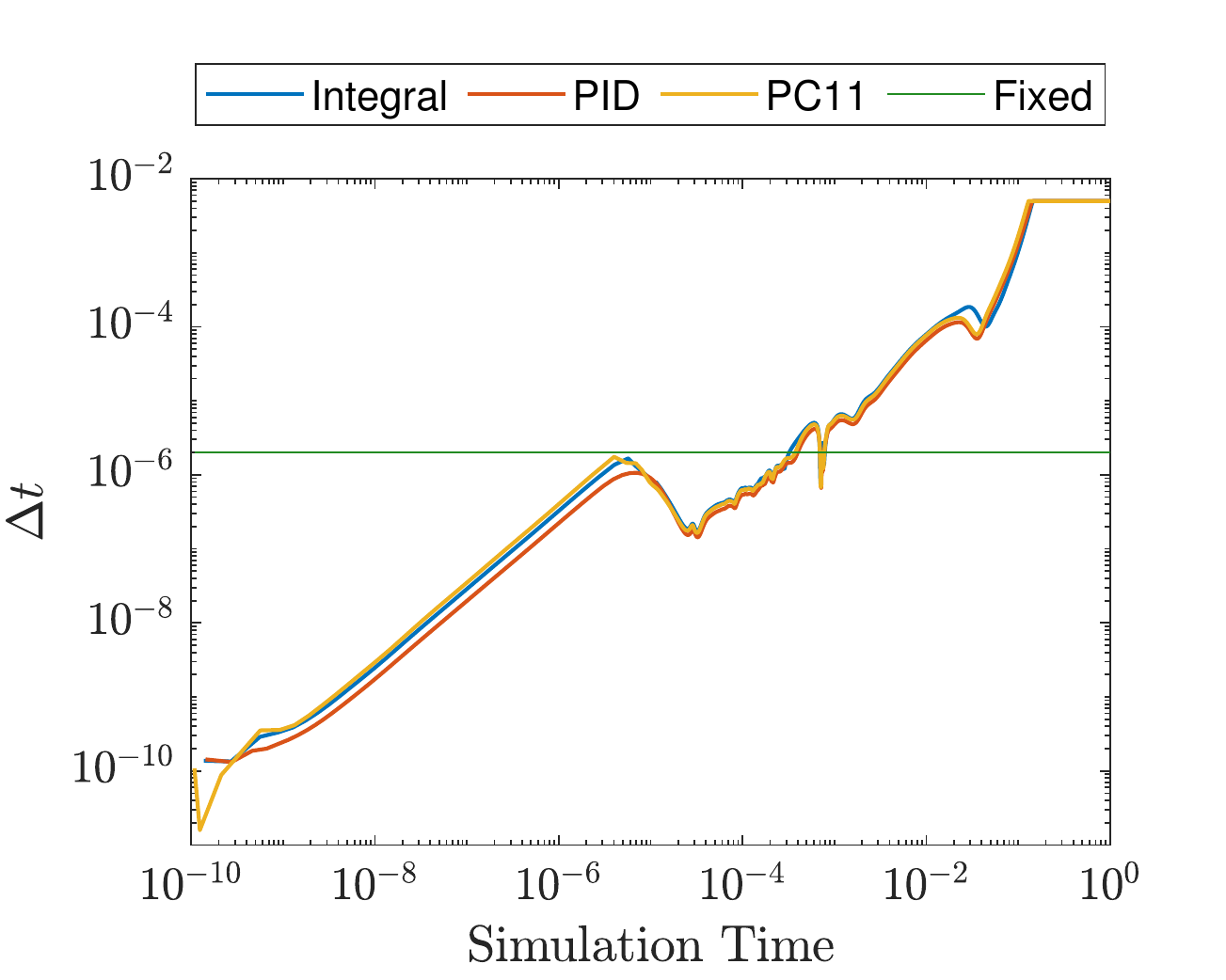}
		\subcaption{Test case C}
	\end{minipage}%
	\caption{Time step size history for the cases described in Fig. \ref{fig:dc2d}.}
	\label{fig:dtcop2d}
\end{figure}

\begin{figure}
	\centering
	\begin{minipage}{.49\textwidth}
		\centering
		\includegraphics[width=.95\linewidth]{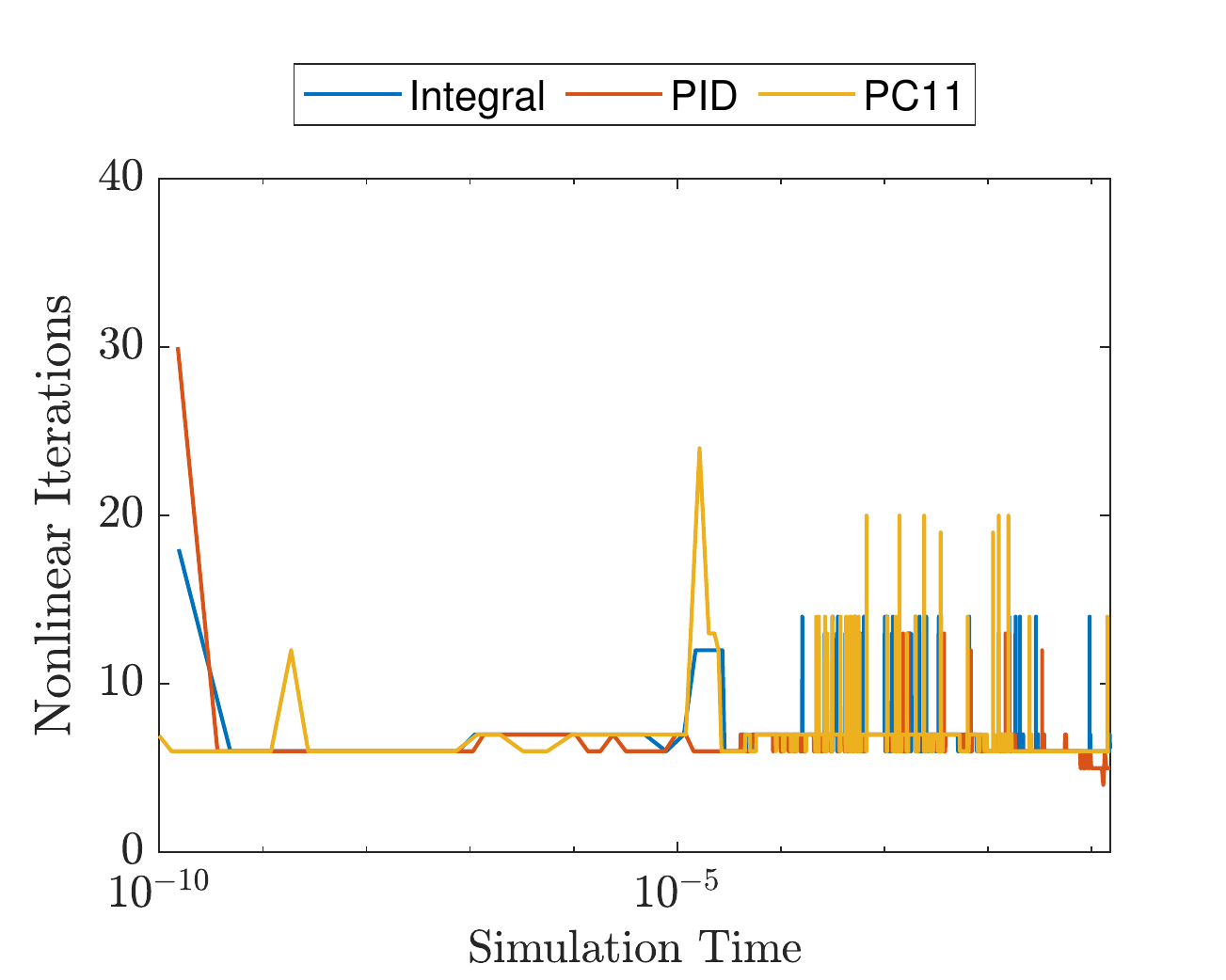}
		\label{fig:nitexA}
	\end{minipage}
	\begin{minipage}{.49\textwidth}
		\centering
		\includegraphics[width=.95\linewidth]{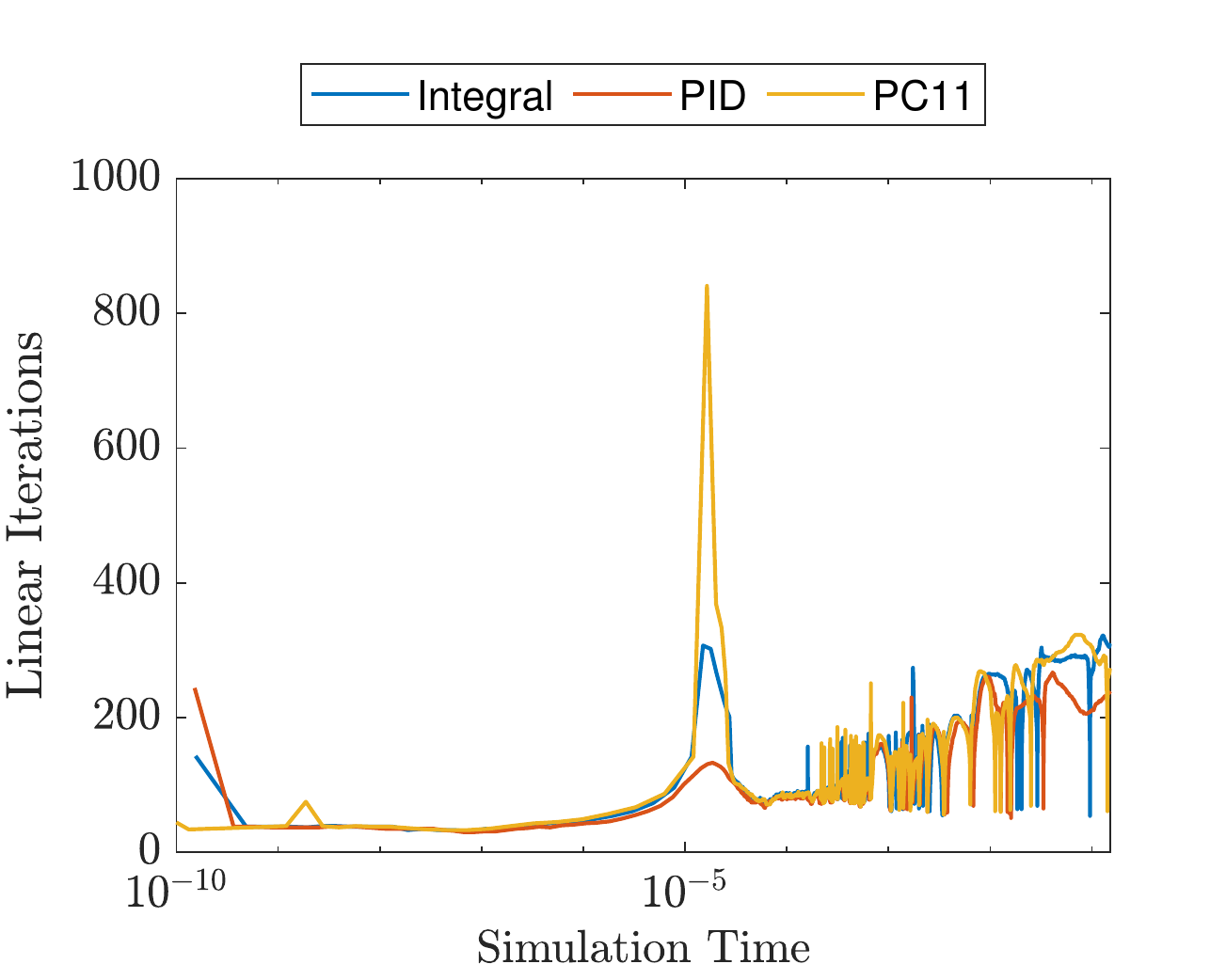}
		\label{fig:litexA}
	\end{minipage}\\
	\begin{minipage}{.49\textwidth}
		\centering
		\includegraphics[width=.95\linewidth]{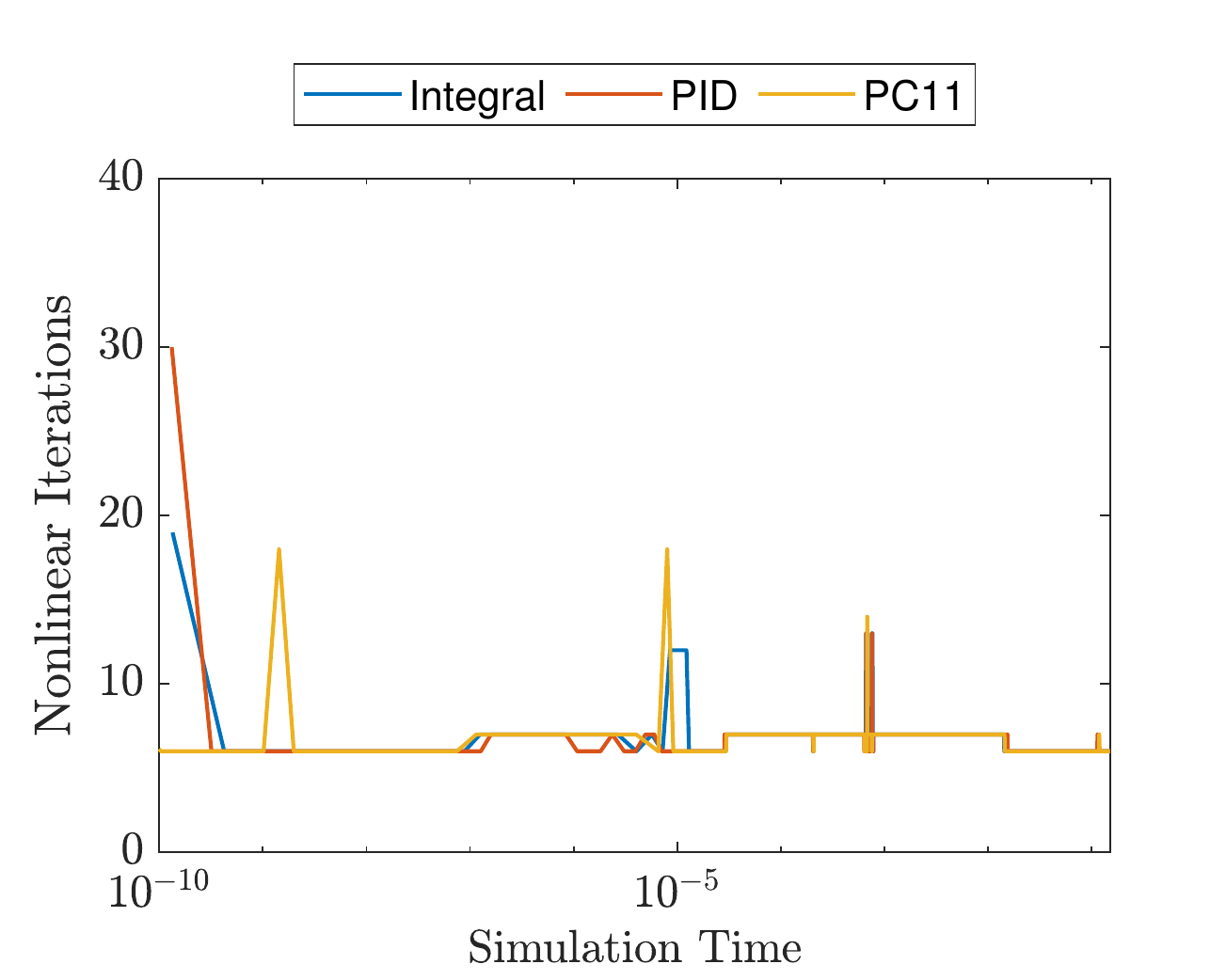}
		\label{fig:nitexB}
	\end{minipage}
	\begin{minipage}{.49\textwidth}
		\centering
		\includegraphics[width=.95\linewidth]{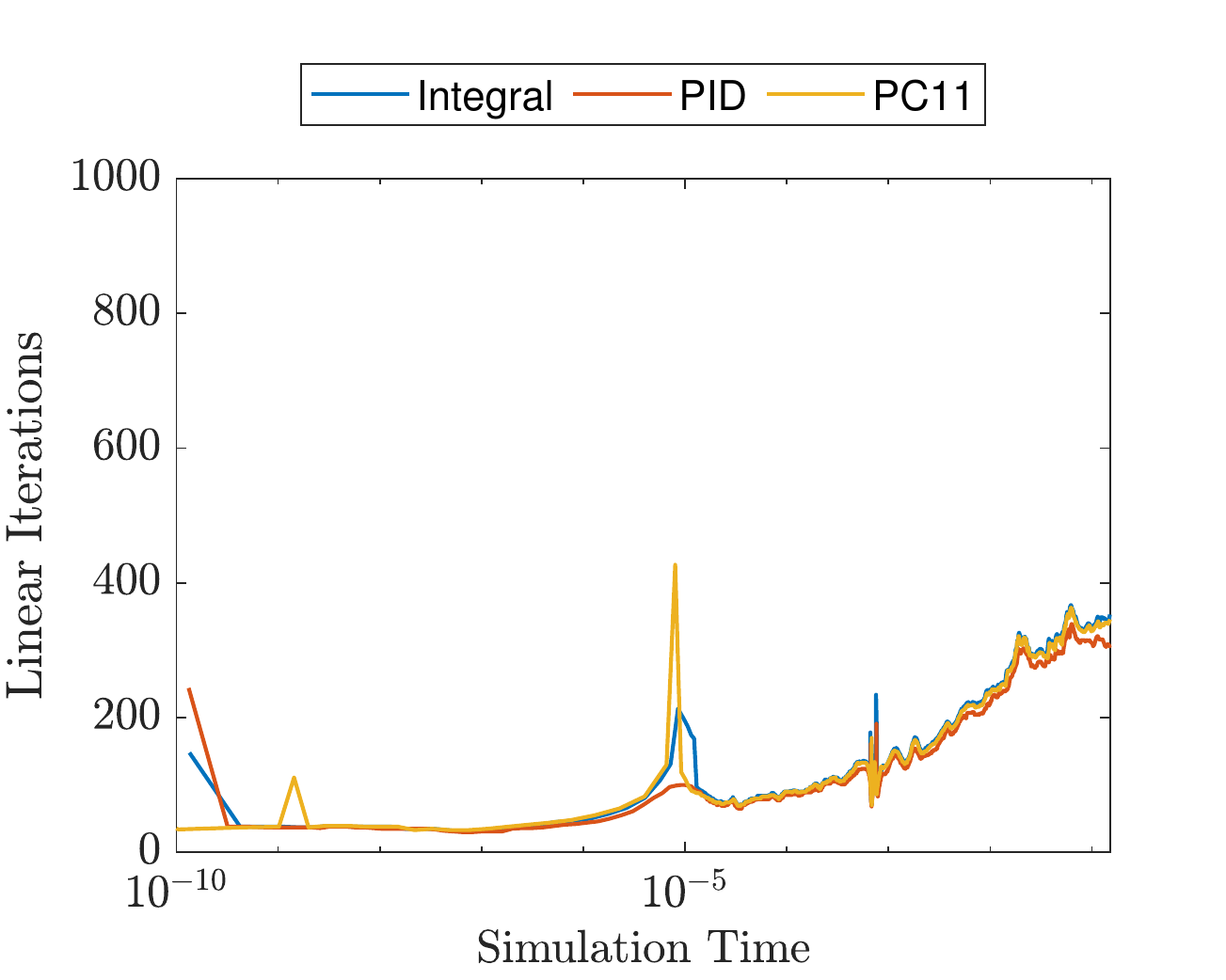}
		\label{fig:litexB}
	\end{minipage}\\
	\begin{minipage}{.49\textwidth}
		\centering
		\includegraphics[width=.95\linewidth]{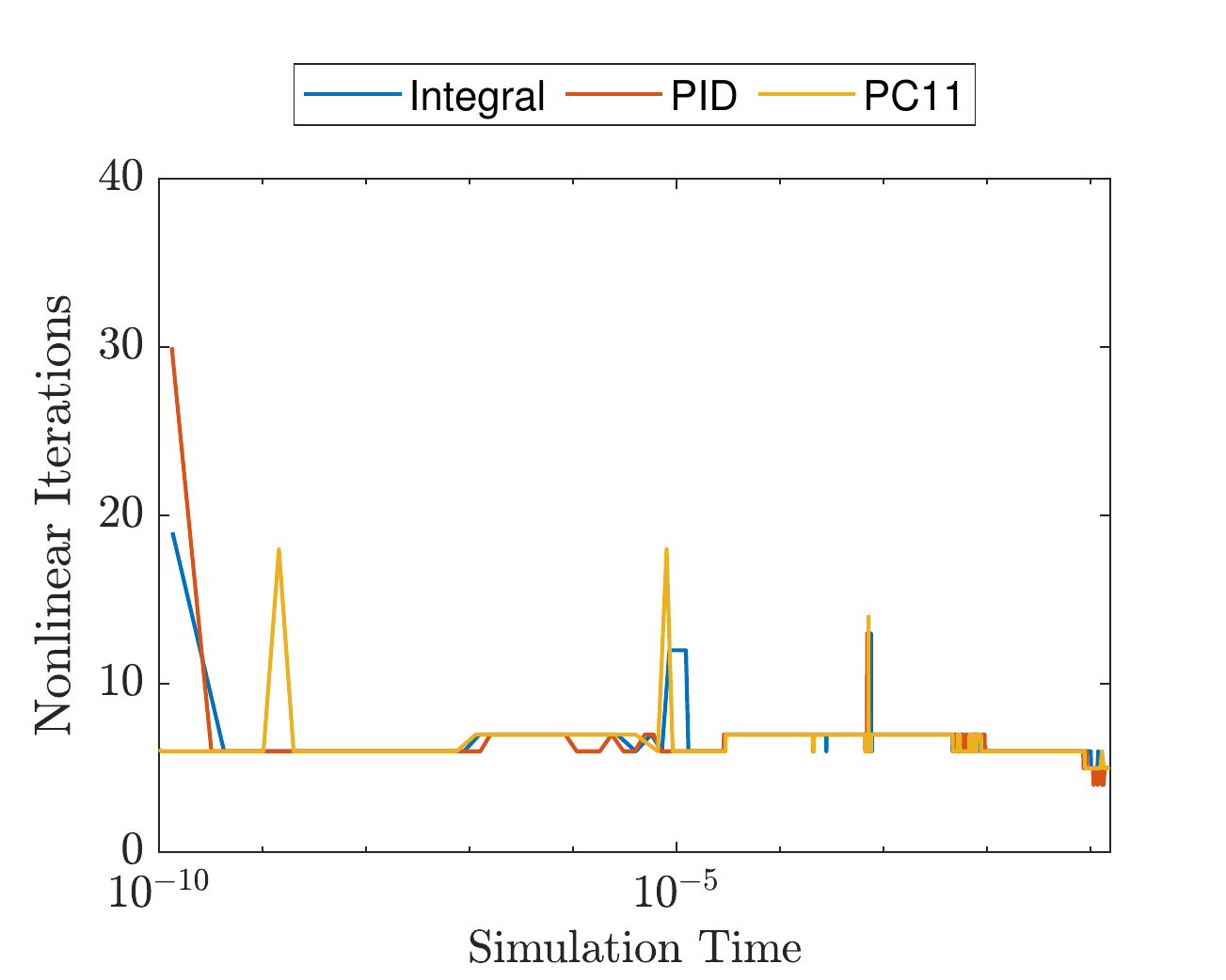}
		\label{fig:nitexC}
	\end{minipage}
	\begin{minipage}{.49\textwidth}
		\centering
		\includegraphics[width=.95\linewidth]{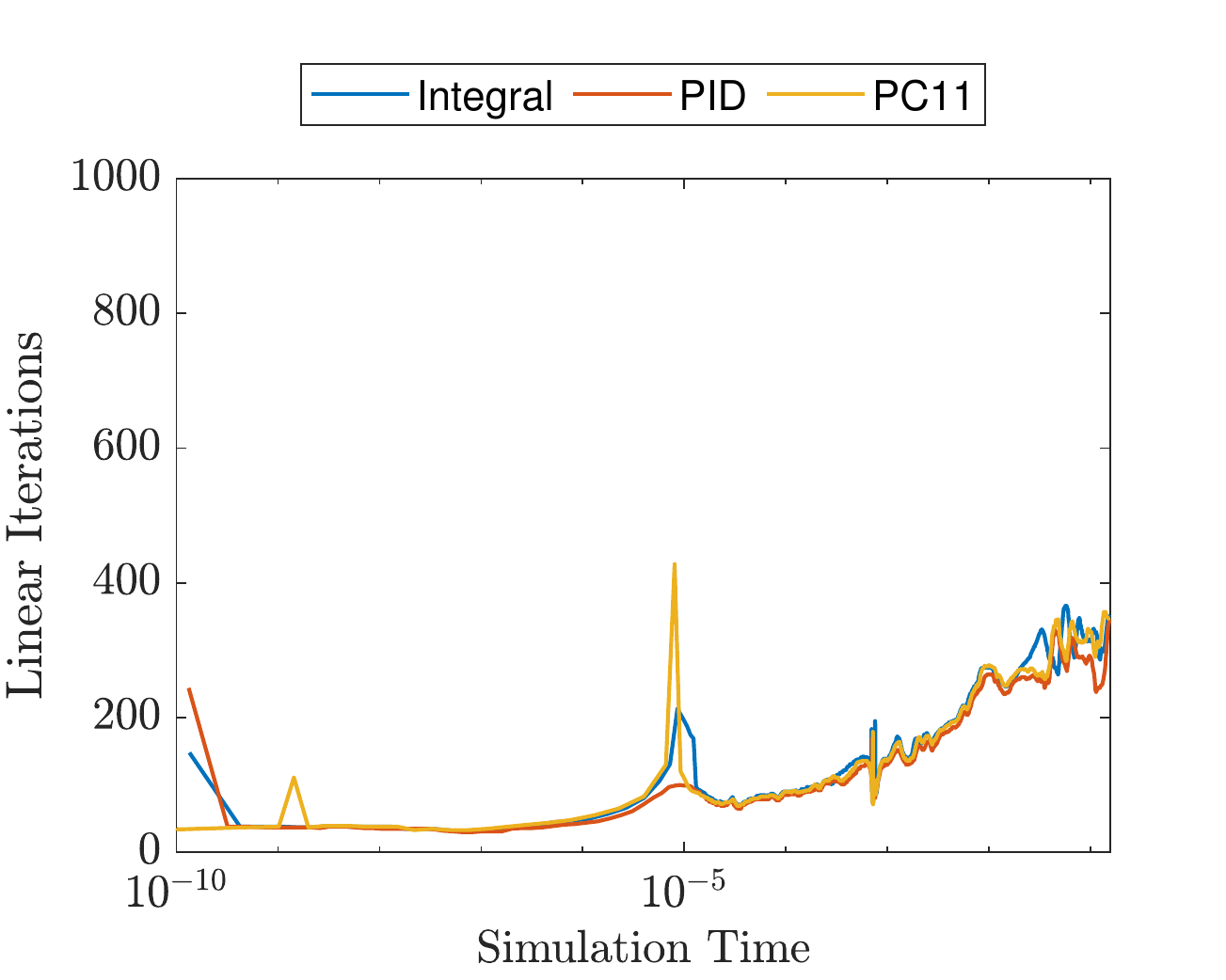}
		\label{fig:litexC}
	\end{minipage}
	\caption{Number of nonlinear (left) and linear iterations (right) for test cases A, B and C, respectively. Rejected steps included. Solver tolerances: $\eta_{NL}=10^{-5}$, $\eta_r=10^{-5}$, $\eta_a=10^{-8}$}.
	\label{fig:nitlit_testcases}
\end{figure}

\begin{table}
	\centering
	\caption{Results for the time adaptivity schemes for each time step controller in the 2D diblock copolymer simulations.}
	\begin{tabular}{|c|c|c|c|c|c|c|}
		\hline
		Test & Time Step           & Accepted      &  Rejected  &  Avg. Nonlinear & Avg. Linear  & Relative CPU\\ 
		Case &Controller     	  &     Steps     &   Steps    &  Iterations     & Iterations   & Effort\\
		\hline
		&I                  &  $2001$  &  $114$  &  $6.9940$ & $131.8300$   & $0.77$\\
		A &PID              &  $2437$  &  $24$   &  $6.6672$ & $121.0061$   & $0.86$\\
		&PC11               &  $2496$  &  $66$   &  $6.6835$ & $136.4883$   & $1.00$\\
		\hline
		&I                  &  $1656$  &  $20$ &  $6.8061$ & $173.3097$  & $0.88$\\
		B &PID              &  $2030$  &  $6$  &  $6.7487$  & $160.7867$ & $1.00$\\
		&PC11               &  $1731$  &   $7$ &  $6.7567$ & $170.4344$  & $0.90$\\		
		\hline
		&I                  &  $1623$ &  $15$ &  $6.6531$ & $178.9383$  & $0.92$\\
		C &PID              &  $1952$ &  $5$  &  $6.6393$ & $162.0035$  & $1.00$\\
		&PC11               &  $1669$ &   $7$ &  $6.6231$ & $170.7717$  & $0.90$\\	
		\hline
	\end{tabular}
	\label{tab:cop2d}
\end{table}

\begin{figure}
	\centering
		\begin{minipage}{.45\textwidth}
		\centering
		\includegraphics[width=.95\linewidth]{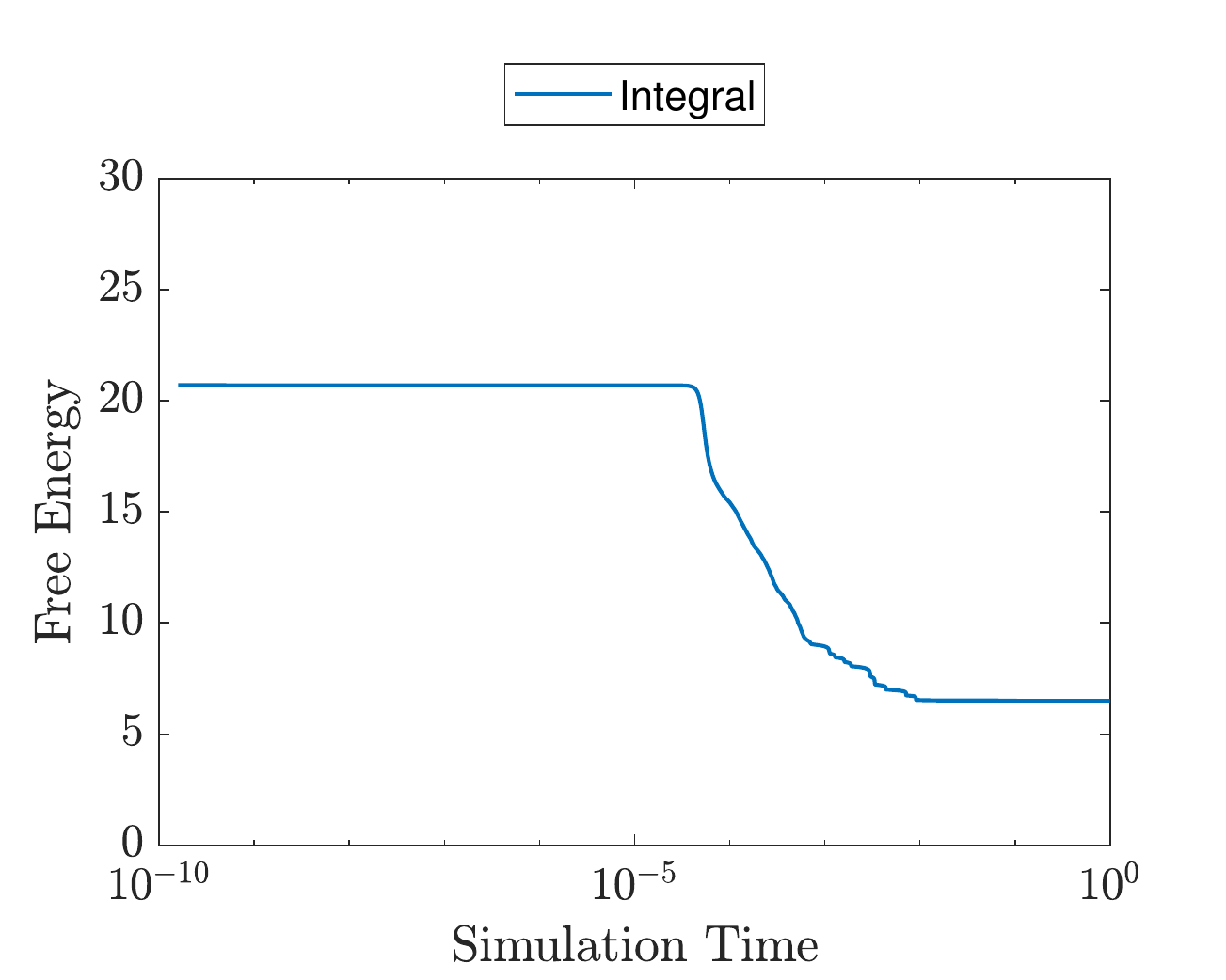}
	\end{minipage}
	\begin{minipage}{.45\textwidth}
		\centering
		\includegraphics[width=.95\linewidth]{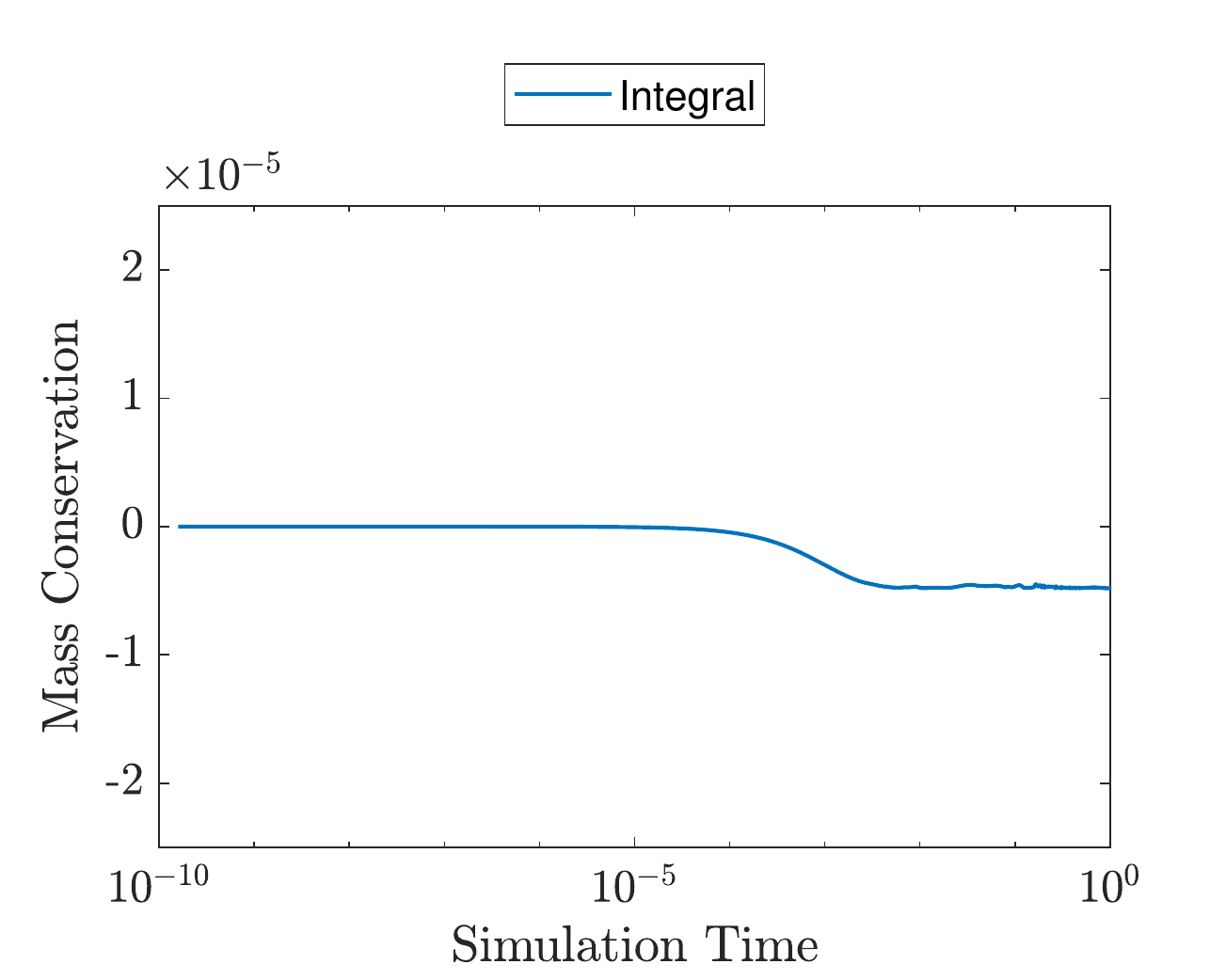}
	\end{minipage}\\
	\begin{minipage}{.45\textwidth}
		\centering
		\includegraphics[width=.95\linewidth]{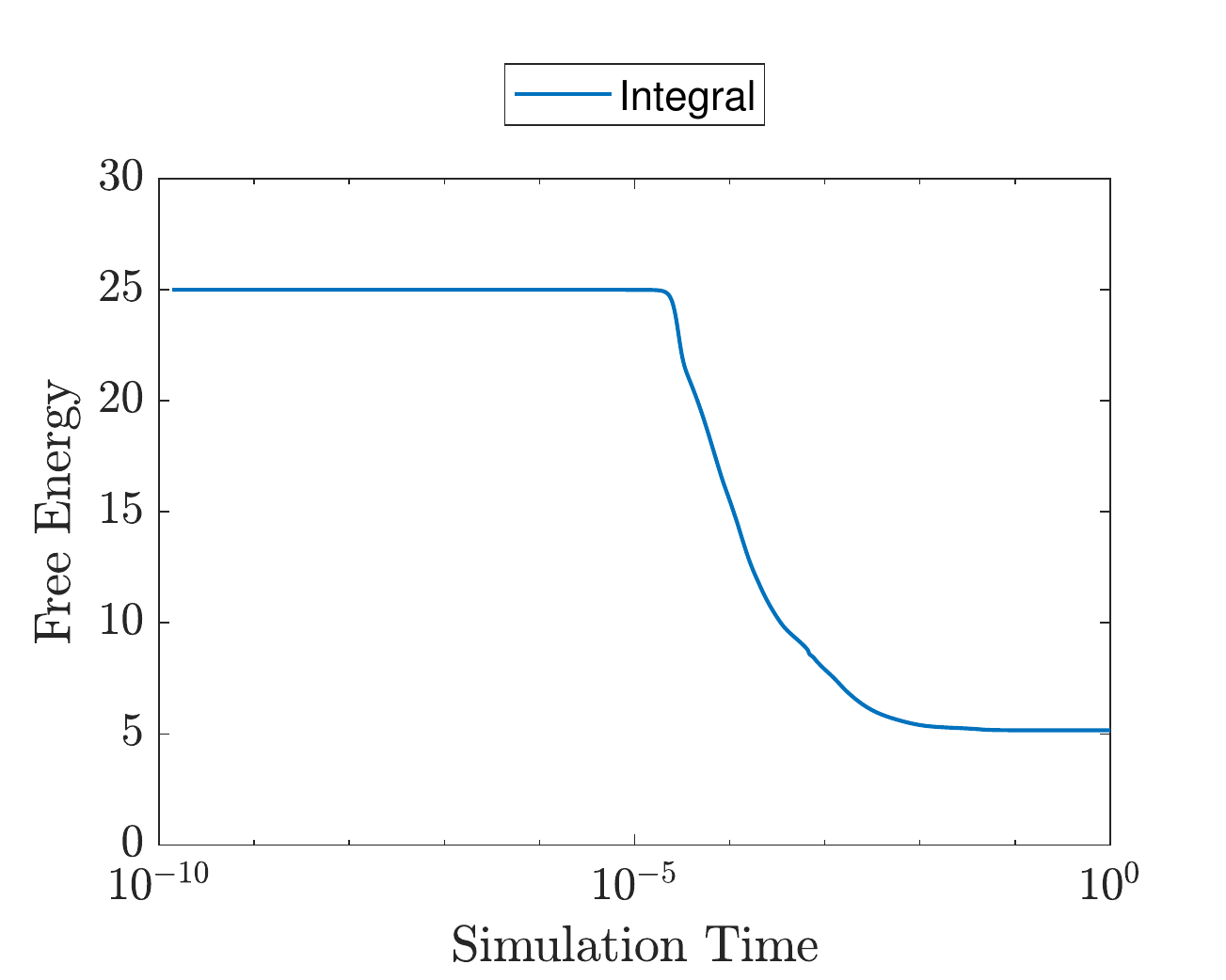}
	\end{minipage}%
	\begin{minipage}{.45\textwidth}
		\centering
		\includegraphics[width=.95\linewidth]{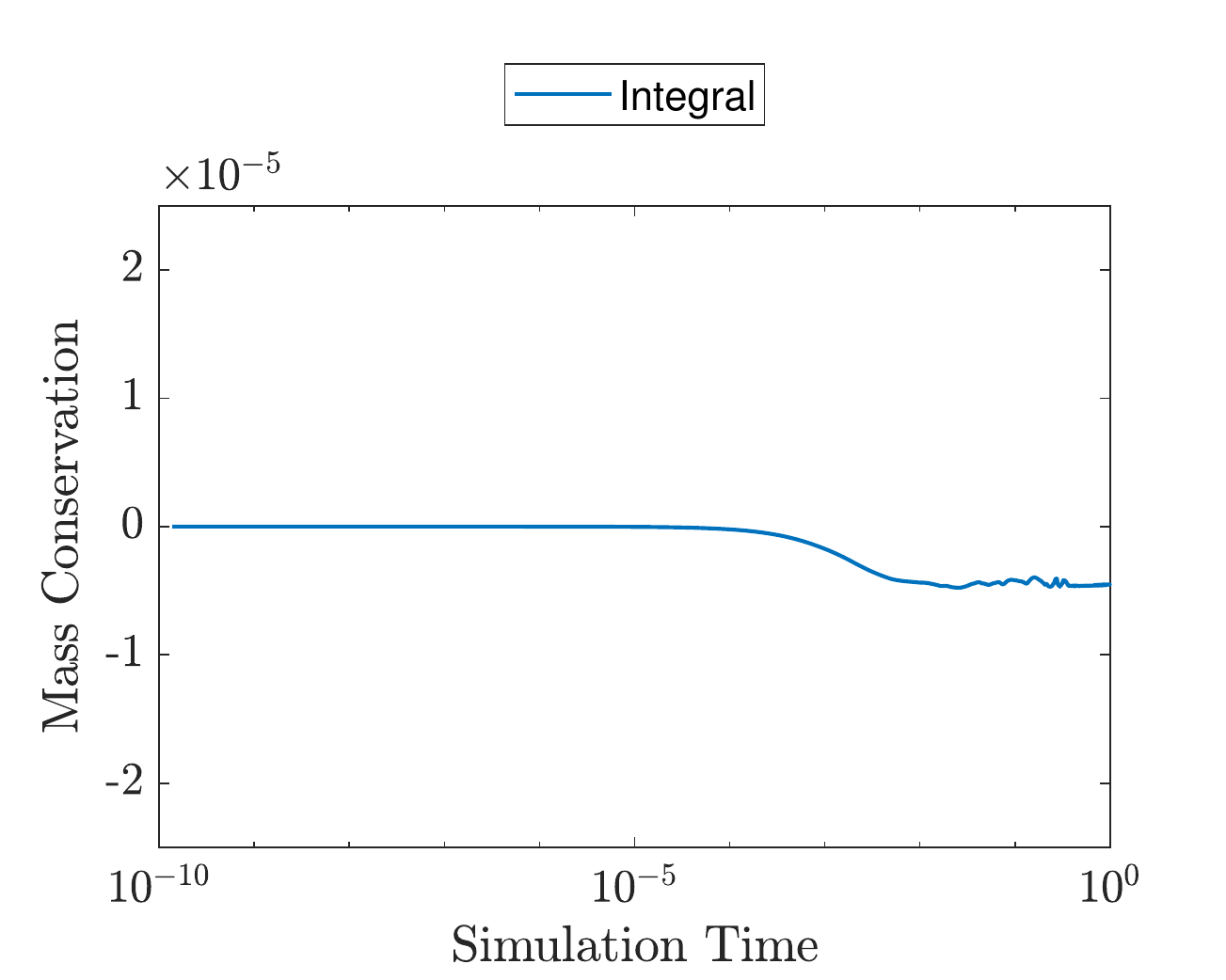}
	\end{minipage}\\
	\begin{minipage}{.45\textwidth}
		\centering
		\includegraphics[width=.95\linewidth]{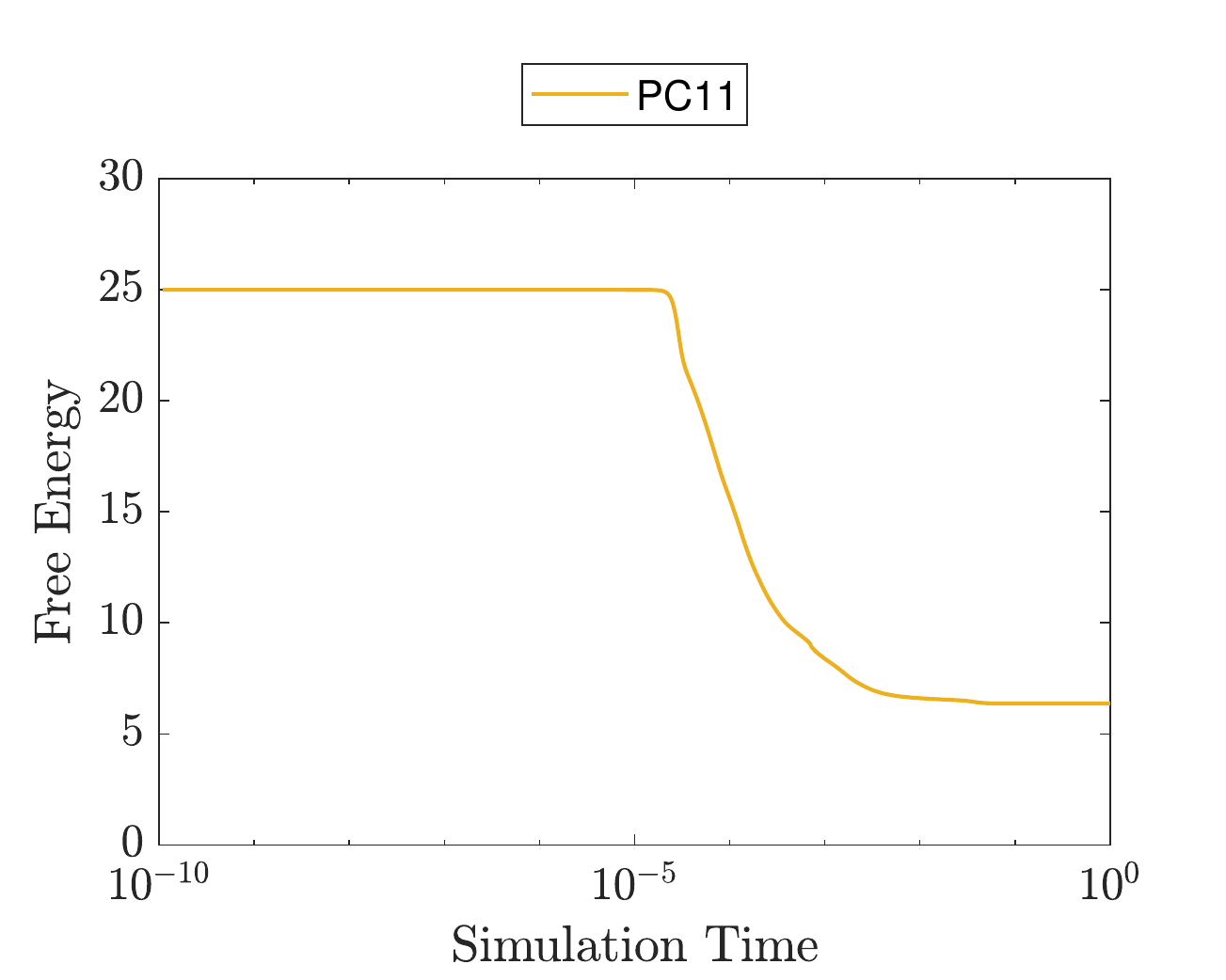}
	\end{minipage}
	\begin{minipage}{.45\textwidth}
		\centering
		\includegraphics[width=.95\linewidth]{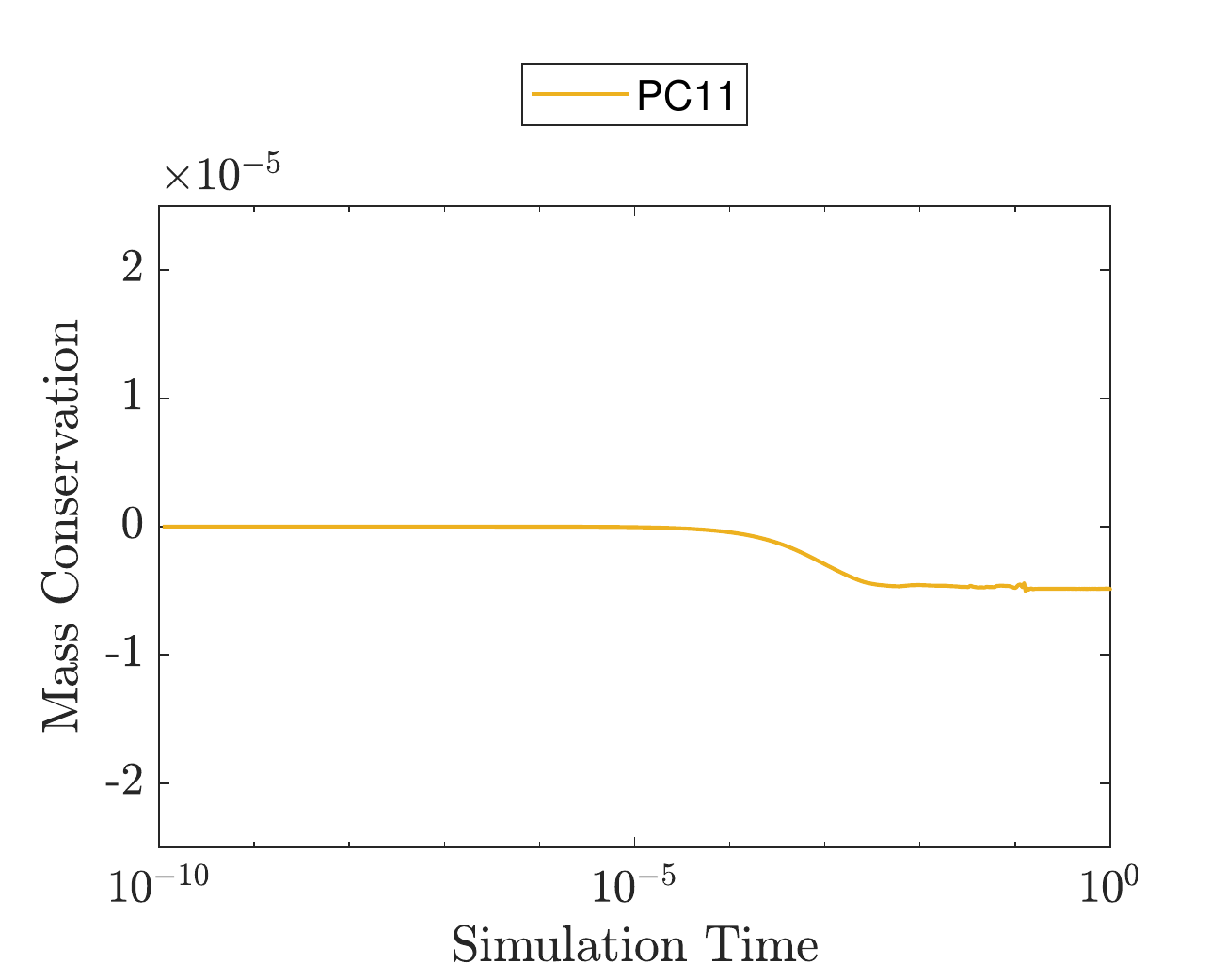}
	\end{minipage}%
	\caption{Free energy and mass conservation for the test cases A, B and C, from top to bottom, for the best performing parameters.}
	\label{fig:massfecop2}
\end{figure}

Figure \ref{fig:dtcop2d} shows the time step size history for these simulations. We see that in test case A the PC11 controller presents a sharp decrease in the time step size in the last simulation stages, but as the simulation approaches the steady-state, the time step size increases, returning to values of the same order of the other two controllers. Table \ref{tab:cop2d} and Figure \ref{fig:nitlit_testcases} show for all test cases the performance results for the three controllers. 
The I controller is the best in test cases A and B, while PC11 is the best performing controller in test case C. We also notice a smaller number of accepted and rejected time steps for simulations with $\bar{\phi} = 0.0$, that is, test cases B and C, in comparison with simulations where $\bar{\phi} = 0.3$ (test case A). The number of linear iterations is smaller for the three controllers in case A, but the number of nonlinear iterations is of the same order in all test cases for the three controllers. The PC11 controller has the larger average number of linear iterations on Test Case A, but the same behavior from the previous simulations is seen on Test Cases B and C, which is, the I controller has the largest number of average linear iterations while the PID presents the least. We note in Figure \ref{fig:nitlit_testcases} that the solutions with PC11 present a sharp increase in the number of linear iterations around $t=1 \times 10^{-5}$, particularly for test case A. The PID solutions at the same time interval exhibit the lower number of iterations. In all three test cases, the number of linear iterations increases when we approach the steady-state. We then evaluate the free energy decay and mass conservation, as shown in Fig. \ref{fig:massfecop2} for the best performing controllers. We see that the free energy decays for all test cases. Mass is for practical purposes conserved, with losses of order $10^{-5}$. Recall that the tolerance for the nonlinear solver is $\eta_{NL}=10^{-5}$, and for the linear solver $\eta_r=10^{-5}$ and $\eta_a=10^{-8}$. Thus the values for mass conservation are compatible with the accuracy obtained in each time step solution. 

\par
\begin{figure}
	\centering
	\begin{minipage}{.75\textwidth}
		\centering
		\includegraphics[trim={5.00cm 0 2.00cm 0},clip,width=.95\linewidth]{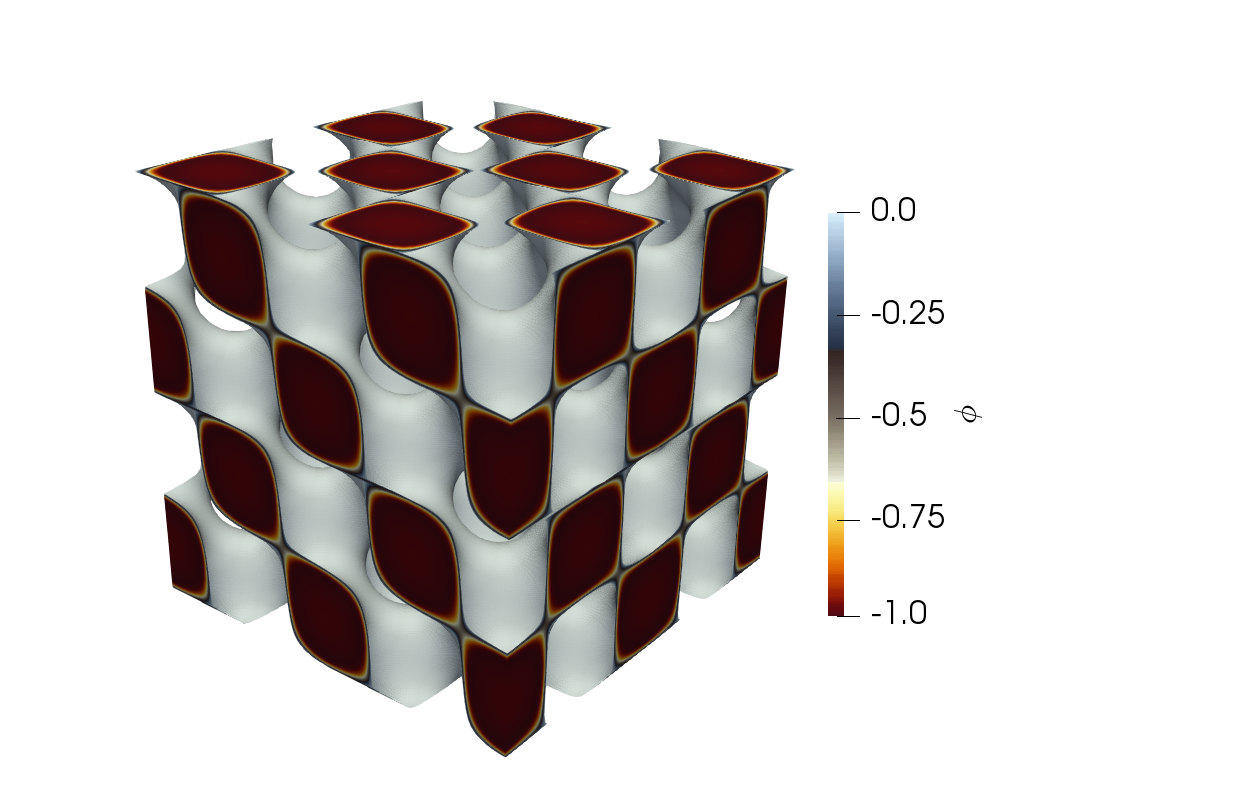}
	\end{minipage}\\
	\begin{minipage}{.75\textwidth}
		\centering
		\includegraphics[trim={2.00cm 0 5.00cm 0},clip,width=.95\linewidth]{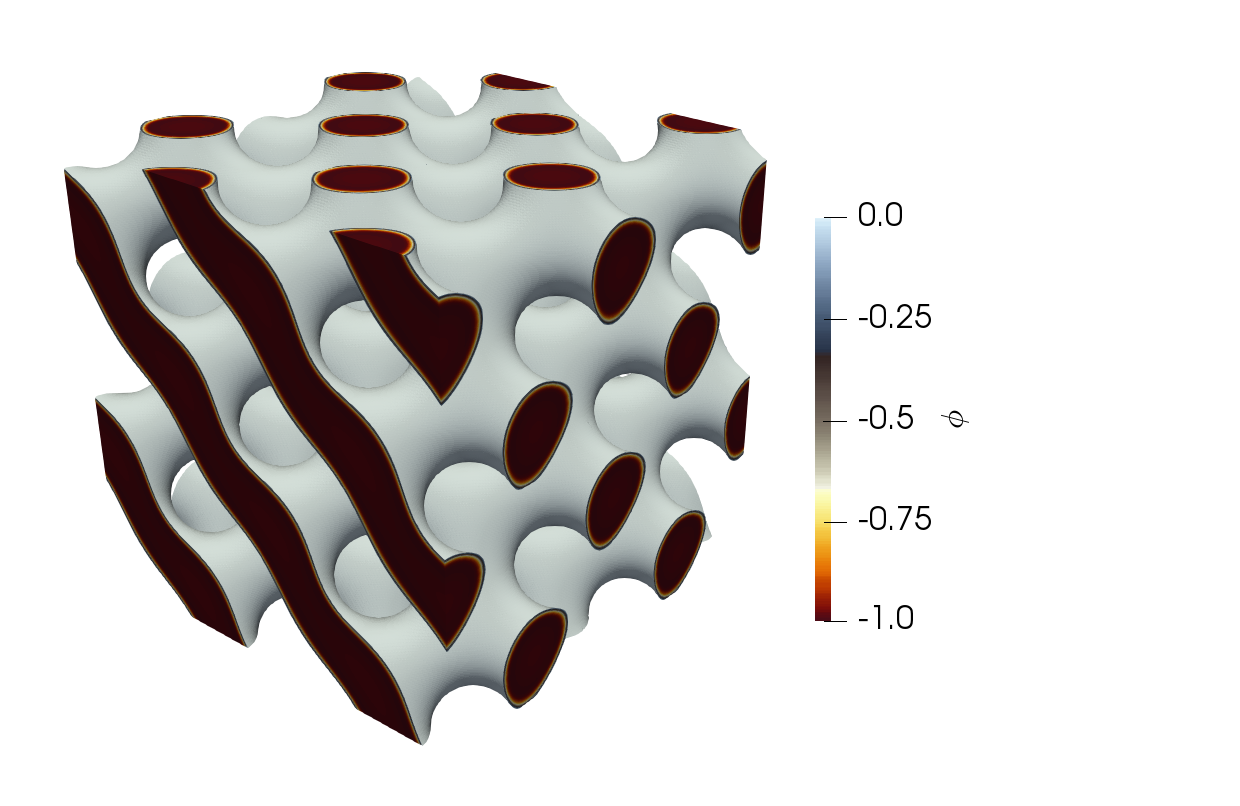}
	\end{minipage}
	\caption{Structure of a monomer on the generated diblock copolymer melts from the 3D simulations for $\bar{\phi} = 0.0$ (top) and $0.3$ (bottom).}
	\label{fig:cop3d}
\end{figure}

\begin{figure}
	\centering
	\begin{minipage}{.75\textwidth}
		\centering
		\includegraphics[clip,width=.95\linewidth]{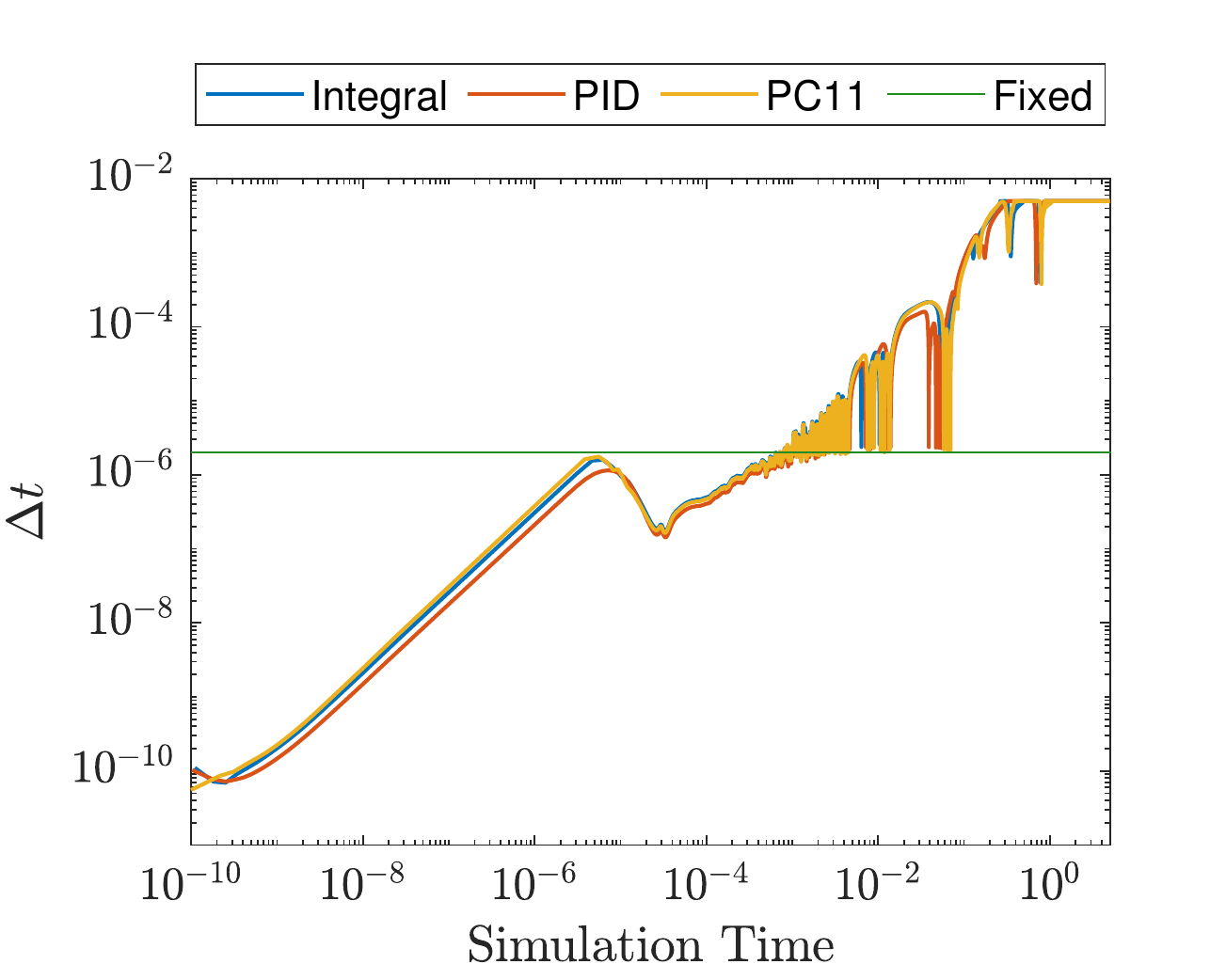}
	\end{minipage}\\
	\begin{minipage}{.75\textwidth}
		\centering
		\includegraphics[clip,width=.95\linewidth]{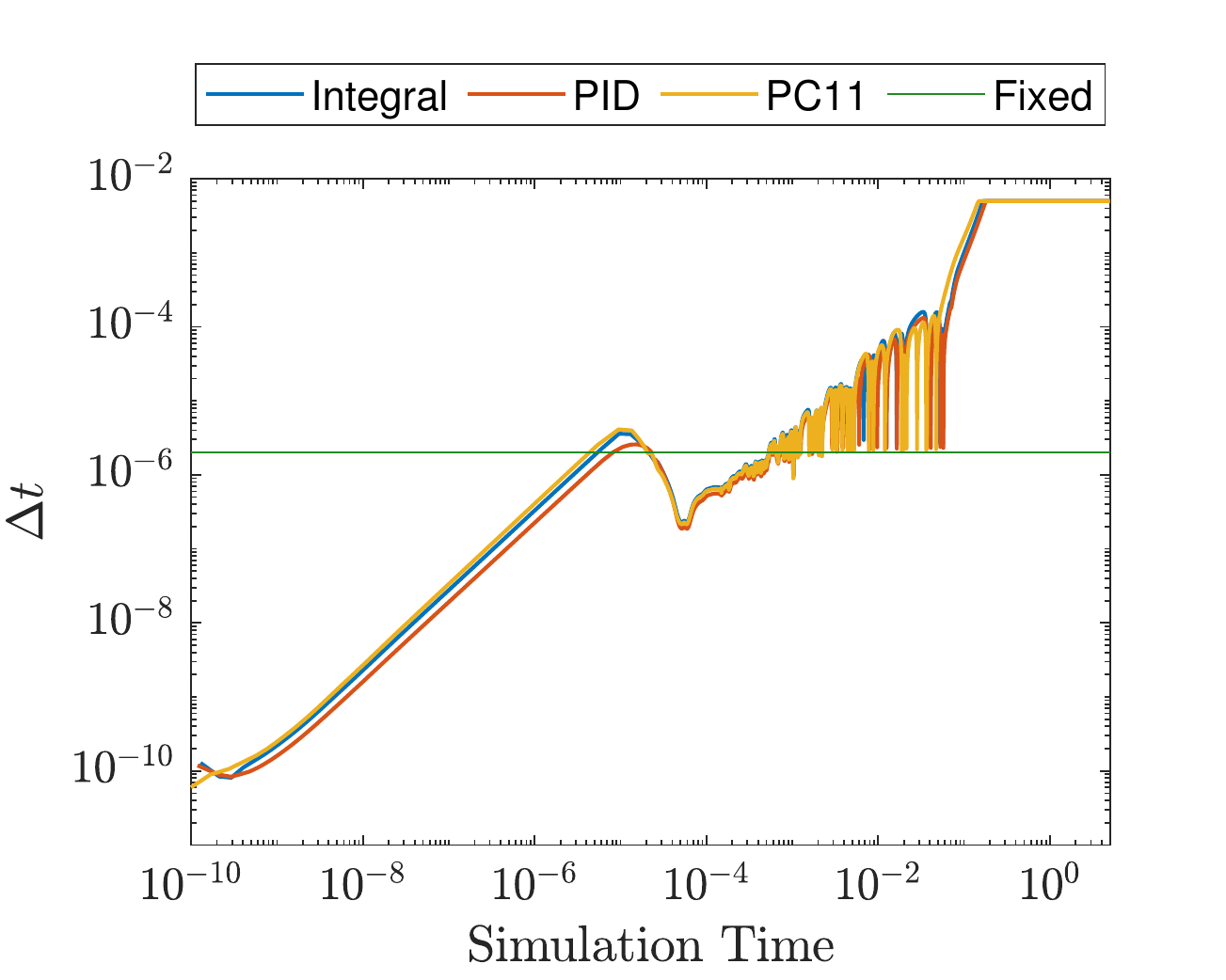}
	\end{minipage} 
	\caption{Time step size history for the three controllers. Top, $\bar{\phi} = 0.0$ and bottom $\bar{\phi}=0.3$}
	\label{fig:dt_cop3d}
\end{figure}

\begin{figure}
	\centering
	\begin{minipage}{.45\textwidth}
		\centering
		\includegraphics[clip,width=.95\linewidth]{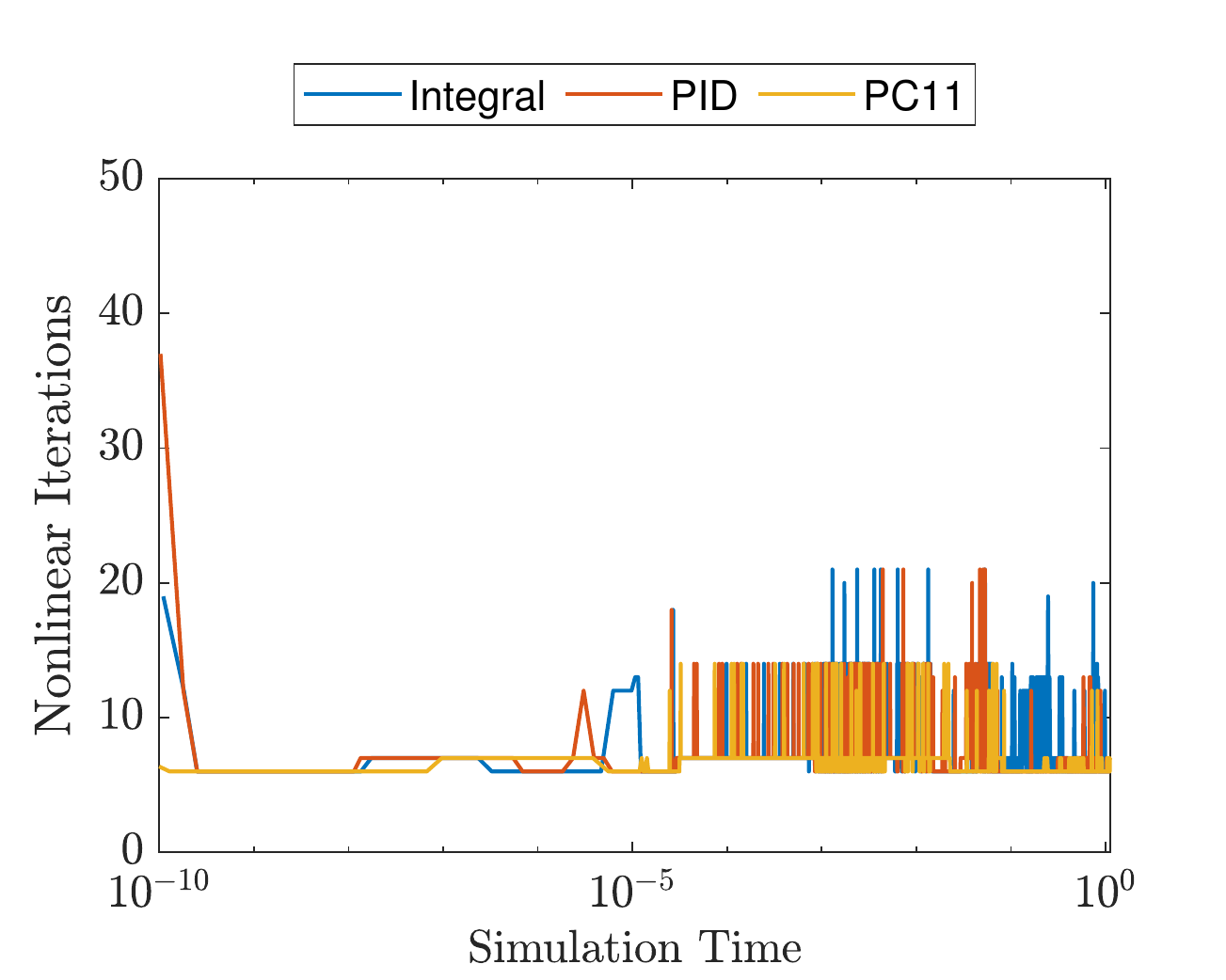}
	\end{minipage}
	\begin{minipage}{.45 \textwidth}
		\centering
		\includegraphics[clip,width=.95\linewidth]{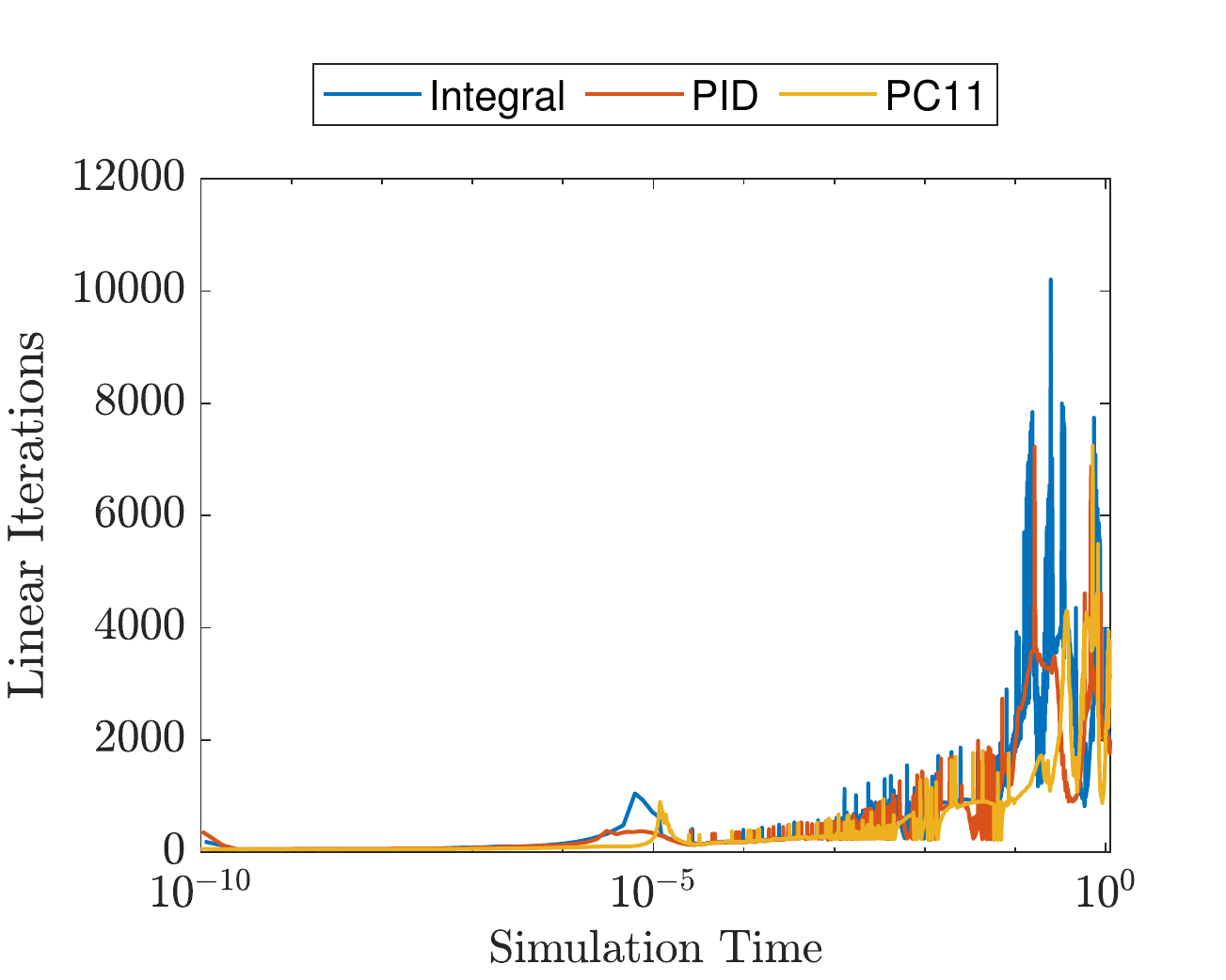}
	\end{minipage} \\
	\begin{minipage}{.45\textwidth}
		\centering
		\includegraphics[clip,width=.95\linewidth]{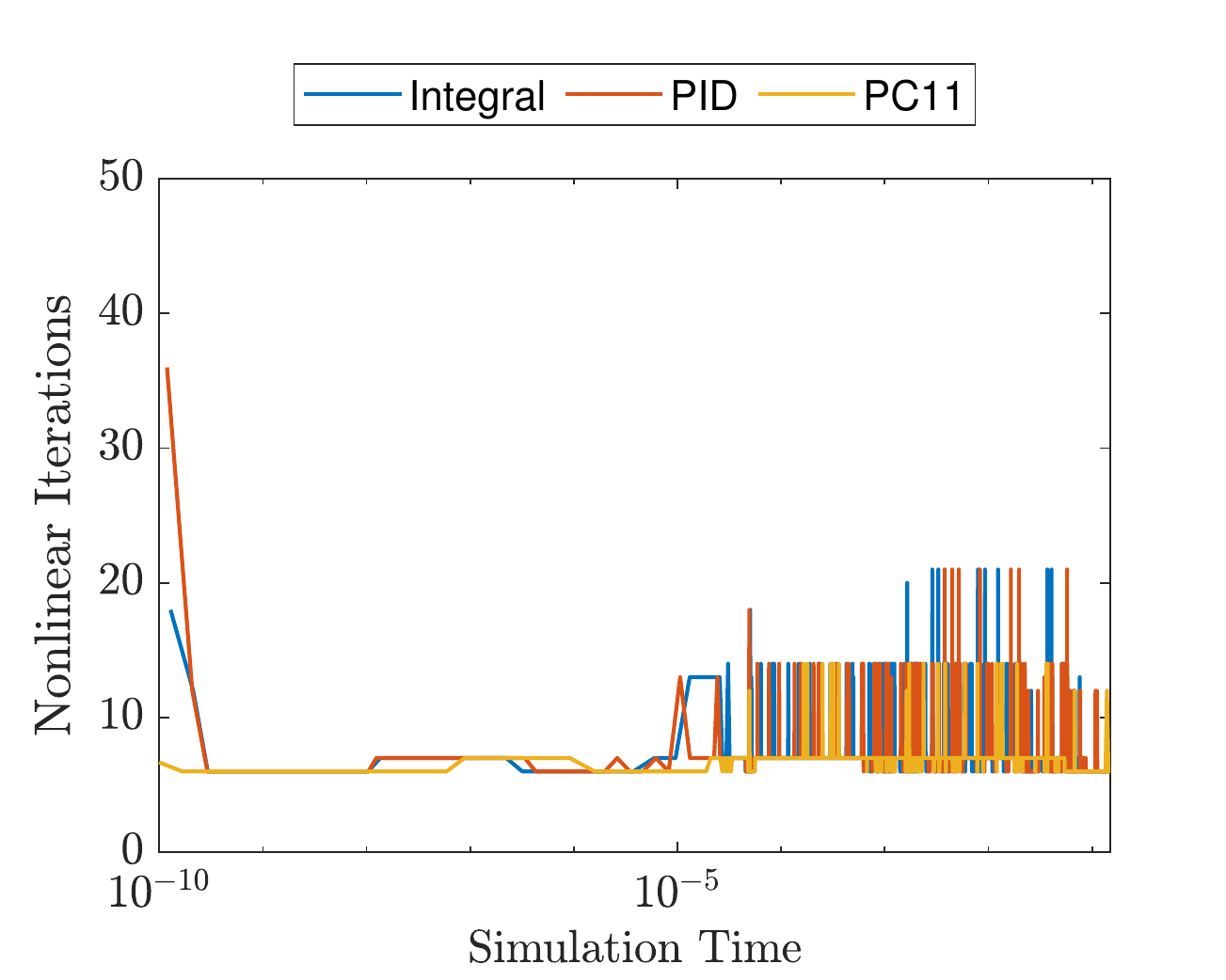}
	\end{minipage}
	\begin{minipage}{.45 \textwidth}
		\centering
		\includegraphics[clip,width=.95\linewidth]{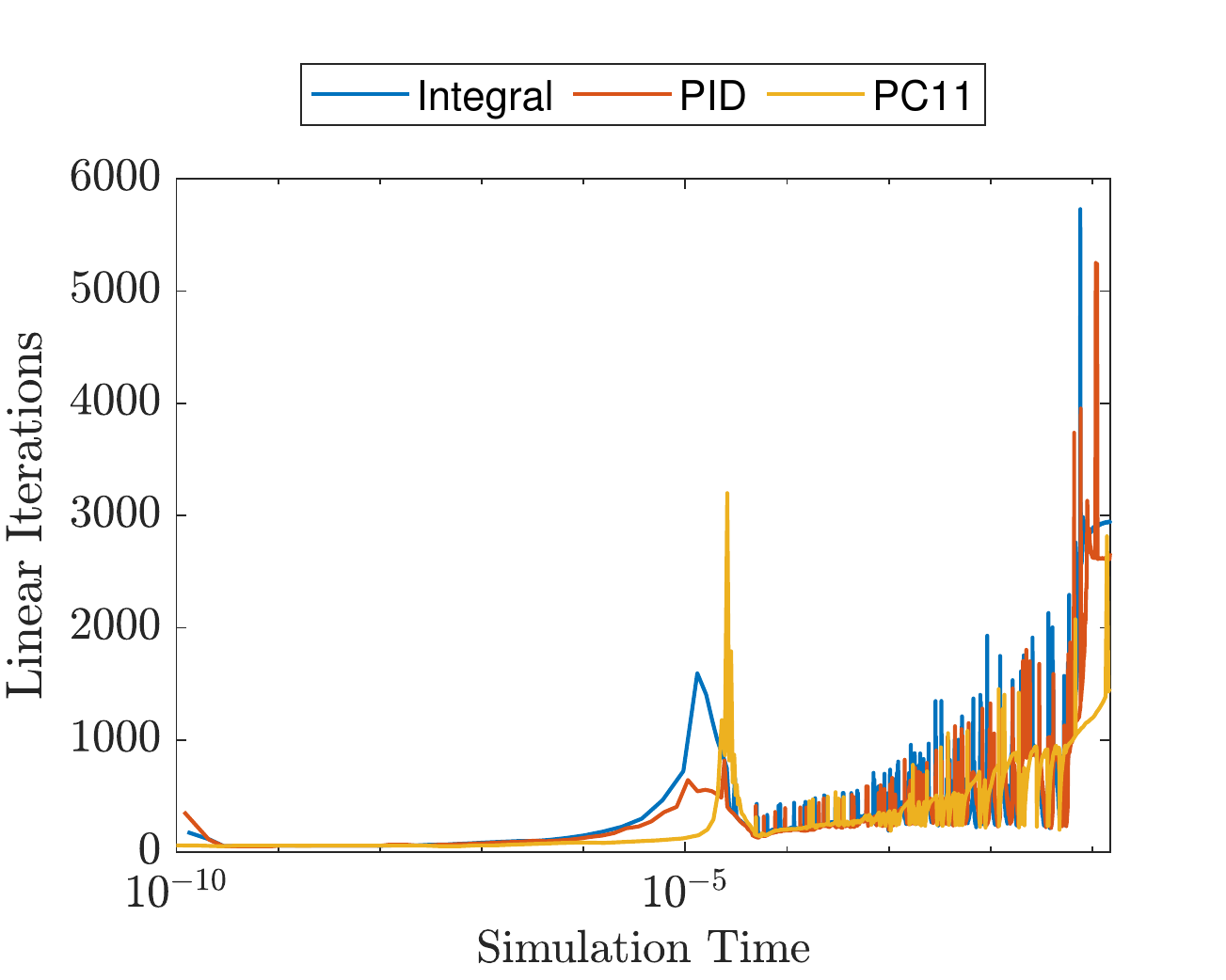}
	\end{minipage}
	\caption{Number of nonlinear and linear iterations during simulation for the three controllers. Rejected steps included. Top $\bar{\phi} = 0.0$; bottom, $\bar{\phi}=0.3$. Solver tolerances: $\eta_{NL}=10^{-5}$, $\eta_r=10^{-5}$, and $\eta_a=10^{-8}$.}
	\label{fig:nitlit3d}
\end{figure}

\begin{figure}
	\centering
	\begin{minipage}{.45\textwidth}
		\centering
		\includegraphics[clip,width=.95\linewidth]{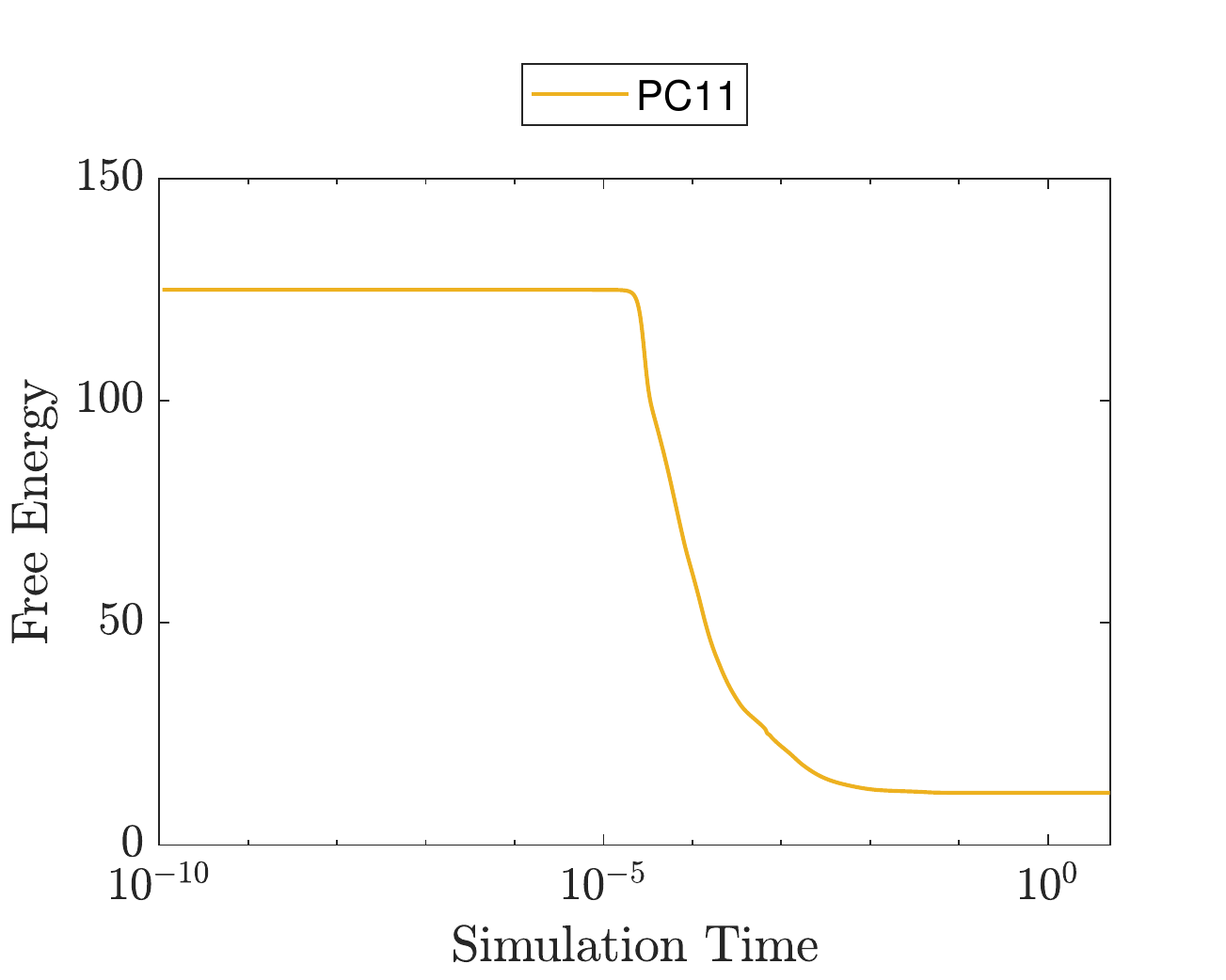}
	\end{minipage}
	\begin{minipage}{.45 \textwidth}
		\centering
		\includegraphics[clip,width=.95\linewidth]{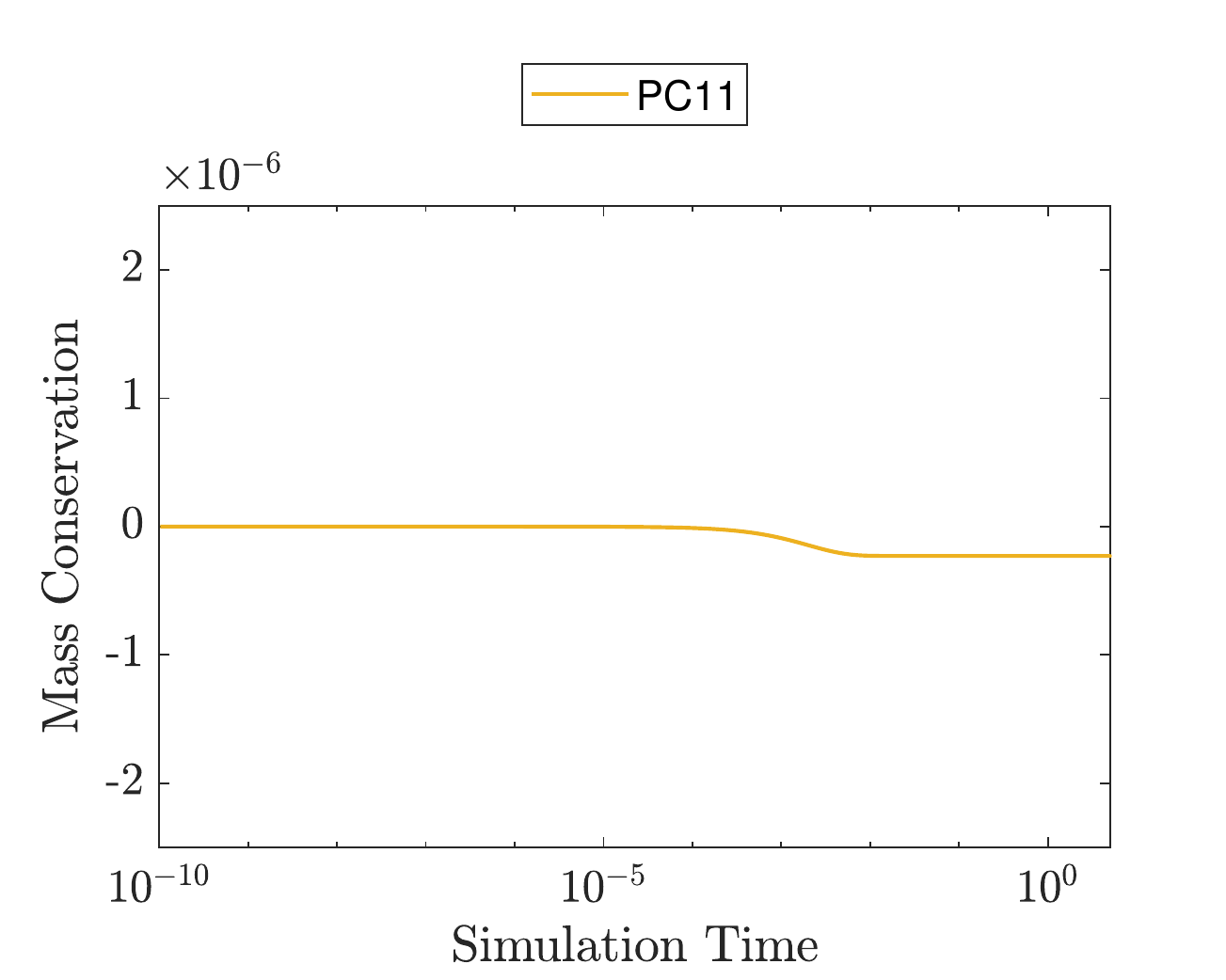}
	\end{minipage} \\
	\begin{minipage}{.45\textwidth}
		\centering
		\includegraphics[clip,width=.95\linewidth]{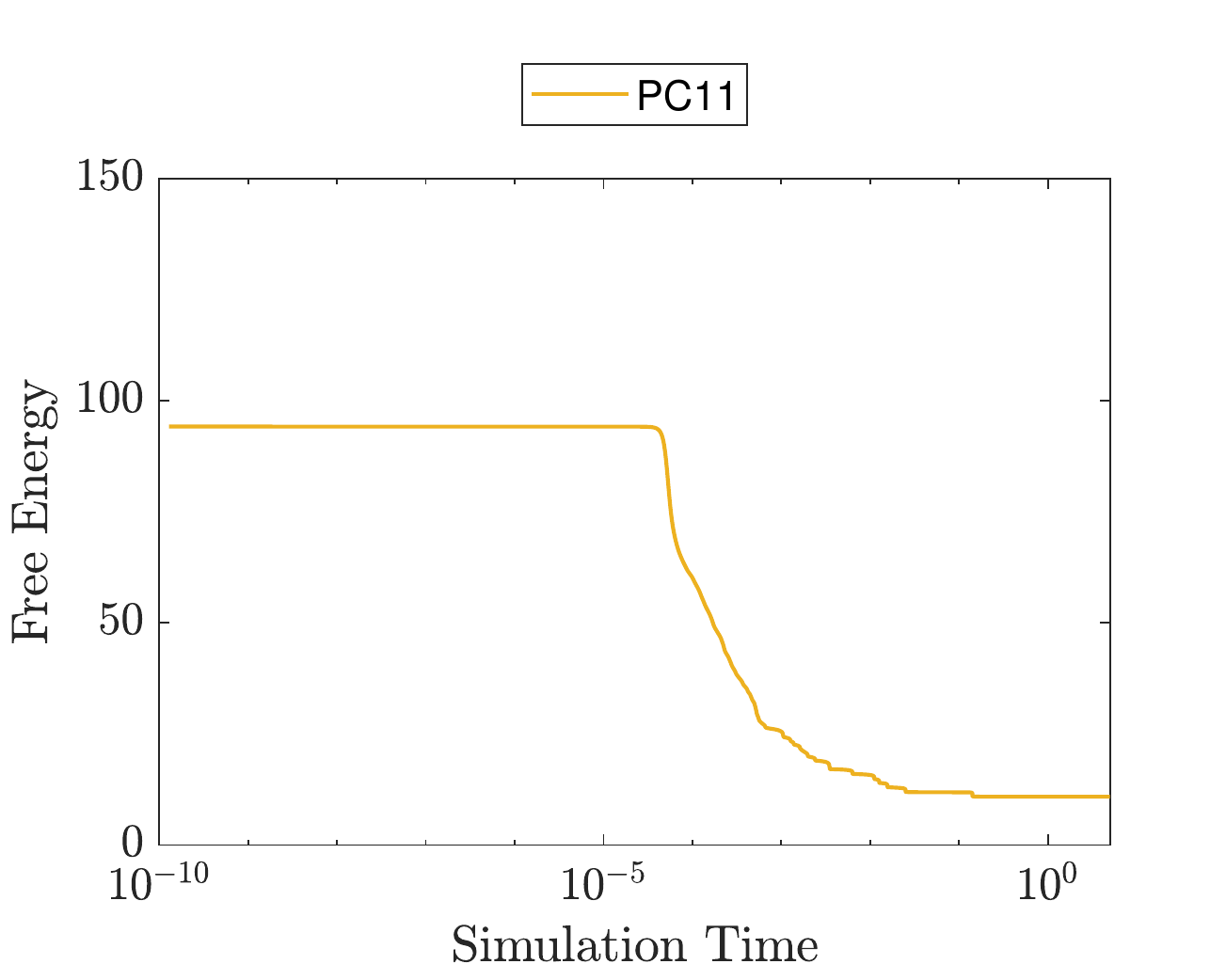}
	\end{minipage}
	\begin{minipage}{.45\textwidth}
		\centering
		\includegraphics[clip,width=.95\linewidth]{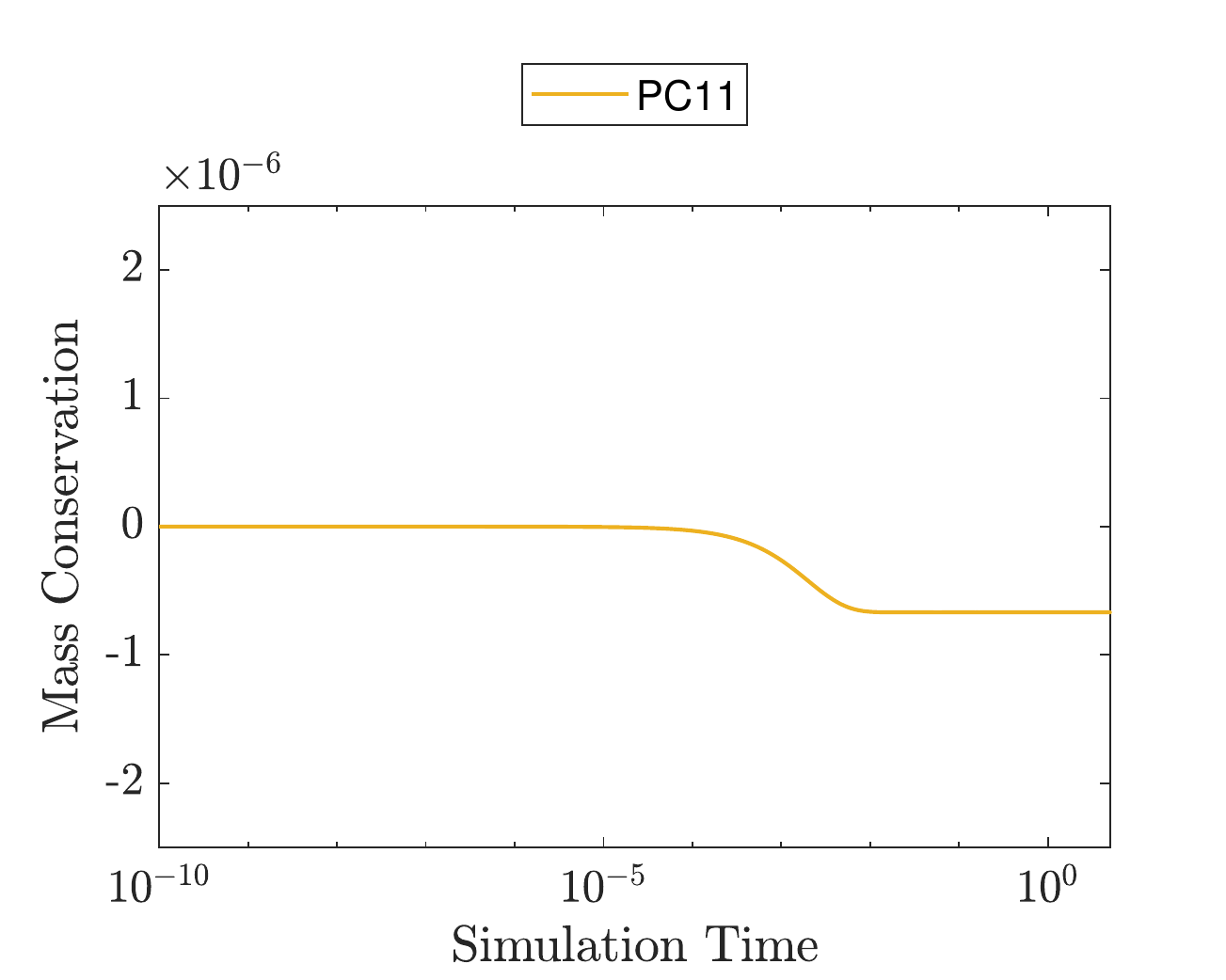}
	\end{minipage}
	\caption{Free energy and mass conservation for the 3D simulations for the best performing controllers. Top $\bar{\phi}=0.0$; bottom, $\bar{\phi}=0.3$.}
	\label{fig:massfecop3d}
\end{figure}

 
We extend our analysis to three dimensions. We consider a cubic domain with $129^3$ nodes, trilinear hexahedral elements, $\sigma = 500$ and evaluate the cases where $\bar{\phi} = 0.0$ and $0.3$. Figure \ref{fig:cop3d} shows the melt structure for the two 3D simulations with the consistent patterns observed in this phenomenon. The chosen parameters lead to stable melts, where the case where $\bar{\phi} = 0.0$ is a bicontinuous melt and $\bar{\phi} = 0.3$ leads to a perforated layer melt \cite{Khandpur1995}. Steady-state is reached on approximately $t = 1.11$ for the case where $\bar{\phi} = 0.0$ and $t = 0.18$ for $\bar{\phi} = 0.3$. Initially, we discuss the time step size histories shown in Figure \ref{fig:dt_cop3d}. We can see that for $\bar{\phi} = 0.0$ and $\bar{\phi} = 0.3$, the time step histories exhibit an initial stage with a fast time step-growth, an intermediate stage, where the time step increases but oscillates, and the final stage where the time step recover a fast growth. For $\bar{\phi} = 0.0$ we also see smaller oscillations in the final stage. We note that the case where $\bar{\phi} = 0.0$ is different from the 2D test cases B and C. In the 3D case, the time step size reveals a much more complex behavior than the 2D case. This behavior is related to the generation of the complex structure melts exhibited in Figure \ref{fig:cop3d}. In terms of efficiency, Fig. \ref{fig:nitlit3d} and Table \ref{tab:cop3d} show the performance data and the time history of the number of nonlinear and linear iterations for the 3D cases. We note that the number of accepted and rejected steps and the average number of nonlinear and linear iterations increased compared to the 2D test cases, reflecting the higher complexity existent in 3D copolymer simulations. We observe that the I controller has the largest number of linear iterations again while the PID controller the smaller. Also, we can see in Fig. \ref{fig:nitlit3d} that for both cases ($\bar{\phi} = 0.0$ and $\bar{\phi} = 0.3$) the number of linear iterations exhibits a remarkable growth as the solution approaches the steady-state. This growth is exceedingly large for the I and PID controllers. We observe that the PC11 controller exhibits the best performance, saving around $20\%$ of the computational effort required by the I and PID controllers for the case where $\bar{\phi} = 0.0$ and almost a $30\%$ gain for $\bar{\phi} = 0.3$. In terms of accuracy, we observe the free energy decay and mass conservation properties in Fig. \ref{fig:massfecop3d} for the best performing controllers for each case. Again, mass conservation is within the accuracy obtained at each time step solve.

\begin{table}
	\centering
	\caption{Results for the time adaptivity schemes for each time step controller in the 3D diblock copolymer simulations.}
	\begin{tabular}{|c|c|c|c|c|c|c|}
		\hline
		$\bar{\phi}$& Time Step           &  Accepted      &  Rejected  &  Avg. Nonlinear & Avg. Linear  & Relative CPU\\ 
		&Controller     	                &     Steps     &   Steps    &  Iterations     & Iterations   & Effort\\
		\hline
		&I                  &  $4339$ &  $313$  &  $7.5168$ & $722.6998$    & $1.00$\\
		$0.0$ &PID          &  $5017$ &  $157$  &  $7.1780$ & $607.4357$    & $0.95$\\
		&PC11               &  $4481$ &  $47$   &  $6.8601$ & $593.4483$    & $0.79$\\
		\hline
		&I                  &  $3762$  &  $224$   &  $7.5472$  & $587.0125$  & $0.99$\\
		$0.3$ &PID          &  $4380$  &  $166$   &  $7.3134$  & $520.3438$   & $1.00$\\
		&PC11               &  $4339$  &   $23$   &  $6.9294$  & $463.0867$   & $0.70$\\		
		\hline
	\end{tabular}
	\label{tab:cop3d}
\end{table}

\section{Conclusions}
\par This paper presents time adaptivity schemes for the nonlocal Cahn-Hilliard equation derived from the Ohta-Kawasaki free energy functional. We have unified the time adaptivity schemes under the linear feedback control theory. The error estimate for the time adaptivity schemes results from an extrapolation method proposed in ~\cite{vignals}, avoiding solving the problem twice at the same time step. We test our scheme on simple examples such as a 2D phase separation with constant mobility and more challenging ones, the diblock copolymer self-assembly in two and three dimensions. We evaluate the accuracy and performance of three time step size controllers. The simulations reveal the PID controller's conservative behavior, which required the computation of more times steps than the other controllers to reach the steady-state and the I controller's aggressive behavior, with many rejected time steps. Our simulations for the CH equation suggest that the PC11 controller is the best performing controller. However, for the 2D NCH simulations, the results vary. Initially, we assessed the values for the tolerance and safety coefficient and observed that the I controller presents the best performance compared to the other controllers when the tolerance is stricter and the safety coefficient smaller. For the standard values used in the literature, the results obtained are accurate, and the PC11 yields the best performance. Besides, we consider different copolymer parameters and evaluate their influence on the controllers' performance. We note that in some cases, the PC11 and the I controllers have a better performance. Subsequently, we consider two 3D NCH simulations. In both, PC11 is the most efficient controller. Furthermore, in terms of accuracy, our numerical results with the second-order energy stable time integration method introduced in ~\cite{vignals} coupled with temporal adaptivity show numerical of evidence mass conservation and free energy decay for the nonlocal case, as anticipated theoretically in \cite{Gal2018}.
\par 
We note that the controllers reveal a subtle interplay between the time step size and the nonlinear and linear system solves. For most of the simulations presented in this study, the I controller has the largest average number of linear iterations while the PID controller the least. This effect is an important metric in the sense that the average number of linear and nonlinear iterations must be taken into account with the total number of times steps to evaluate the performance of a time step size controller. For instance, for the 3D case where $\Bar{\phi} = 0.3$, the I controller presents a smaller number of total time steps (accepted and rejected) than the PC11, but the latter has a smaller overall number of linear iterations. This observation is important because we observe a $30\%$ performance gain by using the PC11 controller in this time and resource-demanding simulation. In terms of accuracy, we also notice that the time adaptivity scheme reproduces all different physics, such as those found in the local and the nonlocal Cahn-Hilliard equation in two and three dimensions. An essential statement obtained from this study is that one should always consider temporal adaptivity for the nonlocal Cahn-Hilliard equations since the gains in the computational effort are substantial.  

\section*{Acknowledgements}
This research was financed in part by the Coordena\c{c}\~ao de Aperfei\c{c}oamento de Pessoal de N\'ivel Superior - Brasil (CAPES) - Finance Code 001. This research has also received funding from CNPq and FAPERJ. Computer time in Lobo Carneiro supercomputer was provided by the High Performance Computer Center at COPPE/Federal University of Rio de Janeiro, Brazil.

\bibliographystyle{Paper01}
\bibliography{Paper01.bib}

\begin{thebibliography}{10}
\providecommand \doibase [0]{http://dx.doi.org/}%

\bibitem{CH1}
Cahn JW, Hilliard JE. {F}ree {e}nergy of a {n}onuniform {s}ystem. I.
  {I}nterfacial {F}ree {E}nergy. {\it {T}he {J}ournal of {C}hemical {P}hysics}
  1958\string; 28(2)\string: 258-267.

\bibitem{CH2}
Cahn JW, Hilliard JE. {F}ree {e}nergy of a {n}onuniform {s}ystem. II.
  {T}hermodynamic {B}asis. {\it {T}he {J}ournal of {C}hemical {P}hysics}
  1958\string; 30(5)\string: 1121-1124.

\bibitem{Kim}
Kim J, Lee S, Choi Y, Lee S, Jeong D. {Basic Principles and Practical
  Applications of the Cahn-Hilliard Equation}. {\it Mathematical Problems in
  Engineering} 2016(ID 9532608)\string: 1-11.

\bibitem{ohtakawasaki}
Ohta T, Kawasaki K. {Equilibrium Morfology of Block Copolymer Melts}. {\it
  Macromolecules} 1986\string; 19(10)\string: 2621-2632.

\bibitem{choksi}
Choksi R, Peletier MA, Williams JF. {On the Phase Diagram for Microphase
  Separation of Diblock Copolymers: An Approach via a Nonlocal Cahn-Hilliard
  Functional}. {\it SIAM Journal on Applied Mathematics} 2009\string;
  69(6)\string: 1712–-1738.

\bibitem{bertozzi}
Bertozzi AL, Esedoglu S, Gillette A. {Inpainting of binary images using the
  Cahn-Hilliard equation}. {\it IEEE Transactions on Image Processing}
  2007\string; 16(1)\string: 285-291.

\bibitem{hohenberg}
Hohenberg P, Halperin B. {Theory of dynamic critical phenomena}. {\it Reviews
  of Modern Physics} 1977\string; 49(3)\string: 436-479.

\bibitem{gurtin}
Gurtin ME, Polignone D, Vinals J. {Two-phase binary fluids and immiscible
  fluids described by an order parameter}. {\it Mathematical Models and Methods
  in Applied Sciences} 1996\string; 06(06)\string: 815-831.

\bibitem{lowengrubtruskinovsky}
Lowengrub J, Truskinovsky L. {Quasi-incompressible Cahn-Hilliard fluids and
  topological transitions}. {\it Proceedings of the Royal Society A:
  Mathematical, Physical and Engineering Sciences} 1998\string;
  454(1978)\string: 2617-2654.

\bibitem{abels2012}
Abels H, Garcke H, Gr\"un G. {Thermodynamically consistent, frame indifferent
  diffuse interface models for incompressible two-phase flows with different
  densities}. {\it Mathematical Models and Methods in Applied Sciences}
  2012\string; 22(3)\string: 1-40.

\bibitem{adriano}
Espath LFR, Sarmiento AF, Vignal P, et al. {Energy exchange analysis in droplet
  dynamics via the Navier-Stokes-Cahn-Hilliard model}. {\it Journal of Fluid
  Mechanics} 2016\string; 797\string: 389-430.

\bibitem{hughesfracture}
Borden MJ, Clemens VV, Scott MA, Hughes TJR, Landis CM. {A phase-field
  description of dynamic brittle fracture.}. {\it Computer Methods in Applied
  Mechanics and Engineering} 2012\string; 217\string: 77-95.

\bibitem{duda}
Silva~Jr. MN, Duda FP, Fried E. {Sharp-crack limit of a phase-field model for
  brittle fracture}. {\it Journal of the Mechanics and Physics of Solids}
  2013\string; 61(11)\string: 2178-2195.

\bibitem{wise}
Wise SM, Lowengrub JS, Cristini V. {An adaptive multigrid algorithm for
  simulating solid tumor growth using mixture models}. {\it Mathematical and
  Computer Modelling} 2011\string; 53(1-2)\string: 1-20.

\bibitem{wu}
Wu X, {van Zwieten} GJ, {van der Zee} KG. {Stabilized second-order convex
  splitting schemes for Cahn-Hilliard models with application to
  diffuse-interface tumor-growth models}. {\it International Journal for
  Numerical Methods in Biomedical Engineering} 2014\string; 30(2)\string:
  180-203.

\bibitem{zhou}
Zhou S, Wang MY. {Multimaterial structural topology optimization with a
  generalized Cahn-Hilliard model of multiphase transition}. {\it Structural
  and Multidisciplinary Optimization} 2007\string; 33\string: 89-111.

\bibitem{shin}
Shin J, Lee HG, Lee JY. {Unconditionally stable methods for gradient flow using
  Convex Splitting Runge–Kutta scheme}. {\it Journal of Computational
  Physics} 2017\string; 347\string: 367-381.

\bibitem{giacomin1}
Giacomin G, Lebowitz JL. {Phase segregation dynamics in particle systems with
  long range interactions. I. Macroscopic limits}. {\it Journal of Statistical
  Physics} 1997\string; 87\string: 37-61.
\newblock \href {\doibase 10.1007/BF02181479} {doi: 10.1007/BF02181479}

\bibitem{giacomin2}
Giacomin G, Lebowitz JL. {Phase segregation dynamics in particle systems with
  long range interactions II: Interface motion}. {\it SIAM Journal on Applied
  Mathematics} 1998\string; 58(6)\string: 1707-1729.
\newblock \href {\doibase 10.1137/S0036139996313046} {doi:
  10.1137/S0036139996313046}

\bibitem{Gajewski2003}
Gajewski H, Zacharias K. {On a nonlocal phase separation model}. {\it Journal
  of Mathematical Analysis and Applications} 2003\string; 286(1)\string: 11-31.
\newblock \href {\doibase 10.1016/S0022-247X(02)00425-0} {doi:
  10.1016/S0022-247X(02)00425-0}

\bibitem{copbook}
Hamley IW. {\it The Physics of Block Copolymers}. 19.
\newblock Oxford University Press: New York, NY, USA .
\newblock 1998.

\bibitem{kimcop}
Kim H, Park S, Hinsberg W. {Block Copolymer Based Nanostructures: Materials,
  Processes, and Applications to Electronics}. {\it Chem. Rev.} 2010\string;
  110\string: 146-177.

\bibitem{Alberti2008}
Alberti G, Choksi R, Otto F. {Uniform energy distribution for an isoperimetric
  problem with long-range interactions}. {\it Journal of the American
  Mathematical Society} 2008\string; 22(2)\string: 569-605.
\newblock \href {\doibase 10.1090/s0894-0347-08-00622-x} {doi:
  10.1090/s0894-0347-08-00622-x}

\bibitem{Ren2003}
Ren X, Wei J. {On energy minimizers of the diblock copolymer problem}. {\it
  Interfaces and Free Boundaries} 2003\string; 5(2)\string: 193-238.
\newblock \href {\doibase 10.4171/IFB/78} {doi: 10.4171/IFB/78}

\bibitem{ohnsh}
Ohnishi I, Nishiura Y, Imai M, Matsushita Y. {Analytical solutions describing
  the phase separation driven by a free energy functional containing a
  long-range interaction term}. {\it Chaos} 1999\string; 9(2)\string: 329-341.

\bibitem{Cristoferi2018}
Cristoferi R. {On periodic critical points and local minimizers of the
  Ohta–Kawasaki functional}. {\it Nonlinear Analysis} 2018\string;
  168\string: 81-109.
\newblock \href {\doibase 10.1016/j.na.2017.11.004} {doi:
  10.1016/j.na.2017.11.004}

\bibitem{vanderberg}
{Van den Berg} JB, Williams JF. {Validation of the bifurcation diagram in the
  2D Ohta-Kawasaki problem}. {\it Nonlinearity} 2017\string; 30(4)\string:
  1584-1638.
\newblock \href {\doibase 10.1088/1361-6544/aa60e8} {doi:
  10.1088/1361-6544/aa60e8}

\bibitem{Choksi2011}
Choksi R, Maras M, Williams JF. {2D phase diagram for minimizers of a
  Cahn-Hilliard functional with long-range interactions}. {\it SIAM Journal on
  Applied Dynamical Systems} 2011\string; 10(4)\string: 1344-1362.
\newblock \href {\doibase 10.1137/100784497} {doi: 10.1137/100784497}

\bibitem{jeong2}
Jeong D, Shin J, Li Y, et al. {Numerical analysis of energy-minimizing
  wavelengths of equilibrium states for diblock copolymers}. {\it Current
  Applied Physics} 2014\string; 14(9)\string: 1263-1272.
\newblock \href {\doibase 10.1016/j.cap.2014.06.016} {doi:
  10.1016/j.cap.2014.06.016}

\bibitem{Farrell2017}
Farrell PE, Pearson JW. {A preconditioner for the Ohta-Kawasaki equation}. {\it
  SIAM Journal on Matrix Analysis and Applications} 2017\string; 38(1)\string:
  217-225.
\newblock \href {\doibase 10.1137/16M1065483} {doi: 10.1137/16M1065483}

\bibitem{cueto}
Cueto-Felgueroso L, Peraire J. {A time-adaptive finite volume method for the
  Cahn-Hilliard and Kuramoto-Sivashinsky equations}. {\it Journal of
  Computational Physics} 2008\string; 227(24)\string: 9985-10017.

\bibitem{gomezhughes}
G{\'{o}}mez H, Calo VM, Bazilevs Y, Hughes TJR. {Isogeometric analysis of the
  Cahn-Hilliard phase-field model}. {\it Computer Methods in Applied Mechanics
  and Engineering} 2008\string; 197(49-50)\string: 4333--4352.

\bibitem{vignals}
Vignal P, Collier N, Dalcin L, Brown DL, Calo VM. {An energy-stable
  time-integrator for phase-field models}. {\it Computer Methods in Applied
  Mechanics and Engineering} 2017\string; 316\string: 1179--1214.

\bibitem{wodo}
Wodo O, Ganapathysubramanian B. {Computationally efficient solution to the
  Cahn-Hilliard equation: Adaptive implicit time schemes, mesh sensitivity
  analysis and the 3D isoperimetric problem}. {\it Journal of Computational
  Physics} 2011\string; 230(15)\string: 6037-6060.

\bibitem{stogner}
Stogner RH, Carey GF, Murray BT. {Approximation of Cahn-Hilliard diffuse
  interface models using parallel adaptive mesh refinement and coarsening with
  C1 elements}. {\it International Journal for Numerical Methods in
  Engineering} 2008\string; 76(5)\string: 636--661.

\bibitem{Parsons}
Parsons Q. {\it Numerical Approximation of the Ohta-Kawasaki Functional}.
  {M.Sc.} thesis. Kellogg College, University of Oxford, UK;  2012.

\bibitem{Gal2018}
Gal CG. {Doubly nonlocal Cahn–Hilliard equations}. {\it Annales de l'Institut
  Henri Poincare (C) Analyse Non Lineaire} 2018\string; 35(2)\string: 357-392.

\bibitem{Li2018}
Li RX, Liang ZZ, Zhang GF, Liao LD, Zhang L. {A note on preconditioner for the
  Ohta–Kawasaki equation}. {\it Applied Mathematics Letters} 2018\string;
  85\string: 132-138.
\newblock \href {\doibase 10.1016/j.aml.2018.06.006} {doi:
  10.1016/j.aml.2018.06.006}

\bibitem{Choksi2008}
Choksi R. {Nonlocal Cahn–Hilliard and isoperimetric problems: Periodic phase
  separation induced by competing long- and short-term interactions}. {\it CRM
  Proceedings and Lecture Notes: Singularities in PDE and the Calculus of
  Variations, American Math. Society} 2006\string: 33--45.

\bibitem{elliott3}
Elliott CM, Songmu Z. {On the Cahn-Hilliard equation}. {\it Archive for
  Rational Mechanics and Analysis} 1986\string; 96(4)\string: 339-357.

\bibitem{eyre}
Eyre DJ. {Unconditionally Gradient Stable Time Marching the Cahn-Hilliard
  Equation}. {\it MRS Proceedings} 1998\string; 529(39).

\bibitem{elliot}
Elliott CM, Stuart AM. {The global dynamics of discrete semilinear parabolic
  equations}. {\it SIAM Journal on Numerical Analysis} 1993\string;
  30(6)\string: 1622-1663.

\bibitem{elliot2}
Elliot CM, French DA, Milner FA. {A second order splitting method for the
  Cahn-Hilliard equation}. {\it Numerische Mathematik} 1989\string;
  54(5)\string: 575-590.

\bibitem{he}
He Y, Liu Y, Tang T. {On large time-stepping methods for the Cahn-Hilliard
  equation}. {\it Applied Numerical Mathematics} 2007\string; 57(5-7)\string:
  616-628.

\bibitem{wells}
Wells GN, Kuhl E, Garikipati K. {A discontinuous Galerkin method for the
  Cahn-Hilliard equation}. {\it Journal of Computational Physics} 2006\string;
  218(2)\string: 860-877.

\bibitem{Li2020}
Li RX, Zhang GF, Liang ZZ. {Fast solver of optimal control problems constrained
  by Ohta-Kawasaki equations}. {\it Numerical Algorithms} 2020\string;
  85\string: 132-138.
\newblock \href {\doibase 10.1007/s11075-019-00837-0} {doi:
  10.1007/s11075-019-00837-0}

\bibitem{fenicsbook}
Logg A, Mardal KA, Wells GN. {\it Automated Solution of Differential Equations
  by the Finite Element Method}.
\newblock Springer .
\newblock 2012.

\bibitem{fenics2}
Aln{\ae}s MS, Blechta J, Hake J, et al. {The FEniCS Project Version 1.5}. {\it
  Archive of Numerical Software} 2015\string; 3(100).

\bibitem{gomezzee}
Gomez H, {van der Zee} KG. Encyclopedia of Computational Mechanics Second
  Edition. {\it Computational Phase-Field Modeling} 2017\string: 1-35.

\bibitem{eyre97}
Eyre DJ. {An Unconditionally Stable One-Step Scheme for Gradient Systems}. {\it
  Unpublished article} 1997.

\bibitem{calo2020}
{Calo} V, Minev P, Puzyrev V. {Splitting schemes for phase-field models}. {\it
  {Applied Numerical Mathematics}} 2020\string; 156\string: 192-209.

\bibitem{guillen}
Guill{\'{e}}n-Gonz{\'{a}}lez F, Tierra G. {Second order schemes and time-step
  adaptivity for Allen-Cahn and Cahn-Hilliard models}. {\it Computers and
  Mathematics with Applications} 2014\string; 68(8)\string: 821-846.

\bibitem{zhang}
Zhang Z, Qiao Z. {An adaptive time-stepping strategy for the cahn-hilliard
  equation}. {\it Communications in Computational Physics} 2012\string;
  11(4)\string: 1261-1278.

\bibitem{soderlind1}
S{\"{o}}derlind G. {Automatic control and adaptive time-stepping}. {\it
  Numerical Algorithms} 2002\string; 31(1-4)\string: 281-310.

\bibitem{soderlind2}
S{\"{o}}derlind G. {Digital filters in adaptive time-stepping}. {\it ACM
  Transactions on Mathematical Software} 2003\string; 29(1)\string: 1-26.

\bibitem{soderlind3}
S{\"{o}}derlind G. {Time-step selection algorithms: Adaptivity, control, and
  signal processing}. {\it Applied Numerical Mathematics} 2006\string;
  56(3-4)\string: 488-502.

\bibitem{hairer}
Hairer E, N{\o}rsett S, Wanner G. {\it Solving Ordinary Differential Equations
  {I} Nonstiff Problems}.
\newblock Berlin: Springer-Verlag Berlin Heidelberg .
\newblock 1993.

\bibitem{hairer2}
Hairer E, Wanner G. {\it Solving Ordinary Differential Equations {II} Stiff and
  Differential-Algebraic Problems}.
\newblock Berlin: Springer-Verlag Berlin Heidelberg .
\newblock 1996.

\bibitem{valli}
Valli AMP, Carey GF, Coutinho ALGA. {Control strategies for timestep selection
  in finite element simulation of incompressible flows and coupled
  reaction-convection-diffusion processes}. {\it International Journal for
  Numerical Methods in Fluids} 2005\string; 47(3)\string: 201-231.

\bibitem{ahmed}
Ahmed N, John V. Adaptive time step control for higher order variational time
  discretizations applied to convection-diffusion-reaction equations. {\it
  Computer Methods in Applied Mechanics and Engineering} 2015\string;
  285\string: 83-101.

\bibitem{Cross1993}
Cross MC, Hohenberg PC. {Pattern formation outside of equilibrium}. {\it
  Reviews of Modern Physics} 1993\string; 65(3)\string: 851-1112.
\newblock \href {\doibase 10.1103/RevModPhys.65.851} {doi:
  10.1103/RevModPhys.65.851}

\bibitem{Swift1977}
Swift J, Hohenberg PC. {Hydrodynamic fluctuations at the convective
  instability}. {\it Physical Review A} 1977\string; 15(1)\string: 319-328.
\newblock \href {\doibase 10.1103/PhysRevA.15.319} {doi:
  10.1103/PhysRevA.15.319}

\bibitem{Khandpur1995}
Khandpur AK, Foerster S, Bates FS, et al. Polyisoprene-Polystyrene Diblock
  Copolymer Phase Diagram near the Order-Disorder Transition. {\it
  Macromolecules} 1995\string; 28(26)\string: 8796-8806.
\newblock \href {\doibase 10.1021/ma00130a012} {doi: 10.1021/ma00130a012}

\end{thebibliography}

\end{document}